\documentclass{article}

\pdfoutput=1

\usepackage{amsmath, amssymb, bm, bbm} 

\usepackage{graphicx} 
\usepackage{wrapfig} 
\usepackage[labelfont=bf]{caption}
\usepackage[labelformat=simple]{subcaption} 
\usepackage{placeins}

\usepackage[margin=1in]{geometry}
\usepackage{authblk} 
\usepackage{booktabs}
\usepackage{microtype}
\usepackage{hyperref}
\usepackage{cleveref}
\usepackage[T1]{fontenc}

\newcommand{\myemph}[1]{\emph{#1}}
\DeclareMathOperator*{\argmax}{arg\,max}
\newcommand*{\horzbar}{\rule[.5ex]{2.5ex}{0.5pt}} 

\def\iw{3.8in}
\def\thone{1.2in}
\def\bhone{2.2in}
\def\thtwo{1.6in}
\def\tradeoffheadingone{{\footnotesize \hspace{22mm} \textbf{ITR} \hspace{33mm} \textbf{Error-free accuracy} \hspace{25mm} \textbf{Absolute deviation}} \vspace{2mm}}
\def\tradeoffheadingtwo{{\footnotesize \hspace{22mm} \textbf{ITR} \hspace{33mm} \textbf{Error-free accuracy} \hspace{25mm} \textbf{Absolute deviation}} \vspace{2mm}}
\def\tradeoffheadingthree{{\footnotesize \hspace{17mm} \textbf{Alphabet accuracy} \hspace{24mm} \textbf{Alphabet deviation} \hspace{14mm} \textbf{Normalized alphabet deviation}} \vspace{2mm}}

\graphicspath{{./images/}}

\title{A Low-complexity Brain-computer Interface for High-complexity Robot Swarm Control}

\date{}

\author[1,2]{Gregory Canal}
\author[1,3]{Yancy Diaz-Mercado}
\author[4]{Magnus Egerstedt}
\author[1]{Christopher Rozell}
\affil[1]{School of Electrical and Computer Engineering\protect\\
Georgia Institute of Technology, Atlanta, GA, 30332\protect\\
\texttt{crozell@gatech.edu}}
\affil[2]{(Current) University of Wisconsin-Madison, Madison, WI, 53715\protect\\
	\texttt{gcanal@wisc.edu}}
\affil[3]{(Current) Department of Mechanical Engineering\protect\\
	University of Maryland, College Park, MD, 20742\protect\\
	\texttt{yancy@umd.edu}}
\affil[4]{Department of Electrical Engineering and Computer Science\protect\\
	University of California, Irvine, Irvine, CA 92697\protect\\
	\texttt{magnus@uci.edu}}

\begin{document}

\maketitle

\begin{abstract}
	A brain-computer interface (BCI) is a system that allows a human operator to use only mental commands in controlling end effectors that interact with the world around them \cite{wolpaw2002brain}. Such a system consists of a measurement device to record the human user's brain activity, which is then processed into commands that drive a system end effector. BCIs involve either invasive measurements which allow for high-complexity control but are generally infeasible, or noninvasive measurements which offer lower quality signals but are more practical to use. In general, BCI systems have not been developed that efficiently, robustly, and scalably perform high-complexity control while retaining the practicality of noninvasive measurements. Here we leverage recent results from feedback information theory \cite{shayevitz2011optimal,omar2010feedback} to fill this gap by modeling BCIs as a communications system and deploying a human-implementable interaction algorithm for noninvasive control of a high-complexity robot swarm. We construct a scalable dictionary of robotic behaviors that can be searched simply and efficiently by a BCI user, as we demonstrate through a large-scale user study testing the feasibility of our interaction algorithm, a user test of the full BCI system on (virtual and real) robot swarms, and simulations that verify our results against theoretical models. Our results provide a proof of concept for how a large class of high-complexity effectors (even beyond robotics) can be effectively controlled by a BCI system with low-complexity and noisy inputs.
\end{abstract}

\section{Introduction}

Brain-computer interfaces (BCI) are systems that consist of hardware to measure a human user's brain activity, an interaction algorithm to map the user's mental commands to control signals, and an end effector that the user operates via these control signals. This direct link between brain and effector provides a means for paralyzed users to circumvent muscular pathways and interact with everyday devices \cite{Soekadar2016hybrid} as well as an augmented interface for healthy users. There are several tradeoffs involved in the design of BCIs, including whether measurements are taken invasively or noninvasively, how many mental commands are needed to drive the effector to a desired behavior, how scalable the system is to effectors of varying complexity, and how robust the system is to user error and system noise in measurement processing. Although BCIs with invasive neural measurements have had experimental success in controlling high-complexity effectors (e.g., robotic arms \cite{collinger2013high, hochberg2012reach, hochberg2006neuronal, wodlinger2014ten}) with many degrees of freedom, such BCIs are only available in research settings and require a surgical procedure for electrode implantation. BCIs with noninvasive measurements (e.g., scalp electrode recordings via an electroencephalogram (EEG)) are more widely implementable due to their relative ease of use and lower cost, but are limited to controlling comparatively simpler effectors (e.g., basic wheelchair control \cite{leeb2007self,galan2008brain,iturrate2009noninvasive,rebsamen2007controlling,huang2012electroencephalography,li2013hybrid,long2012hybrid,li2013design,carlson2013brain,zhang2016control,muller2010brain,rebsamen2010brain}, cursor control \cite{blankertz2007non,wolpaw2004control,xia2013mental,mcfarland2010electroencephalographic,long2012target,trejo2006brain}) with few degrees of freedom due to lower signal-to-noise ratios. In recent years there has been an emerging interest in improving these tradeoffs for neurotechnology in commercial and clinical applications, with aims to both broaden intended uses and engineer higher quality BCI devices \cite{imagining2020, musk2019integrated, nonsurgical2018, pratt2021kernel}.

Despite this increased interest, there remains a large gap between the complexity of potential end effectors and the capabilities of interaction algorithms that map the user's mental commands from \emph{noninvasive} interfaces to control signals. There are several specifications required for a noninvasive interaction algorithm to meet this need. First and foremost, any such algorithm must be implementable by humans through mental commands easily learned with training. Furthermore, such interaction algorithms must be scalable so that they remain tractable with a minimal increase in user overhead when controlling more complex effectors. Similarly, increases in effector complexity should not result in a need for increased measurement capabilities (e.g., additional EEG features). Because even the most advanced BCI measurements are susceptible to errors, an interaction algorithm must be robust to such errors. Finally, due to the wide range of applications that can benefit from BCIs, an ideal interaction algorithm should be designed for general use and be easily adaptable to a variety of specific tasks.

Currently, interaction algorithms for noninvasive BCIs fall into two broad categories that only achieve a subset of these specifications. In the first category, the user selects discrete effector behaviors from a finite set of options displayed on an interface, such as choosing waypoints for a motorized wheelchair  \cite{rebsamen2010brain} or selecting letters on a virtual keyboard  \cite{scherer2004asynch,birbaumer1999spelling,kubler2009brain,cheng2002design}. Although this type of interaction is easy to use, it scales poorly since it becomes increasingly tedious for a user to select their desired behavior as the number of options (i.e., the precision) increases. In the second category, continuous features from measured brain activity are directly mapped to continuous control over effector action spaces with arbitrary precision (e.g., robotic arm control \cite{meng2016noninvasive, edelman2019noninvasive}, quadcopter control \cite{lafleur2013quadcopter}, cursor control in up to three-dimensional space  \cite{blankertz2007non,wolpaw2004control, xia2013mental, mcfarland2010electroencephalographic,long2012target,trejo2006brain}). Unlike discrete selection, continuous control allows a user to navigate an effector’s action space with arbitrary precision in a scalable method. However, this type of interaction is severely limited in that each additional effector degree of freedom requires an independent, continuous measurement feature, which scales poorly and typically limits an effector to a maximum of three degrees of freedom for EEG-based BCIs.

The main contribution of this paper is a robust interaction algorithm that reaps the benefits of both discrete selection and continuous control while addressing the critical disadvantages of each. The key innovation of our information-theoretic approach is that each new input is used in conjunction with closed-loop feedback to the user to efficiently refine the entire effector state simultaneously through a sequence of simple and tractable decisions. While our approach builds off of techniques used in prior work \cite{omar2010feedback,akce2010remote,tantiongloc2017information, akce2013brain}, it utilizes a new method for effector parameterization that significantly expands the class of controllable systems. We test our approach on human control of a mobile robot swarm (a large collection of robots, as depicted in \Cref{subfig:gritsbotswarm}), where a human operator issues high-level, global commands which are executed by the swarm in a distributed fashion (individual robot depicted in \Cref{subfig:gritsbotind}).

Robot swarm control serves as an ideal testbed for our approach, since robot swarms are high-complexity cyber-physical systems that can be naturally parameterized beyond three degrees of freedom and have been previously tested in a BCI setting \cite{karavas2017hybrid}. Part of what makes robot swarm control complex is the necessity to coordinate the individual robot motion to avoid collisions while attempting to achieve their objectives, e.g., reach a target formation. Robot swarms typically consist of weak robots which possess limited computation, sensing, and communication capabilities. Thus, in order to achieve the desired behavior, the control must rely on local sensing information and scale well in complexity with the number of robots in the swarm. These local rules result in the desired global emergent behavior. When humans are involved, the swarm formation must be achieved quickly and be cohesive enough to provide the human operator with clear visual feedback to aid in the decision-making.

Over the last couple of decades, there have been many developments in large classes of coordination algorithms and abstractions that support the required mapping from low-complexity, high-level commands to highly complex coordinated swarm behaviors~\cite{mesbahi2010graph}. Recent advances in coverage control \cite{diazmercado2015distributed,xu2020multi} provide an excellent approach to perform this mapping for formation control. The algorithms allow for a human operator to broadcast reference swarm spatial densities and boundaries in the robot domain that encode desired formations. The robots in the domain can then coordinate their motion with nearby robots to robustly achieve the commanded density distributions in real time in a scalable, distributed manner. In this paper, we show through an array of human trials and simulations that refining the entire state space is an effective approach for BCI swarm control, thereby demonstrating the potential and flexibility of our method for controlling high-complexity end effectors with low-complexity inputs.

\section{Refining End Effector Behavior}

To understand our interaction algorithm at a high-level, first consider the task of finding a word in an English dictionary. A natural strategy is for the user to repeatedly bisect the remaining pages depending on whether their desired word comes before or after the current page. Our interaction algorithm is analogous to this efficient search procedure: the BCI user selects an effector behavior from an ordered dictionary of candidate behaviors through a sequence of bisections. Specifically, suppose that the BCI user learns a lexicographical ordering rule for the set of effector behaviors, which determines a total order of behaviors organized as a dictionary. At each round of interaction, the effector presents to the user the behavior that bisects the remainder of the dictionary. The user indicates to the effector (via  a binary mental command) if their desired behavior precedes or succeeds the candidate behavior, and the dictionary scope is narrowed based on their reply. Rather than strict elimination of half of the dictionary at each step, the algorithm uses a probabilistic weighting over the dictionary to  account for possible noise in the user's input (see \Cref{sec:PMimp}). Eventually, the user will have provided enough refinements for the end effector to correctly converge to the user's desired behavior. Importantly, this procedure does not involve the adjustment of individual effector parameters, but instead only requires the user to decide on the precedence of their total desired behavior with respect to the current candidate behavior. Although each dictionary bisection affects every effector parameter, the user only has to make a simple binary decision at each round, regardless of the number of effector parameters; this is distinct from a brute-force approach where the user adjusts each parameter individually.

While this interaction algorithm is intuitively satisfying, it is also endowed with rigorous performance guarantees that become apparent when the entire interface is framed as a feedback communications system: the human user acts as a ``transmitter'' by encoding their desired effector behavior (the ``message'') through a sequence of binary BCI inputs (``codes''). These inputs are sequentially decoded by the end effector to refine a new estimate of the user's desired behavior, which is fully observable to the user as ``noiseless feedback'' and informs the choice of their next input. Because there is some chance that the user's binary input will be misclassified or that the user will make a decision error, the sequence of classification results can be modeled as outputs of a noisy \myemph{binary symmetric channel} (BSC) with a \myemph{crossover probability} equal to the misclassification probability. When framed as such a communications system, our interaction algorithm is mathematically equivalent to the \myemph{posterior matching} coding scheme \cite{shayevitz2011optimal}. Posterior matching is an optimal capacity-achieving code \cite{cover2006elements}, meaning that this interaction algorithm communicates the user's desired behavior to the effector with as few binary inputs as possible for a given error rate.

In previous work, posterior matching has been used as an interaction algorithm in noninvasive BCIs for tasks such as text entry or vehicle path planning \cite{omar2010feedback,akce2010remote,tantiongloc2017information, akce2013brain}. In these cases, a dictionary of ordered effector behaviors can be formed by constructing each dictionary element, or \emph{string}, as a concatenation of \emph{characters} from a fixed \emph{alphabet}. For example, in text entry and path planning a string is constructed as a concatenation of English language letters and arc segments, respectively. In either of these cases, the precedence between two strings can be determined by identifying the first character that differs between the strings (referred to here as the \emph{critical character}), and assigning precedence to the string whose critical character comes earliest in the character alphabet (e.g., `a' precedes `z,' arcs angled left precede arcs angled right). We refer to such dictionaries as \emph{homogeneous} since in each case a single alphabet is used for all character positions in the behavior string. Tasks such as text entry or path planning can be adequately modeled by homogeneous dictionaries, since each additional effector parameter (e.g., letter or arc segment) is of the same type.

Unlike the tasks described above, many high-complexity effectors cannot be described with homogeneous dictionaries by concatenating characters from a single alphabet. For example, in robot swarm control, each swarm configuration is characterized by varied parameters describing position, shape, and size. To model these high-complexity effectors we design a \emph{heterogeneous} dictionary, where a different alphabet is used for each character position in the behavior string. To our knowledge, posterior matching has not been deployed as an interaction algorithm using heterogeneous dictionaries, and it was previously unknown if BCI users can successfully learn and apply a heterogeneous dictionary to posterior matching control of a high-complexity effector. As we detail below, we demonstrate in a large-scale interface study that people can learn such a heterogeneous dictionary with little training and make pairwise string comparisons with high proficiency.

\begin{figure}[htb]
	\def\lc{0.6}
	\def\rc{0.4}
	\def\robw{2in}
	\def\robspace{5mm}
	\centering
	\begin{minipage}{\lc\textwidth}
		\centering
		\begin{subfigure}[t]{\textwidth}
			\centering
			\includegraphics[width=\iw]{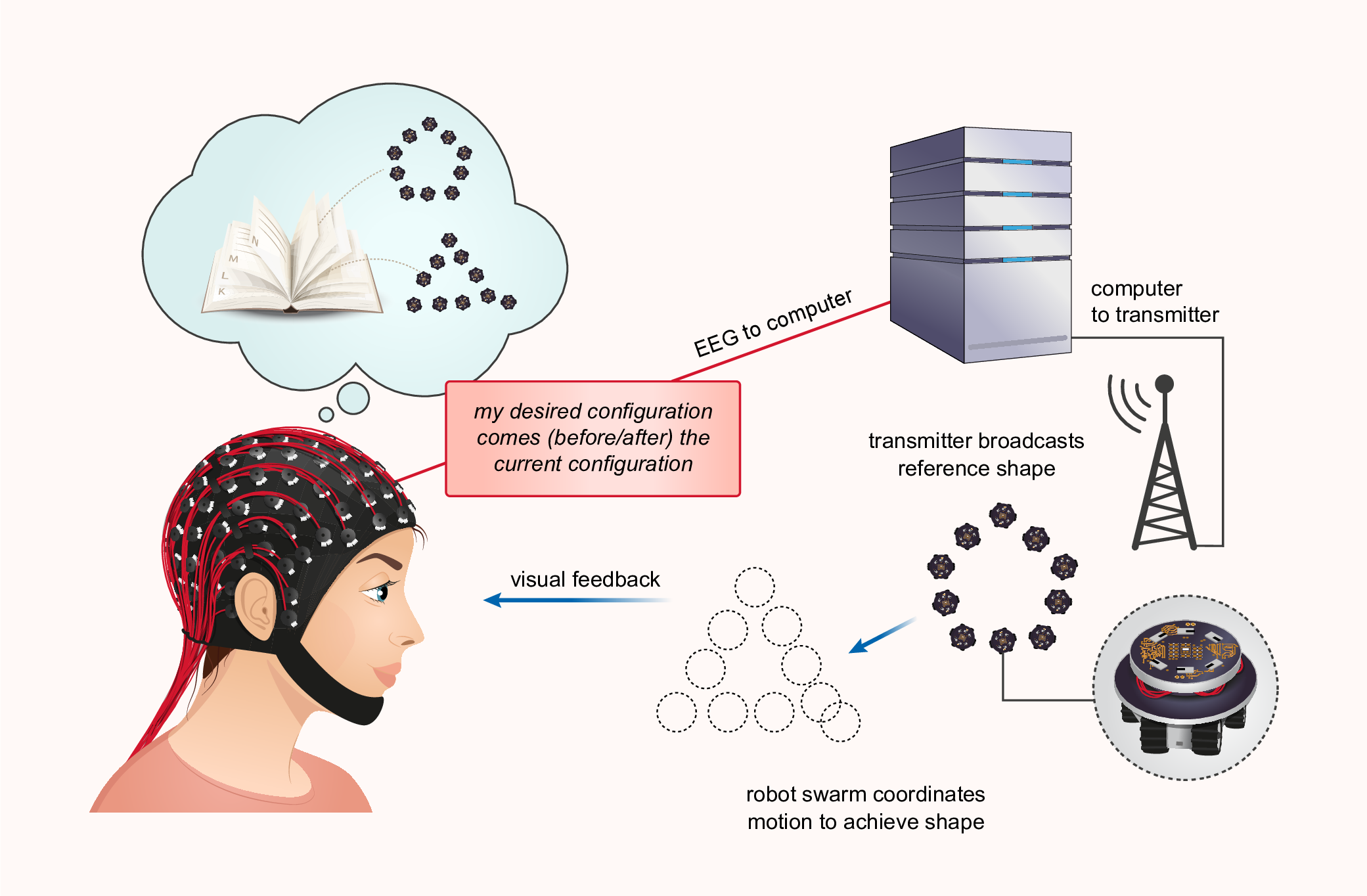}
			\caption{}
			\label{subfig:helmet}
		\end{subfigure}%
		\\
		\begin{subfigure}[t]{\textwidth}
			\centering
			\includegraphics[width=\iw]{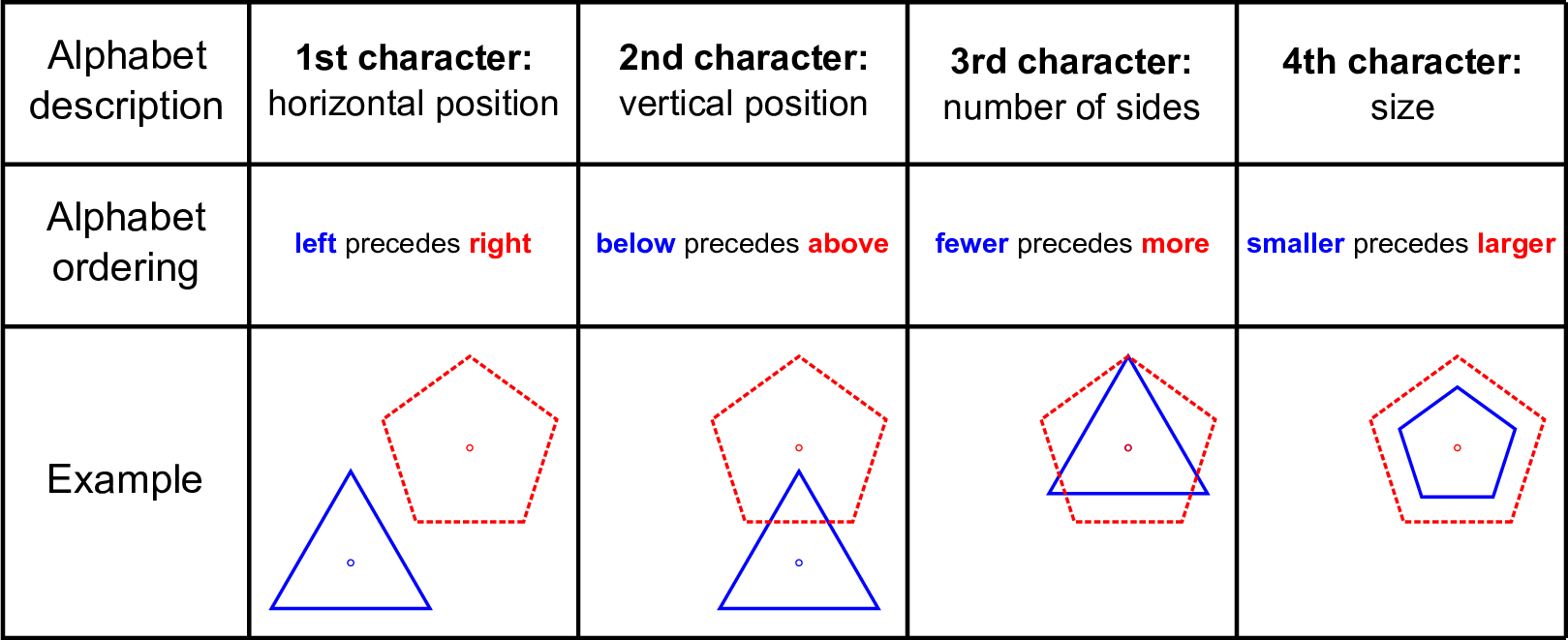}
			\caption{}
			\label{subfig:dictionary}
		\end{subfigure}%
	\end{minipage}%
	\hfill
	\begin{minipage}{\rc\textwidth}
		\centering
		\begin{subfigure}[t]{\textwidth}
			\centering
			\includegraphics[width=\robw]{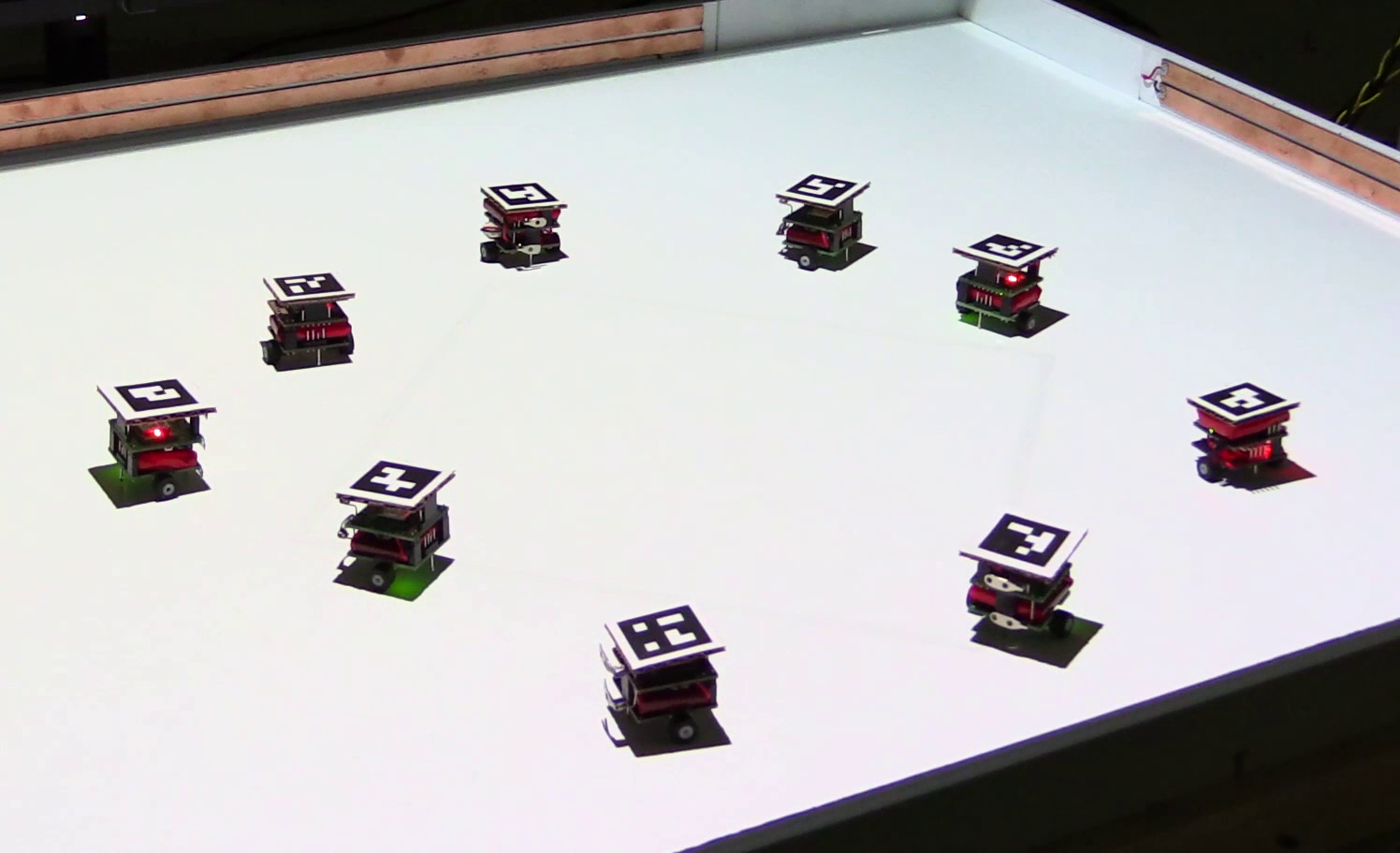}
			\caption{}
			\label{subfig:gritsbotswarm}
		\end{subfigure}%
		\\
		\vspace{\robspace}
		\begin{subfigure}[t]{\textwidth}
			\centering
			\includegraphics[width=\robw]{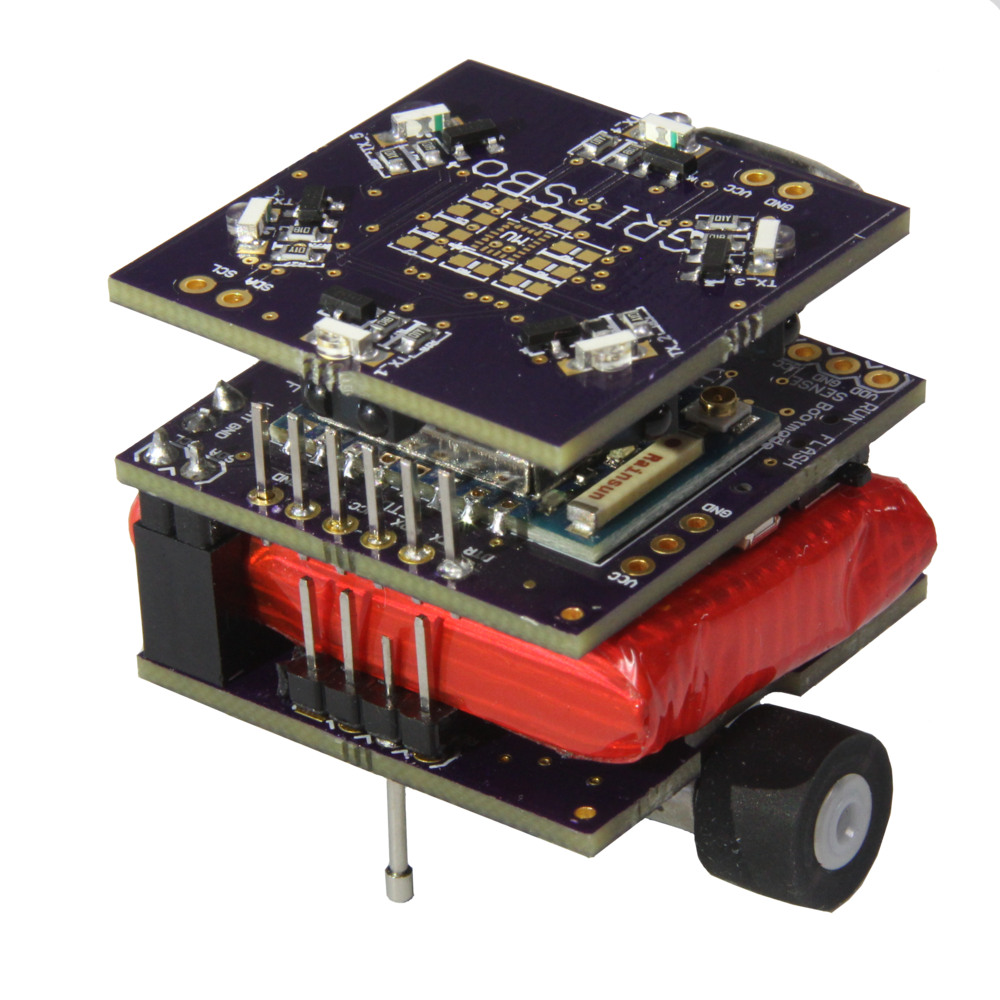}
			\caption{}
			\label{subfig:gritsbotind}
		\end{subfigure}%
	\end{minipage}
	\caption[Refining effector behavior through configuration sorting.]{\textbf{Refining effector behavior through configuration sorting.} Effector behavior is determined through iterative refinement from the BCI user.
		\textbf{a,} In the example of robot swarm configuration refinement, the BCI user indicates through a mental command (e.g., binary motor imagery) if their desired configuration comes before or after the current configuration in the swarm dictionary (see \textbf{b}). A computer decodes the input by classifying scalp electrode recordings from the user, and updates a posterior distribution over the configuration dictionary. The median of the updated distribution is selected as a new configuration guess and transmitted to the swarm through a global update. Each individual robot then adjusts its position locally so that the overall configuration conforms to the new guess in a distributed manner.
		\textbf{b,} In the swarm configuration dictionary, character alphabets are defined in order from first to last as follows, with alphabet precedence in parentheses: horizontal position of the configuration center (centers to the left preceding centers to the right); vertical position of the configuration center (centers below preceding centers above); number of sides (fewer sides preceding more sides); configuration size as the radius from the center to each vertex (smaller radii preceding larger radii). Each example panel depicts a pair of strings whose critical character corresponds to the panel column, with blue (solid) configurations preceding red (dashed) configurations in the overall dictionary ordering.
		\textbf{c,} Example of a robot swarm coordinating to form a globally specified configuration by using only local information. 
		\textbf{d,} Close-up view of an individual mobile robot used in our demonstrations. A robot swarm (as in \textbf{c}) consists of several such robots collectively performing global actions in a distributed manner.}
	\label{fig:dictionary_overall}
\end{figure}

While one might conceive of a variety of heterogeneous dictionaries to describe swarm configurations, here we adopt a dictionary of regular polygons as a proof of concept. Each polygon string is parameterized by characters including horizontal position, vertical position, number of sides, and size, with distinct alphabets for each character position (\Cref{subfig:dictionary}). To search this polygon dictionary with posterior matching, the BCI user issues hand motor imagery (MI) inputs detected via EEG measurements to indicate if the desired behavior comes before or after the currently demonstrated behavior in the dictionary. MI tasks are a well-studied and popular binary input modality where the user mentally visualizes wrist flexions of either their left or right hand and the resulting changes in EEG frequencies are detected by a binary classifier \cite{pfurtscheller1997motor, blankertz2008optimizing}. To refine the swarm, the user determines the first character where their desired configuration differs from the current configuration and issues a left-hand (right-hand) MI input if their desired polygon preceded (succeeds) the current polygon at the critical character. As the complexity of the dictionary increases, the sequential scan to find the critical character may take marginally more time, but the decision by the user is ultimately based only on a simple evaluation of that character (despite each user input potentially updating all characters). Note that this approach is not limited to EEG-based MI, and is compatible with any binary input mechanism including inputs detected by invasive BCIs. We refer to this combination of a heterogeneous swarm dictionary with binary input posterior matching as SCINET: Swarm Control via Interactive Neural Teleoperation (illustrated in \Cref{subfig:helmet}).

\subsection{Dictionary Construction}
\label{sec:suppdict}
We constructed the swarm dictionary with the following characters in each configuration string, in order of character precedence: horizontal position, vertical position, number of sides, and size. Horizontal position and vertical position refer to the coordinates of the center of each polygon, respectively (see \Cref{fig:polyparam}). Size refers to the distance between the polygon center and each vertex (this value is the same for each vertex since the polygons are regular). The number of characters in each alphabet is as follows: 5 horizontal positions; 2 vertical positions; 3 numbers of sides; and 2 polygon sizes. Characters in the horizontal position alphabet were chosen to uniformly span the robot arena (virtual or physical), as were the characters in the vertical position alphabet. The ``number of sides'' alphabet has characters given by 3, 4, or 5 sides, with the polygon rotation set by fixing a vertex at the ``12 o'clock'' position of each shape. The two size distances were tuned such that size differences were visually discernible, while not causing robots to overflow outside the span of the arena. With an arena of width 1.5 and height 1 (specified in units relative to the arena height), these specifications translate to the following alphabet characters: horizontal position $\{0.4,0.575,0.75,0.925,1.10\}$; vertical position $\{0.4,0.6\}$; number of sides $\{3,4,5\}$; size $\{0.3,0.4\}$. These values (except for number of sides) are specified in abstract units relative to the arena height, and are scaled at runtime to the physical dimensions of the actual swarm arena; for instance, if the physical swarm arena is 2.5 feet in height, then the first horizontal position character is $0.4 \times 2.5 = 1$ foot from the left arena edge. In total, this combination of alphabets produces a dictionary with $5 \times 2 \times 3 \times 2 = 60$ total possible polygons, and hence 60 possible swarm behaviors.

\FloatBarrier

\section{Dictionary Sorting Proficiency}
Although in theory SCINET is capable of controlling an arbitrary number of degrees of freedom (i.e., string characters), this scalability is limited in practice by the ability and ease by which the BCI operator can sort strings according to the swarm dictionary ordering. To be a useful approach, a typical human user must be able to quickly learn the swarm dictionary and subsequently sort any pair of strings, with high proficiency when the critical character is located at any position in the string. To evaluate these user capabilities in an isolated manner from the rest of the BCI system, we conducted a user study where participants ($n=150$) used a point-and-click interface to select between configurations on a screen. Each participant was first presented with a set of graphical and text instructions explaining the polygon dictionary ordering and how to use it to sort a given string pair. Each participant was then presented with 150 randomly selected shape pairs from the dictionary (\Cref{subfig:pair_example}), and asked to indicate which shape precedes the other in the dictionary ordering. We provided each participant with a visual aid to use as a reference during the task (\Cref{subfig:info_graphic}); such an aid could also be presented to a BCI operator in a practical setting.

Overall, participants were able to sort shape pairs with high accuracy. When evaluating sorting accuracy over all pairs of strings (\Cref{subfig:acc_overall}), most subjects sorted with nearly perfect accuracy (median 99.3\% accuracy). Furthermore, response accuracy does not appear to decrease as the position of the critical character appears later in the string (median 100\% accuracy for all characters, see \Cref{subfig:acc_crit}). When evaluating critical character performance for each individual participant, we also find that most participants exhibit non-decreasing or only modestly decreasing performance as character depth increases (\Cref{fig:userregression}). Although a user's capacity for learning and memorizing a dictionary ordering may create a performance bottleneck, this can be mitigated by providing users with a mnemonic aid to assist in recalling the ordering, as was done in our study. These results suggest that users can rapidly learn and apply string sorting in our heterogeneous dictionary, and that adding more characters (i.e., effector parameters) does not hinder a user's ability to effectively compare pairs of strings across multiple parameters simultaneously.

\subsection{Online User Study Details}
\begin{figure}[htb]
	\def\th{\thone}
	\def\bh{\bhone}
	\centering
	\begin{subfigure}[t]{0.5\textwidth}
		\centering
		\includegraphics[height=\th]{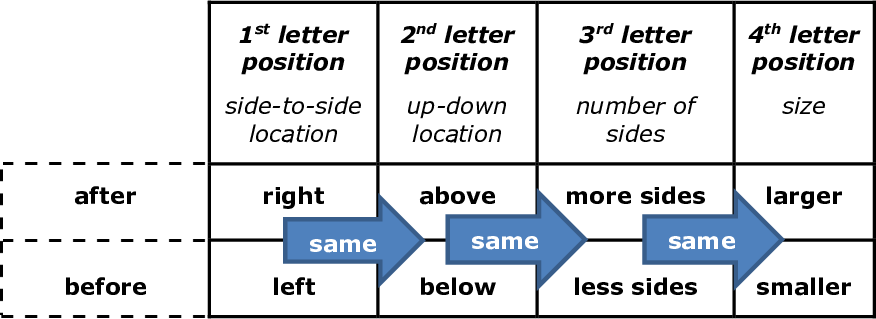}
		\caption{}
		\label{subfig:info_graphic}
	\end{subfigure}%
	\hfill
	\begin{subfigure}[t]{0.5\textwidth}
		\centering
		\includegraphics[height=\th]{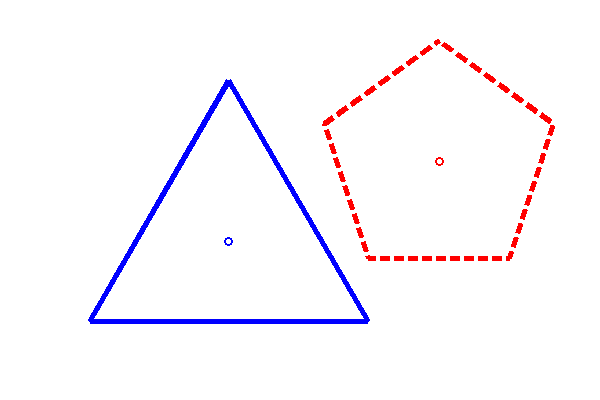}
		\caption{}
		\label{subfig:pair_example}
	\end{subfigure}%
	\\
	\begin{subfigure}[t]{0.5\textwidth}
		\centering
		\includegraphics[height=\bh]{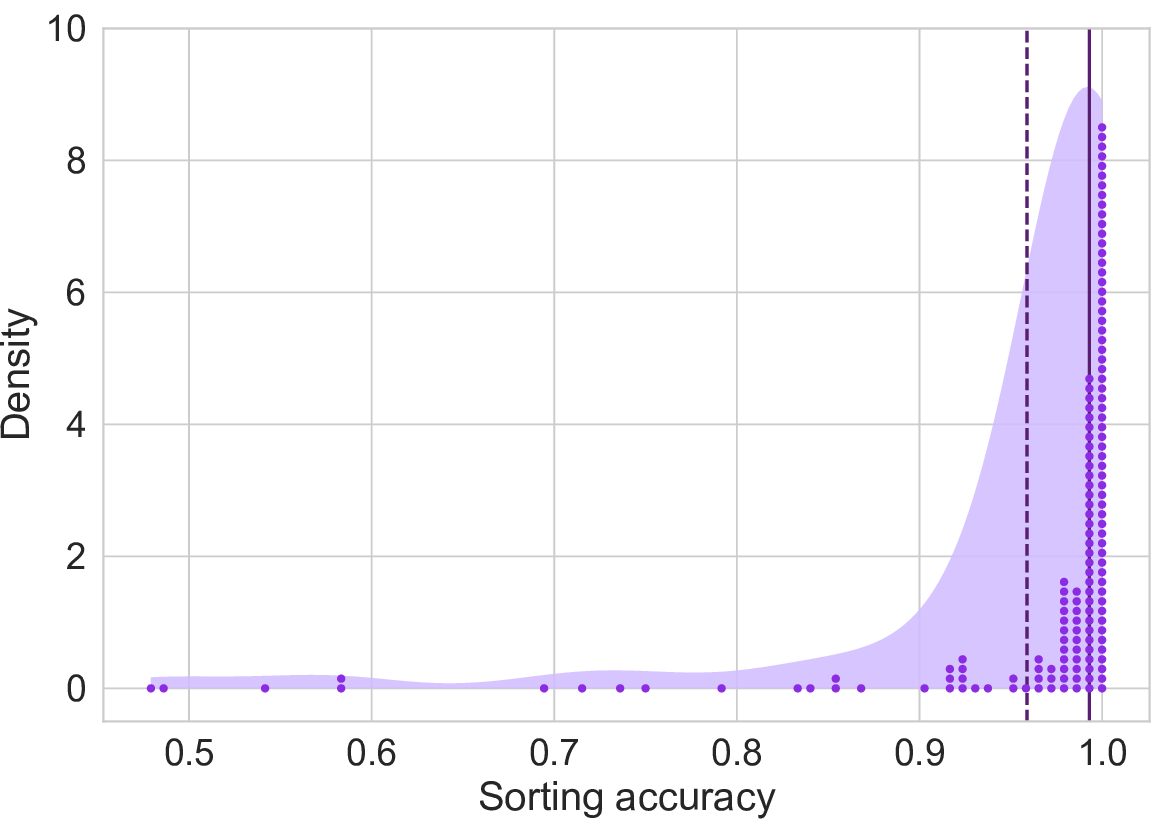}
		\caption{}
		\label{subfig:acc_overall}
	\end{subfigure}%
	\hfill
	\begin{subfigure}[t]{0.5\textwidth}
		\centering
		\includegraphics[height=\bh]{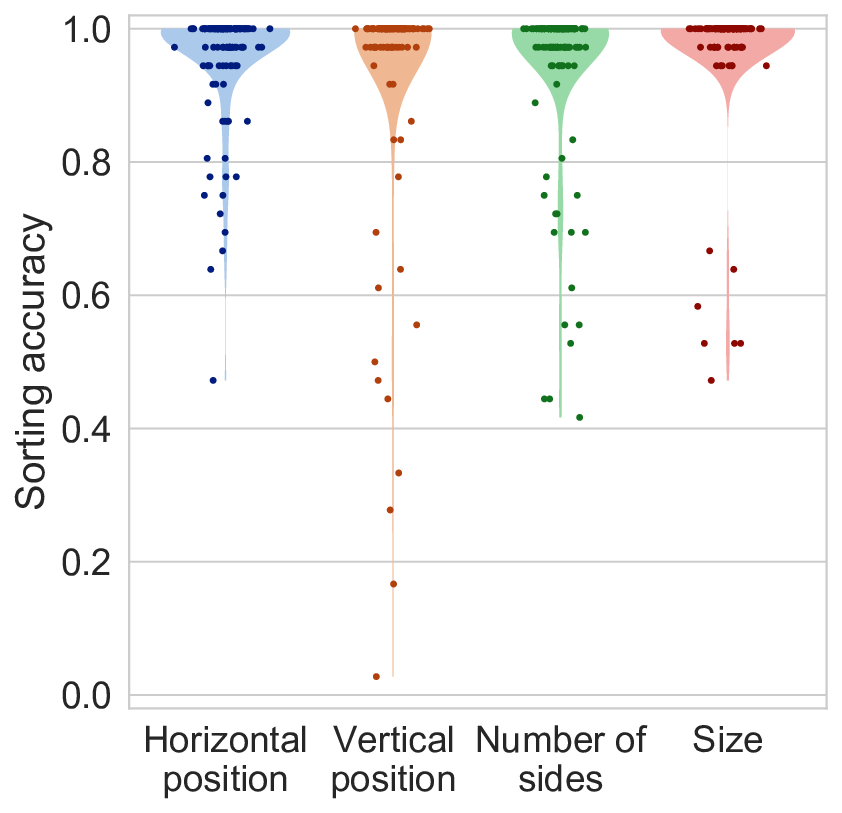}
		\caption{}
		\label{subfig:acc_crit}
	\end{subfigure}%
	\caption[Evaluating configuration sorting proficiency in a user study.]{\textbf{Evaluating configuration sorting proficiency in a user study.}
		\textbf{a,} Mnemonic aid for dictionary ordering recall using plain language, provided to each user study participant ($n=150$).
		\textbf{b,} Example shape pair presented to each participant; in this case, the correct sorting is that the blue (solid) triangle precedes the red (dashed) pentagon, since horizontal position is the critical character and the triangle's center is located further to the left.
		\textbf{c,} Estimated distribution of overall dictionary sorting accuracy across all participants, where each participant is represented by a dot. The vast majority of participants were able to correctly sort configurations with high accuracy, with a mean accuracy (dashed vertical) of 95.8\% and a median accuracy (solid vertical) of 99.3\%.
		\textbf{d,} Dictionary sorting accuracy for each critical character. Each column shows accuracy across all participants (represented as dots, with added horizontal jitter for visual clarity) when calculated only for queried shape pairs with the respective critical character. As the depth of the critical character position increases, sorting ability does not decrease, as exhibited by a median accuracy of 100\% for all characters. These results support the scalability of sorting heterogeneous dictionary strings as an interaction mechanism, since accuracy does not decrease as additional characters are added to the dictionary.}
	\label{fig:user-study}
\end{figure}

We conducted the online user study\footnote{Protocols for both the online user study (Protocol H16266) and robot control portions (Protocol H10263) were approved by the Georgia Tech Institutional Review Board. Both studies complied with ethical regulations set by the Review Board, including online user study participants providing informed consent.} via Amazon Mechanical Turk\footnote{https://www.mturk.com/} by creating a Human Intelligence Task (HIT) for participant submission. The HIT contained both a set of graphical and text instructions teaching the swarm dictionary to the participant, followed by a set of 150 shape pair queries (see ancillary files for full study instructions). Once a participant accepted a HIT task, they proceeded to read the instructions, answer all queries, and submit their responses. In total, 150 participants were recruited in the study, corresponding to 150 submitted and accepted HITs. Each shape pair query presented a blue, solid shape and a red, dashed shape as in \Cref{subfig:pair_example} (polygon outlines were presented rather than actual swarm configurations), and asked ``For the image below, select whether the test shape (red dashed edges) comes after or before the reference shape (blue solid edges), as defined in the instructions above.'', which the participant responded with ``Before'' or ``After.'' During the study, each participant had access to an informational graphic presented in \Cref{subfig:info_graphic} as a visual aid in recalling the dictionary ordering.

The 150 shape pairs were randomly generated in such a way that the critical character determining the order of each pair was evenly distributed across all four letters. Within this query set, 6 ``cheat detection'' pairs were presented each consisting of two identical triangles with all the same parameters except horizontal position, which is an ``easy'' question and is unlikely to be answered incorrectly unless a participant is randomly selecting answers to finish the study as quickly as possible (\Cref{fig:test_shape}); the participants were not told that these pairs were used for cheating detection. Before approving a participant's HIT submission, we evaluated their responses on these cheat detection queries to assess if they were simply selecting answers at random. The remaining 144 queries were evenly distributed between shape pairs where the horizontal position, vertical position, number of sides, or size was the first character to differ between the two configurations in question (36 shape queries per critical character, resulting in $36 \times 4 + 6 = 150$ total pairs).

To generate a shape pair with the desired critical character (36 pairs for each critical character), a character was randomly generated for each alphabet that precedes the critical character, and set for \emph{both} shapes in the pair. This way, the critical character would in fact be the first character that differs between the shapes in question. Next, two \emph{distinct} characters were randomly selected from the critical character alphabet, one for each shape in the pair. Finally, the remaining characters succeeding the critical character were randomly populated separately for each shape in the pair. All generated shape pair queries (including pairs for cheating detection) were then shuffled into a random order. Although the query order was randomly shuffled, each HIT (and therefore each participant) responded to the same fixed order of queries; in other words, query order was not randomized \emph{between} participants.

To qualify for participation in the study, participants must have had a record of at least 1,000 approved HITs from previous tasks on Mechanical Turk, and must have had an overall HIT approval rate of 95\% or greater at the time of submission. After qualifying participants accepted and completed our HIT, they were automatically approved unless flagged as being suspect of randomly selecting answers, in which case they were manually reviewed. Participants were automatically rejected if they did not answer every query, or if they had already completed the HIT previously. The details of this process are presented in \Cref{sec:cheat}. Each approved participant was paid \$8 for completing all pair orderings, and was awarded a \$4 bonus if they achieved an accuracy of 95\% or higher of correct pair orderings. Overall, 150 participants were recruited, of which all 150 were approved. Of these, 125 achieved an overall response accuracy of over 95\% and so were awarded a \$4 bonus.

The 6 cheat detection queries were omitted during data analysis, resulting in 144 shape pairs analyzed per participant. Overall sorting accuracy was calculated per subject as the fraction of correct responses to these 144 regular queries. Sorting accuracy was calculated per critical character as the fraction of correctly answered queries among the 36 shape pair queries with the respective critical character. Distributions are plotted in \Cref{subfig:acc_overall,subfig:acc_crit} as kernel density estimates.

\FloatBarrier

\section{Full System Evaluation}

Beyond the interaction algorithm, there are a number of additional factors which can affect SCINET performance in the full system. Namely, the user must not only compare the current swarm configuration against their target string in the dictionary ordering, but must then issue a binary input via a mental command and subsequently observe the real-time changes in the swarm's behavior. Due to practical effects such as user fatigue, the user's error in issuing inputs may stray from the theoretical BSC assumed by the posterior matching algorithm. Since posterior matching assumes a fixed BSC crossover probability, it is unclear if non-ideal input statistics will result in poor system performance, and if such effects can be modeled. To evaluate SCINET in practice, we measure accuracy of a physical SCINET implementation against a simulation model that accounts for these practical effects.

\begin{figure}[htb]
	\centering
	\def\th{\thtwo}
	\def\ts{8pt}
	\def\bhl{1.2in}
	\def\bh{1.8in}
	\def\vs{2ex}
	\begin{subfigure}[t]{\textwidth}
		\centering
		\includegraphics[height=\th]{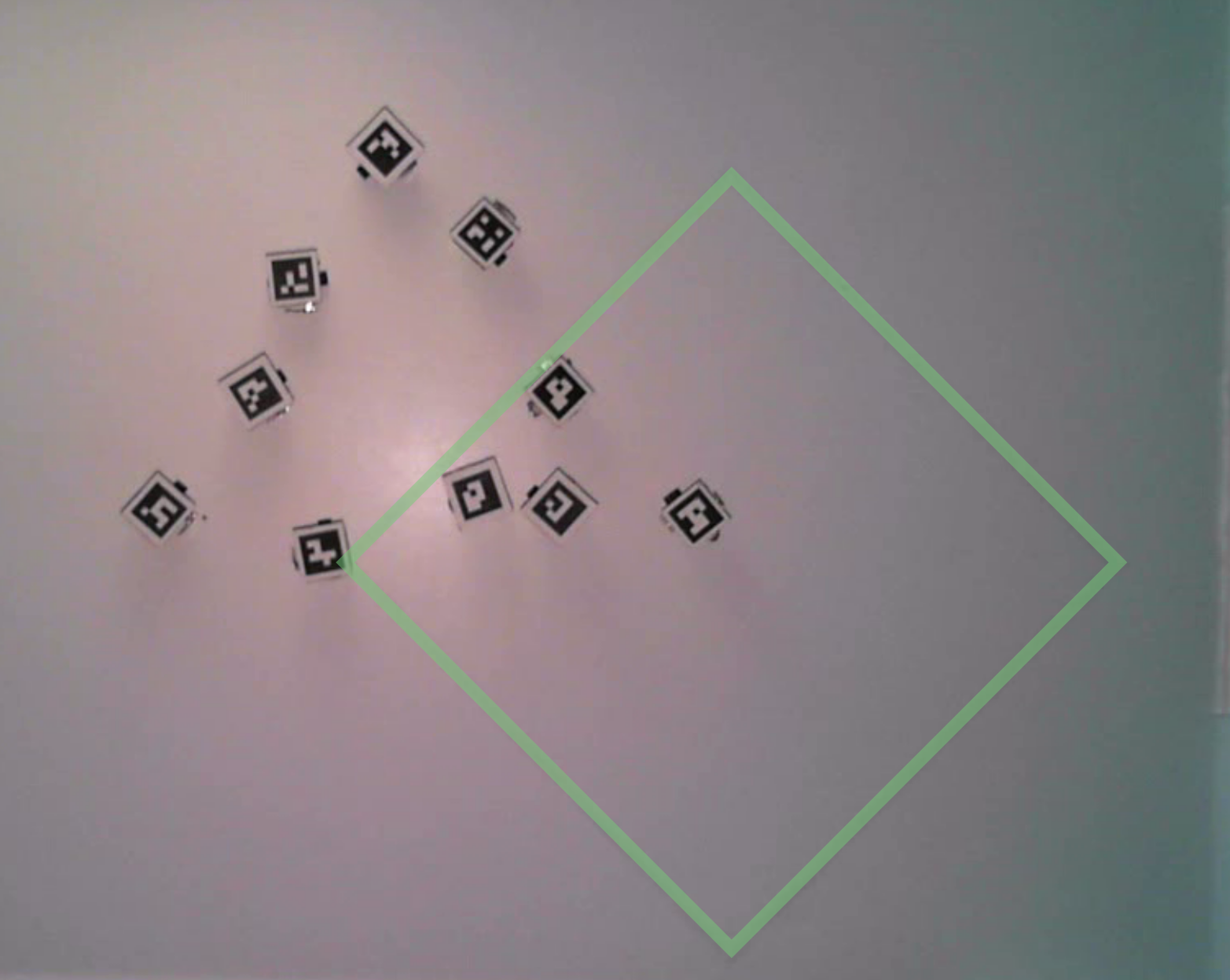}%
		\hspace{\ts}
		\includegraphics[height=\th]{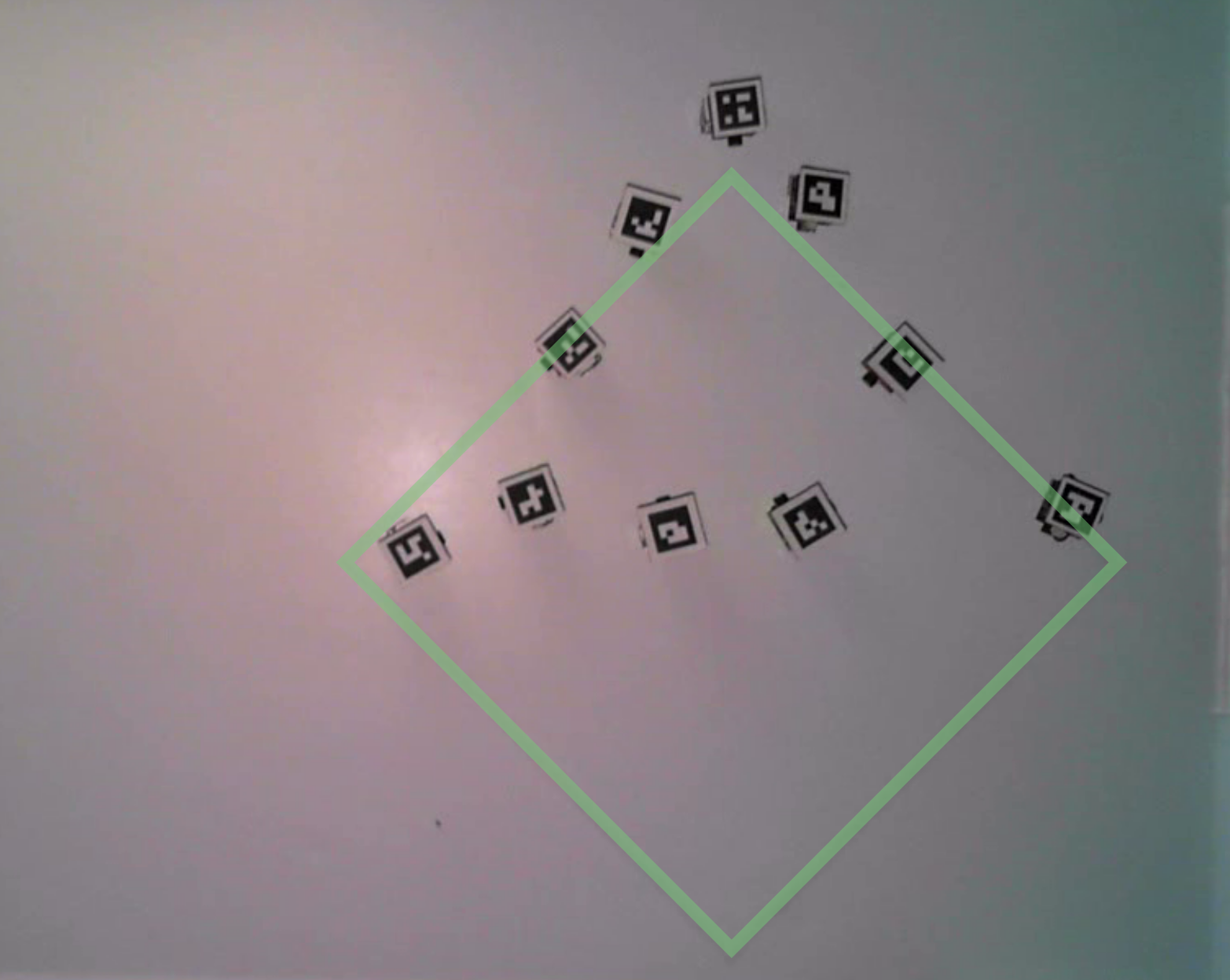}%
		\hspace{\ts}
		\includegraphics[height=\th]{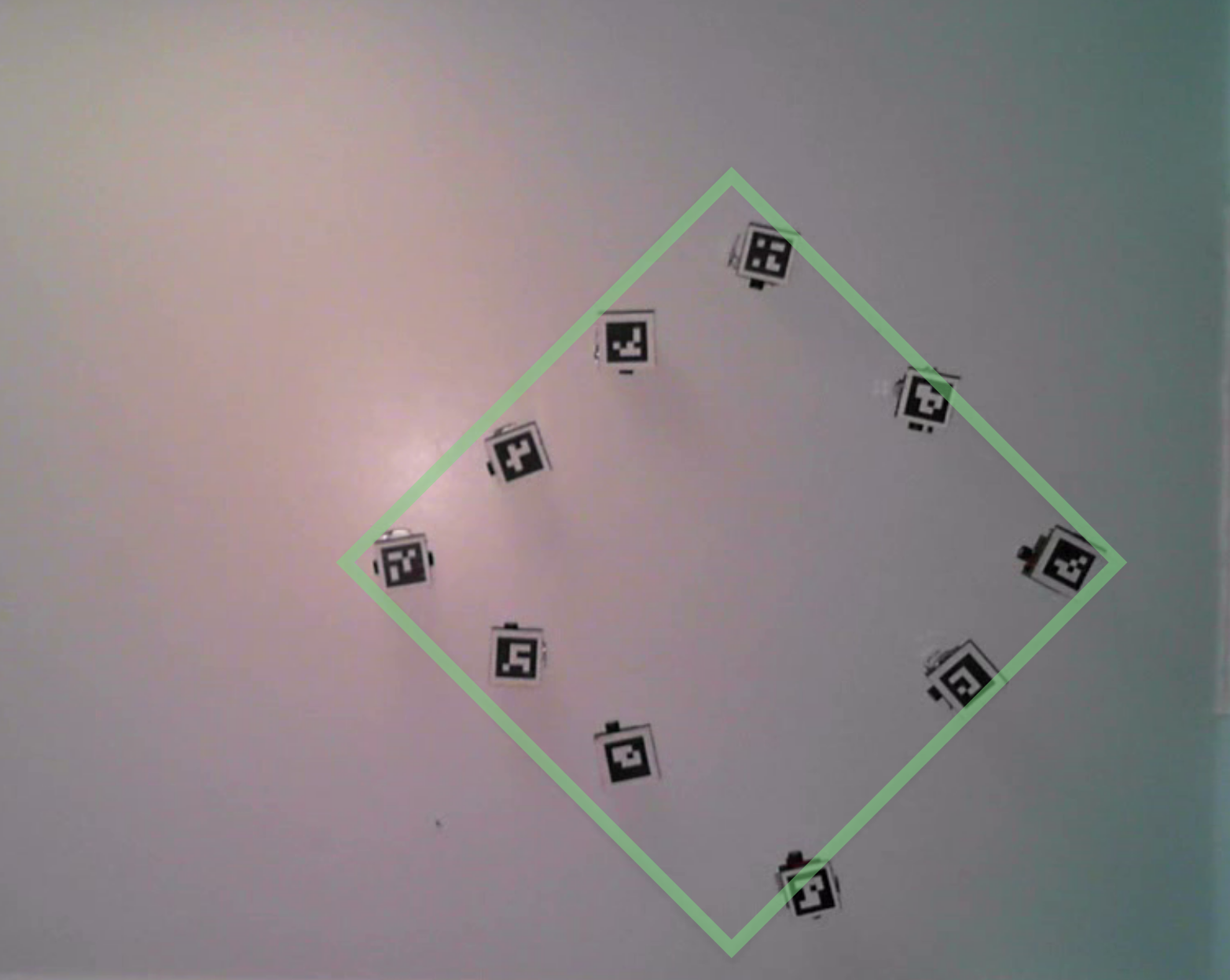}%
		\caption{}
		\label{subfig:sequence}
	\end{subfigure}%
	\\
	\vspace{\vs}
	\begin{subfigure}[t]{0.25\textwidth}
		\centering
		\includegraphics[height=\bhl]{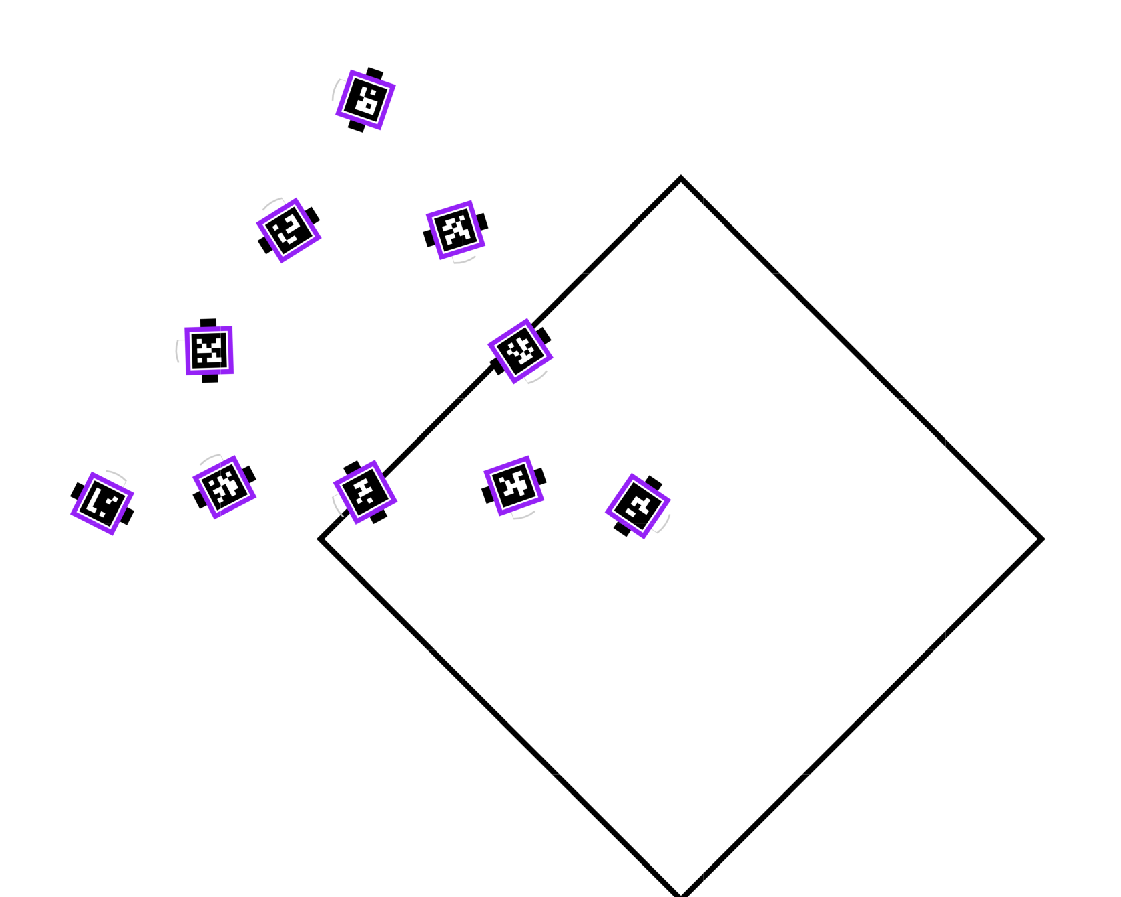}
		\caption{}
		\label{subfig:virtual}
	\end{subfigure}%
	\hfill
	\begin{subfigure}[t]{0.37\textwidth}
		\centering
		\includegraphics[height=\bh]{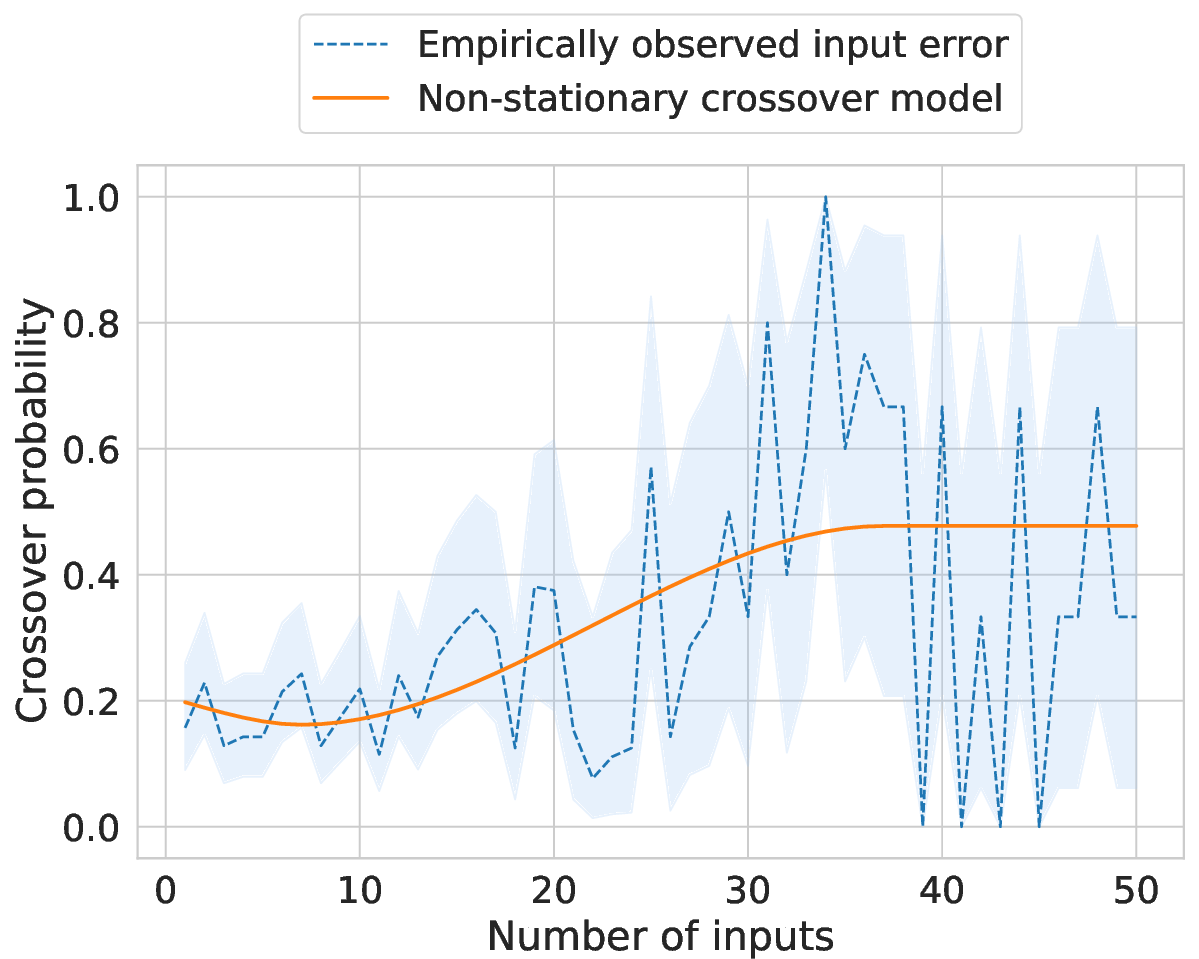}
		\caption{}
		\label{subfig:crossover}
	\end{subfigure}%
	\hfill
	\begin{subfigure}[t]{0.37\textwidth}
		\centering
		\includegraphics[height=\bh]{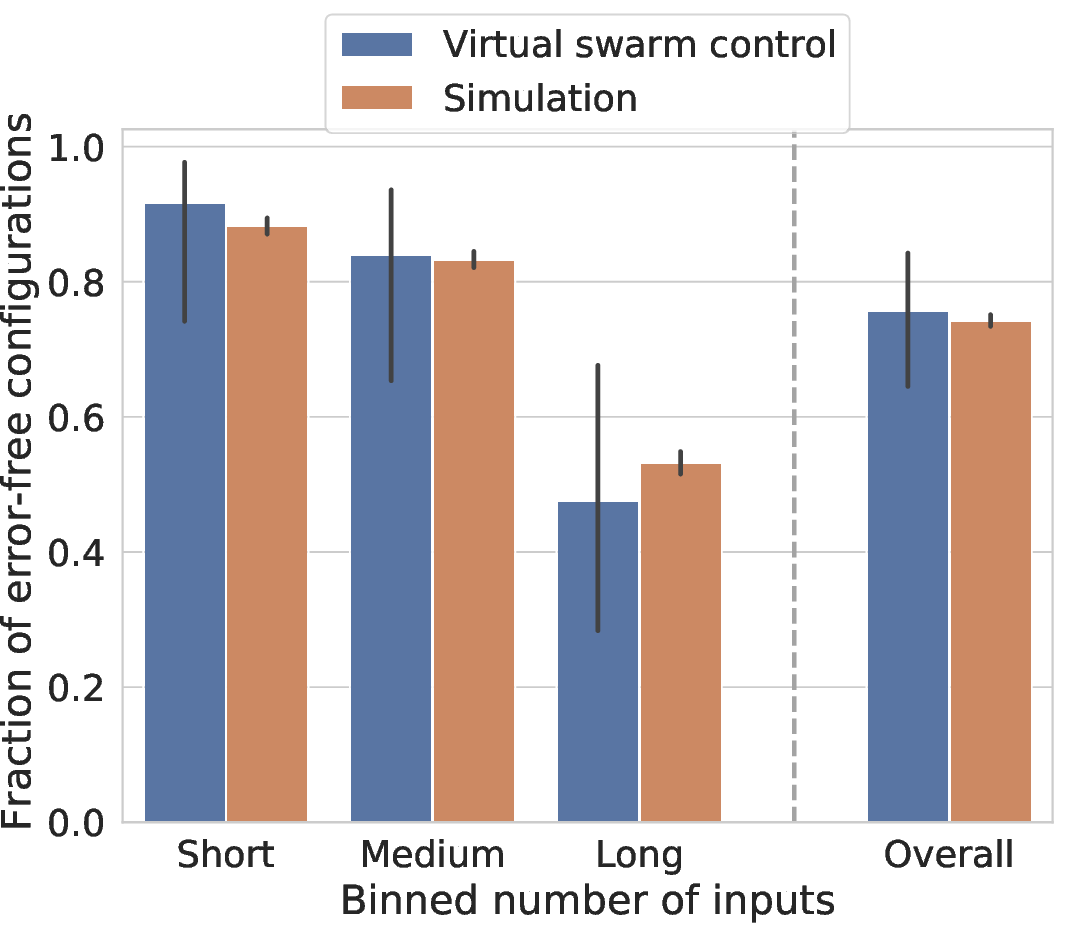}
		\caption{}
		\label{subfig:simcomp}
	\end{subfigure}%
	\caption[End-to-end testing of full system with EEG inputs and swarm control.]{\textbf{End-to-end testing of full system with EEG inputs and swarm control.} SCINET was tested as a full system by the first author for controlling both physical (\textbf{a}) and virtual (\textbf{b}) robot swarms.
	\textbf{a,} When controlling physical robots, the target configuration is presented to the user as an illuminated shape on the robot arena (accentuated here for visibility).
	\textbf{b,} During virtual swarm control, the BCI user views a monitor that presents a target configuration (depicted as a shape outline) alongside the swarm's current configuration. The virtual robots simulate realistic robot motion, and readjust their positions dynamically after each user input.
	\textbf{c,} Non-stationary crossover probability versus number of inputs in virtual swarm control trials ($n=70$). At each input, we estimate the empirical crossover probability (blue, dotted line, with 95\% Wilson confidence interval \cite{lawrence2001}) as the fraction of trials where that input was decoded incorrectly with respect to the target configuration. A modified cubic function was fit to this changing crossover probability (orange, solid line) and used to generate errors in a realistic SCINET simulation.
	\textbf{d,} Comparison of experimental and simulation configuration accuracy as a function of number of inputs until convergence, where the non-stationary crossover profile from \textbf{c} was used for simulating input errors. Results are binned into short, medium, and long trials by selecting bin edges at the 1/3 and 2/3 percentiles of virtual swarm convergence times, such that each bin includes approximately the same number of trials. Configuration accuracy within each bin is computed as the fraction of trials ($n=10,000$) converging successfully, where error bars depict the 95\% Wilson confidence interval. The overall configuration accuracy across all trials (regardless of number of inputs to converge) is also depicted. Experimental accuracy (binned and overall) closely matches that of the simulated model, suggesting that posterior matching (which assumes a fixed crossover probability) with input errors generated by the profile in \textbf{c} is an appropriate model for the observed experimental behavior. This suggests that the end-to-end system is performing as expected (once the input statistics are accounted for), and that it is reasonable to use this simulator to further explore system behavior.}
	\label{fig:realSCINET}
\end{figure}

As a pilot demonstration, the first author trained an EEG MI classifier and used the rules of posterior matching to control a simulated robot swarm (presented visually on a monitor) (\Cref{subfig:virtual}). In a series of repeat trials, target configurations were presented and MI commands were issued to steer the swarm towards the specified configuration. As one might anticipate, over the course of issuing a sequence of MI inputs, the error rate of user inputs (calculated with respect to the correct input according to the rules of posterior matching) varied as additional commands were issued (\Cref{subfig:crossover}). In theory, this error can be attributed to both the user error of issuing the incorrect posterior matching input, as well as classification error due to the MI detection algorithm classifying the input incorrectly. We conclude from the previous dictionary sorting user study that the former error source is small (estimated at 4.2\%, see \Cref{subfig:acc_overall}), and therefore the increasing net input errors are likely due to degrading MI signal feature separation (\Cref{fig:LDAfull}). This effect is possibly due to user fatigue in issuing a sequence of inputs with minimal training, and resulted in an overall input error of 21.8\%. Given previous work on MI inputs, this error rate is likely to be significantly improved with higher-fidelity interfaces and more extensive user training \cite{lotte2007review}.

Despite nontrivial input error rates, this prototype system achieved an overall configuration selection accuracy of 75.7\% (\Cref{subfig:simcomp}), calculated as the fraction of trials where the swarm converged perfectly to the specified target with zero error; this greatly exceeds the accuracy of 1.67\% that would be obtained by chance selection alone. Furthermore, we can account for these observed results with a simple model on the non-stationary input statistics. We fit a piecewise polynomial to the empirical crossover probability (\Cref{subfig:crossover}) and use this profile to generate input errors in a posterior matching simulation that assumes a fixed crossover probability. This simulation model obtains a similar configuration accuracy (74.3\%) to that observed in practice (\Cref{subfig:simcomp}). Additionally, this model matches the observed behavior even when evaluating trials based on their required numbers of inputs to converge, which is a distinguishing element between trials since longer convergence is associated with increasing input errors and, therefore, with decreased performance.

The first author also demonstrated SCINET's capability to be implemented in a (non-virtual) cyber-physical system by successfully steering a physical robot swarm in multiple trials as a proof-of-principle to complement the virtual simulations of swarm behavior (\Cref{subfig:sequence}, see ancillary videos). Taken together, these results collectively demonstrate that SCINET can achieve reasonable configuration accuracy despite the presence of non-stationary input errors, and that performance can be captured by a simple model. Additionally, the availability of a simulator that closely matches observed empirical behavior allows us to explore the performance of SCINET with more general dictionaries.

\FloatBarrier

\subsection{Robot Swarm Setup}
The Robotarium arena and its virtual counterpart, both provided by the Georgia Robotics and InTelligent Systems Laboratory (GRITS), were used as swarm operating spaces. The Robotarium \cite{pickem2017robotarium} is a remotely accessible, multi-robot research facility that provides global position and orientation tracking of fiducial markers placed on each robot, a WiFi communication infrastructure to broadcast information to the robots, and an automatic recharging mechanism. The robot swarm consists of GRITSBots \cite{pickem2015gritsbot}, which are differential-drive wheeled mobile robots with WiFi communication and infrared range-sensing capabilities. These robots may be modeled as unicycles, i.e., for the $i$\textsuperscript{th} robot in the swarm, the planar position $p_i = (x_i,y_i)$ and orientation $\theta_i$ follow the dynamics given by
\begin{align*}
	\begin{bmatrix}
		\dot{x}_i\\\dot{y}_i\\\dot{\theta}_i
	\end{bmatrix} = \begin{bmatrix}
		\cos(\theta_i) & 0\\
		\sin(\theta_i) & 0\\
		0& 1 
	\end{bmatrix}\begin{bmatrix}
		v_i\\\omega_i
	\end{bmatrix},
\end{align*}
where $v_i,\omega_i$ are its linear and angular velocities, respectively. The Robotarium API \cite{wilson2020robotarium} provides a simulator that enables the testing of algorithms in a virtual setting prior to deployment in the real robots.

Each swarm configuration (physical or virtual) consists of ten robots ($n=10$), which collectively conform to a specified coverage density $\phi(q,t)\in(0,\infty)$ which describes the desired distribution for all points $q$ in the space $D\subset \mathbb{R}^2$ at time \cite{diazmercado2015distributed} $t$. The robots achieve this distribution by finding an optimal configuration with respect to the the locational cost \cite{cortes2004coverage} as weighted by the reference density $\phi$, defined as
\begin{align*}
	\mathcal{H}(p_i,t) = \sum_{i=1}^n \int_{V_i}\|q-p_i\|^2\phi(q,t)\,dq,
\end{align*}
where the $V_i\subset D$ form a Voronoi tessellation of the space using the position of the robots as generators, and properly partition $D$. The optimal configuration is achieved through a distributed control law \cite{diazmercado2015distributed} which relies only on nearby neighbor information, given by
\begin{align*}
	\dot{p_i} = \kappa(c_i(p_i,t)-p_i) + \frac{\partial c_i}{\partial t} + \sum_{j\in \mathcal{N}_i(t)} \frac{\partial c_i}{\partial p_j}\left( \kappa(c_j(p_j,t)-p_j) + \frac{\partial c_j}{\partial t}\right),
\end{align*}
where $\kappa>0$ is a tuning parameter, $c_i(p_i,t)$ is the center of mass for $V_i$, and $\mathcal{N}_i(t)$ is the set of robots near robot $i$ at time $t$. This control law is mapped into the unicycle dynamics through a near-identity diffeormorphism \cite{olfati2002near}. Specifically, for $\lambda>0$ the linear and angular velocities are obtained as
\begin{align*}
	\begin{bmatrix}
		v_i\\\omega_i
	\end{bmatrix}
	=
	\begin{bmatrix}
		\cos(\theta_i) & \sin(\theta_i)\\
		-\tfrac{1}{\lambda}\sin(\theta_i) & \tfrac{1}{\lambda}\cos(\theta_i)
	\end{bmatrix}
	\dot{p}_i.
\end{align*}

To use this interface, the abstract polygons in our dictionary need to be translated to a continuous density function describing swarm coverage. This was achieved by constructing a Gaussian mixture model (GMM) from the vertices and edges of a given polygon. Specifically, we placed an isotropic Gaussian distribution at each polygon vertex, and on each edge we placed two Gaussian distributions with means evenly spaced between the edge's vertices, and with a 10/1 ratio of variance parallel to the edge to variance perpendicular to the edge (see \Cref{fig:GMM}). To define this GMM more formally, let $v_1=[x_1,y_1]^T$ and $v_2=[x_2,y_2]^T$ denote two vertex coordinate pairs connected by an edge, and let $w = 2(v_2 - v_1)/3$. An isotropic Gaussian distribution with coordinate-wise variance of $0.007\lVert w \rVert_2$ was placed at each vertex, in units relative to the arena height. Two additional Gaussian distributions with means at $v_1 + w/2$ and $v_1 + w$ were placed on the edge between $v_1$ and $v_2$, each with a covariance matrix of
\[\Sigma = \begin{bmatrix} \frac{w}{\lVert w \rVert_2} & \frac{w^\perp}{\lVert w \rVert_2} \end{bmatrix} \begin{bmatrix} 0.07\lVert w \rVert_2 & 0 \\ 0 & 0.007\lVert w \rVert_2 \end{bmatrix} \begin{bmatrix} \frac{w}{\lVert w \rVert_2} & \frac{w^\perp}{\lVert w \rVert_2} \end{bmatrix}^T \quad \text{where } w^\perp = \begin{bmatrix} 0 & 1 \\ -1 & 0 \end{bmatrix} w.\]
This GMM was then transmitted to the Robotarium using User Datagram Protocol (UDP) packets via WiFi.

\subsection{Motor Imagery Input Classification}

In order for the user to provide a binary input through the use of a mental command detected by EEG, raw signals from scalp electrodes must be processed and subsequently classified into one of two commands. Although EEG is associated with low spatial resolution and high sensitivity to noise, its high temporal resolution can be leveraged to extract simple mental commands from electrical activity. In the case of motor imagery, it has been shown that mental imagery of left or right hand dorsiflexions produces discernible EEG features over different spatial regions on the scalp \cite{pfurtscheller1997motor}. Specifically, left and right hand motor imagery produces a decrease in the power of the mu (8-12 Hz) and beta (18-26 Hz) bands over the contralateral side of the scalp (a phenomenon called \emph{event-related desynchronization}, or ERD), and sometimes produces an increase of power in these bands over the ipsilateral side (called \emph{event-related synchronization}, or ERS) \cite{wolpaw2002brain, pfurtscheller1997motor}. If these signature changes in power spectra can be recognized, then binary classification can be performed to detect left or right hand motor imagery.

To built such a motor imagery classifier with acceptable accuracy, we adopt a procedure that combines protocols from a series of studies related to optimal spatial filtering of EEG signals for motor imagery classification \cite{blankertz2008optimizing, ramoser2000optimal,pfurtscheller2000current, muller1999designing, guger2000real}. At a high level, the method first temporally filters EEG measurements in an ERD/ERS frequency range of interest, then trains spatial filter coefficients that maximize the signal power in one motor imagery class and minimize it in the other. This spatial filtering process, known as Common Spatial Patterns (CSP) filtering, yields features that discriminate between power spectrum changes due to different motor imagery classes. Finally, these filtered and processed features are classified with a linear discriminant analysis (LDA) classifier.

EEG measurements are sampled at 2 kHz from a 32-electrode BioSemi ActiveTwo system. The use of CSP filtering requires the use of at least 18 electrodes over the motor cortex \cite{ramoser2000optimal}; here, we record electrodes F3, Fz, F4, FC5, FC1, FC2, FC6, T7, C3, Cz, C4, T8, CP5, CP1, CP2, CP6, P3, Pz, and P4 based on the International 10/20 system. BioSemi ActiView\footnote{https://www.biosemi.com/download.htm} is used to monitor EEG signal quality during scalp recording setup. Signals are downsampled to 128 Hz and referenced using the Common Average Reference (CAR), which subtracts the mean of all electrodes from each individual signal \cite{mcfarland1997spatial}. Then, signals are temporally filtered with a 3rd order Butterworth notch filter centered at 60 Hz with a band of 57-63 Hz and a pass band ripple of 0.5 dB, a 6th order Butterworth band pass filter with a band of 0.5-50 Hz and a pass band ripple of 0.5 dB, and a 6th order Butterworth bandpass filter with a band of 8-30 Hz and a pass band ripple of 0.5 dB to limit the considered frequencies to the mu and beta ranges \cite{muller1999designing}.

In order to detect the power spectrum changes due to ERD/ERS during motor imagery, the choice of spatial filter coefficients among electrodes must be optimized to maximally discriminate between left and right hand motor imagery. The CSP method is ideal for this type of discrimination since it maximally distinguishes between intraclass signal power, which directly translates to the discrimination of ERD/ERS activity and therefore to the detection of binary motor imagery commands (see \Cref{sec:CSP} for details of CSP training). The two most discriminative CSP filters per class (four filters total) are applied to spatially filter the temporally filtered signals, yielding a signal with four channels. A temporal average of the square of each channel is taken over a window of length $T$ with an offset of $t_0$ seconds (see below for details of parameter selection), resulting in an average power $p_i$ for channel $i \in \{1,2,3,4\}$. The final feature vector $f$ of length four is then constructed by taking the natural log of each channel power $p_i$, normalized by the total power across all channels, i.e.\ $v_i = \ln\left(\frac{p_i}{\sum_{i=1}^4 p_i}\right)$. Finally, this feature vector is passed through a binary linear discriminant analysis (LDA) classifier \cite{hastie2009elements} to extract the issued left or right hand motor imagery command. We summarize this process in the feature extraction portion of \Cref{fig:pipeline}.

CSP filters and LDA classifiers are trained with a procedure adapted from Guger et al.\ \cite{guger2000real}. The BCI user sits in front of a monitor and imagines left or right hand dorsiflexions according to a corresponding left or right arrow cue which appears on screen (\Cref{fig:training}). During a training session, each motor imagery class (left or right hand) is presented for 30 synchronized recorded training points, with all 60 inputs presented in a randomized order. During each synchronized training point recording, a fixation cross appears for 2 seconds, at which point a left or right arrow cue is displayed for 1.25 seconds, prompting the subject to imagine the corresponding movement. The fixation cross remains for 3.75 seconds after, during which the subject continues to imagine the instructed movement. This results in a total training interval of 5 seconds. The cross is then cleared, followed by an inter-stimulus-interval of uniformly randomly length between 1 and 2.5 seconds. Windows at a length of $T=4$ s offset by 0.5 s are extracted from the 5 second training interval (e.g.,\ windows with $t_0=0$ s, $t_0=0.5$ s, or $t_0=1$ s) and used to train CSP filters and LDA classifiers based on the signal processing procedure described previously. 10x10 cross-validation is used to evaluate the accuracy of each 4 second window over all training data, and the best 4 second window is selected to use for synchronous user inputs using during testing.

If a cross-validation accuracy of 0.7 is exceeded for the best 4 second window, the feature extraction and classifier pipeline is considered trained. Otherwise, additional sessions of 15 training points for each class are collected until some subset of training sessions results in filters and a classifier with a cross-validated accuracy of at least 0.7. For instance, suppose a first training set of 30 data points per class, labeled as dataset $S_1$, does not result in a sufficient cross-validation accuracy. Then, a second training session of 15 data points per class is run, resulting in an additional training dataset labeled $S_2$. The same filter, classifier, and window extraction procedure described above is performed individually on $S_2$ and $S_1 \cup S_2$, and the best model saved. If this model's cross-validated accuracy does not exceed 0.7, another training set of 15 data points per class is collected, resulting in training dataset $S_3$. The best model from $S_3, S_1 \cup S_3, S_2 \cup S_3,$ and $S_1 \cup S_2 \cup S_3$ is saved; this procedure continues until a trained model exceeds the 0.7 threshold. The final cross-validated error from the saved model is used to estimate the crossover probability parameter in the posterior matching procedure during testing.

During testing, the distance-to-hyperplane output of the LDA classifier is used to create a feedback bar updated in real-time to aid the user in tuning their motor imagery features \cite{neuper1999enhancement}. The feedback bar points in the direction of the classifier's detected input (left or right) and has a length proportional to the distance from each instantaneous feature vector $f$ to the classifier hyperplane. As we describe in \Cref{sec:virtualeval}, this distance is a direct measure of classification confidence. Feedback is generated over $T=1$ second windows overlapped by 0.0625 seconds, and is displayed continuously during the entire testing phase.

OpenVibe\footnote{http://openvibe.inria.fr/} is used for the real-time collection and processing of EEG signals, with CSP filters and LDA classifier training performed offline in MATLAB. During testing, the lab streaming layer\footnote{https://github.com/sccn/labstreaminglayer} communication protocol is used to interface in real-time between signal acquisition, feature extraction, and feedback presentation in OpenVibe, and  feature classification and posterior matching operation in MATLAB.

\subsection{Swarm Control Trials}

In order to demonstrate SCINET's performance, the first author (henceforth referred to as ``the subject'') learned the dictionary ordering for swarm configurations, trained CSP filters and LDA classifiers with left/right imagined dorsiflexions, and evaluated his swarm control ability using a virtual Robotarium arena over 70 trials. On each day (with one session of trials per day), the subject sat in front of two monitors, one of which presented the visualizations required for training and feedback for testing (run on a PC laptop) and the other ran the Robotarium simulation in MATLAB (run on a MacBook laptop). At the start of each session, the subject trained spatial filters and classifiers using the aforementioned procedure until the specified training threshold was met. Then, the subject performed 10 test trials per day on a Robotarium simulation. For each test trial, a target swarm configuration was selected randomly without replacement from a set of possible targets and displayed in the simulator as a blue outline (as in \Cref{subfig:virtual}). The target set was constructed as a single copy of each string in the dictionary (60 total), plus 40 additional copies that are evenly spaced throughout the dictionary (for a total of 100 configurations).

The subject then issued the appropriate motor imagery commands to steer the swarm to each specified target configuration according to the posterior matching algorithm (see the \Cref{sec:PMimp} for a mathematical algorithm description). For the special case where a target and guess configuration were equal, a ``right'' command was issued. The subject issued each command in a synchronized input window of 5 seconds in length. After each command was issued, a new configuration was broadcast to the robot swarm, and the robots readjusted their positions while the subject waited and observed their movement. After each robot's velocity fell below a prespecified threshold, the swarm controller detected that the swarm had settled on a single configuration and another input was requested from the subject. At this point, the subject heard a single audible beep, which indicated that the swarm had made its guess, and that they should  decide on their next input. After a two-second pause, the subject heard three more beeps, each separated by a single second, to count down to the start of the synchronized input window. A final beep signaled the start of a 5 second input window, during which the subject visually fixated on the real-time feedback bar. A single beep signaled the end of the synchronization window, at which point the subject could stop their command. Feature extraction and classification was performed using the same CSP filters, LDA weights, and timing parameter $t_0$ for extraction of a $T=4$ second window as during training. After each input was issued the system indicated the classification result on-screen with a left or right arrow and the swarm rearranged to its updated configuration, after which a new input window began and the subject observed the swarm as feedback for their next command (\Cref{fig:physical}).

This process iterated until the posterior matching algorithm converged to a final estimate of the subject's configuration, at which point three short, audible beeps were played. Convergence was defined by the algorithm maintaining a posterior distribution for the subject's target configuration, and stopping when any configuration met or exceeded a prespecified posterior threshold. A single trial ended at the sooner of posterior matching converging or the number of issued inputs reaching a maximum of 50 inputs. When the trial ended by either means, the maximum posterior probability configuration was selected as the algorithm's final estimate.

The threshold for the convergence stopping criterion was selected from a lookup table of convergence thresholds specified for various BSC crossover probabilities and desired trial lengths (see ancillary files). For a given crossover probability, 500 posterior matching simulated trials (described below) were performed offline for each of several candidate thresholds, and the corresponding table entry was set as the threshold that achieved the highest convergence accuracy while not having an average number of inputs greater than the specified trial length. Our specified average trial length for threshold lookup was set to 25 inputs, which is an estimated number of synchronous inputs an EEG user can issue before becoming fatigued. The lookup table was constructed by evaluating crossover probabilities from 0 to 50\% at increments of 5\%, and posterior stopping thresholds of 0\% to 100\% at increments of 5\%. If the model's crossover probability (i.e., the trained classifier's cross-validation error) did not appear in the lookup table, the next highest crossover probability in the table was used for lookup.

To compute the configuration accuracy and expected number of input values in the lookup table, each posterior matching simulation trial used the specified crossover probability and candidate stopping threshold. Unlike the simulations described for modeling a non-stationary input error profile, here each crossover probability used to generate input errors was fixed throughout the entire simulation trial, and this generated error crossover was equal to the crossover assumed by posterior matching in its posterior distribution updates. In each simulation trial, a target string was selected at random from the configuration dictionary, and the rules of posterior matching followed to simulate the role of a user. Each simulated user input was passed through a simulated BSC with a fixed crossover probability at the specified value. Each simulated trial was run for a maximum of 50 inputs, as was the case for the virtual swarm control experiments. 

The subject engaged in 7 total days of completed trials spread over the course of 3 weeks, with 10 trials performed per day. On each day, the subject trained the EEG classifier using the aforementioned procedure, completed 5 virtual swarm control trials, took a rest period, and then completed 5 more trials. During one particular session, the subject perceived that the EEG classifier feedback bar was qualitatively deteriorating after the first 5 trials, and added 2 additional training sessions of 15 data points per class to the training set for the second half of the session. On the other 6 days, both sets of 5 trials used the same initially trained classifier. There was an 8th day of trials omitted from this study. On this day, the subject trained the classifier as above and completed 3 trials on this day, but aborted the session due to a feeling of complete loss of ability to issue motor imagery inputs. Upon further investigation, it was found that these 3 sessions had a net EEG input error of 53\%, explaining the lack of control. These two ad hoc adjustments (additional training, aborted session) are justifiable since the purpose of this experiment is to evaluate SCINET's overall performance under the assumption of a reasonably trained and sustainable EEG classifier.  

\subsection{Realistic Simulation Baseline}
To fit a non-stationary crossover probability model to the empirical data in \Cref{subfig:crossover} for use in a realistic SCINET simulation, an input error profile was modeled by first fitting a least-squares cubic regression to the empirical crossover probability curve (\Cref{subfig:bscpoly}). The data points corresponding to one issued input, the minimum point, and maximum point of this cubic function were then used to fit a piecewise cubic Hermite interpolating polynomial (PCHIP), where the maximum point was held until the maximum number of inputs (\Cref{subfig:bscpchip}). The motivation behind this procedure was to generalize the crossover behavior at lower numbers of inputs while enforcing monotonicity as the number of inputs increased, since a decrease in crossover probability would not realistically model factors such as user fatigue increasing with more inputs. The resulting PCHIP was used to generate input errors in our realistic SCINET simulation. Specifically, at input number $i$, the correct posterior matching response was corrupted with a Bernoulli error (statistically independent of all previous and future errors) with bias given by the PCHIP value at input $i$.

Even though input errors were generated according to the PCHIP, during each posterior matching iteration the simulator modeled a binary symmetric channel with a \emph{fixed} crossover probability. This simulates the real-world effect of the trained classifier producing a cross-validation error that is used as the BSC crossover probability estimate for each trial, yet during the trial the BCI's actual input error statistics may change with additional inputs. We set the simulator's fixed crossover estimate as the average input error across all inputs and all virtual swarm trials, which evaluated to 21.8\%. This value serves as an estimate of the aggregate error to be expected over the course of a virtual swarm trial.

Once the PCHIP error generator was fit and the fixed crossover probability set, the posterior matching simulation was run for 10,000 trials. At the start of each trial, a configuration was selected uniformly at random from the dictionary to serve as a target for posterior matching. We implemented the same stopping criteria for each trial as in the virtual swarm trials performed by the subject: the posterior convergence threshold was selected from the same lookup table of thresholds using the same procedure, and a maximum of 50 inputs per simulation trial was enforced. When comparing simulation results against empirical results from virtual swarm control in \Cref{subfig:simcomp}, trials were binned by convergence time as follows: ``Short'' trials converged between 1 and 12 inputs (inclusive); ``Medium'' trials converged between 13 and 18 inputs (inclusive); and ``Long'' trials converged between 19 and 50 inputs (inclusive). The number of trials converged in each bin were 24 Short, 25 Medium, and 21 Long virtual swarm trials, and 2,786 Short, 3,748 Medium, and 3,466 Long simulated trials.

The subject also demonstrated two successful trials on the physical Robotarium system (see ancillary videos), but the quantity of these trials was limited due to laboratory demand for the system and practical considerations such as robot battery life. We implemented the same EEG training and experimental setup procedures as in the virtual swarm sessions, with the only difference being that physical robots responded to user commands rather than virtual robots.

\section{Generalizing Performance Tradeoffs}
Ultimately, the accuracy and number of controllable degrees of freedom (and hence the dictionary size) in SCINET is determined by the error rate of the input mechanism and budget on the allowable number of inputs; increasing the controlled degrees of freedom requires additional inputs to refine effector behavior. To more fully explore this tradeoff, we use different input error profiles and dictionary sizes (corresponding to a variety of end-effector degrees of freedom) to simulate posterior matching as well as a baseline interaction algorithm (called \emph{stepwise search}) that resembles discrete menu selection in existing BCIs. In stepwise search, each binary input updates the swarm's guessed configuration by moving to the next string in the dictionary, in the direction indicated by the user's input.\footnote{This algorithm is similar to the \emph{fixed offset} \cite{omar2008querying} and \emph{sequential-select} \cite{akce2013brain} policies explored in previous work on posterior matching-based BCIs.} Note that the number of steps needed for convergence in stepwise search scales linearly with the size of the dictionary.

In the data collected from a simple interface (with input characteristics reported in \Cref{subfig:crossover}), our proposed interaction mechanism can work well in some scenarios despite the relatively high overall error rate and the non-stationary error profile that nears chance probability (50\%) as the number of inputs increases. However, large dictionaries providing more resolution will suffer a performance bottleneck with this non-stationary error profile because they require more inputs for convergence. While developing high-performance input mechanisms is not the focus of this work, we evaluate SCINET system performance with realistic improved input mechanisms by simulating both posterior matching and stepwise search for a fixed crossover probability of 10\% (comparable to input errors seen in prior work \cite{lotte2007review}) and a variety of dictionary sizes (expressed as an equivalent number of degrees of freedom by subdividing each dictionary with an average alphabet size from our physical system). The \emph{information transfer rate} (ITR) \cite{wolpaw2002brain} (specified in bits per trial) is shown in \Cref{subfig:staticitr}, demonstrating that SCINET, with this simulated input mechanism, can achieve increasingly high information rates with larger dictionaries. The fraction of error-free configurations (i.e., perfectly achieving the desired configuration) also approaches 100\% (\Cref{subfig:staticacc}), even with large dictionaries and non-zero error rates in the user input. Finally, to study the rate of convergence of the estimated configuration to the target in the dictionary (which is not reflected in the fraction of error-free configurations), we also measure the absolute deviation of the estimated configuration from the target and observe that error decays quickly regardless of dictionary size (\Cref{subfig:staticL1}). 

In all metrics, posterior matching vastly outperforms discrete menu selection through a stepwise search approach. While larger dictionaries require more inputs to refine a configuration to a desired level of accuracy, SCINET with a fixed crossover probability still achieves high performance for large dictionaries in a modest number of inputs. We note that with a fixed input profile, SCINET can successfully control upwards of 6 separate degrees of freedom, which (to our knowledge) exceeds the current capabilities of noninvasive continuous control BCIs. We also plot the same performance metrics for simulations where input errors are generated according to the non-stationary profile observed in our physical experiments (\Cref{subfig:dynamicitr,subfig:dynamicacc,subfig:dynamicL1}). Even with this adverse input characteristic, SCINET greatly outperforms stepwise search across all metrics. Although performance degrades for larger dictionary sizes, these larger dictionary sizes correspond to estimated degrees of freedom that lie beyond the control capabilities of typical noninvasive BCIs.

\subsection{Simulator Details}
To generalize the performance of SCINET to arbitrary dictionaries, the simulator from \Cref{subfig:simcomp} was modified slightly. To fully evaluate the tradeoff between achieved configuration accuracy and required number of inputs for each dictionary size, we disabled convergence for both posterior matching and stepwise search (see \Cref{sec:step} for a mathematical description of stepwise search) and instead output an instantaneous configuration estimate after each issued input. After $k$ inputs, this instantaneous estimate was taken as the configuration with maximum posterior probability, i.e., $\bm{z}_{j^*}$ where $j^*= \argmax_{1 \le j \le N_d} \alpha_j(k)$ (see \Cref{sec:PMimp}). This maximum a posteriori (MAP) estimate is distinct from the guess produced during each algorithm interaction with the user, and is used only for analytical purposes to produce an error estimate. By outputting an instantaneous guess after each input and computing its configuration accuracy, we can directly observe the tradeoff between obtainable configuration accuracy and number of inputs for each algorithm and dictionary size.

\begin{figure}[htb]
	\def\sfw{0.3}
	\centering
	\includegraphics[width=0.9\textwidth]{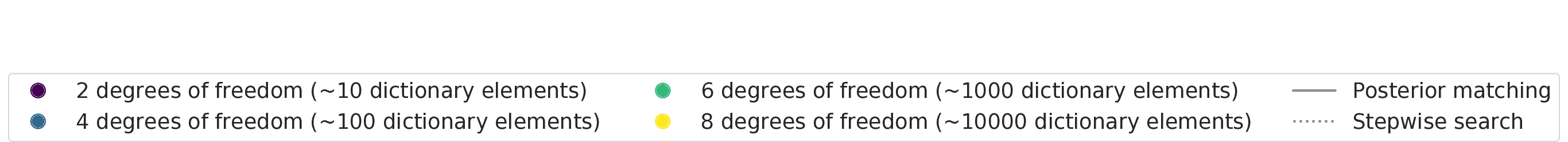}

	\tradeoffheadingone
	
	\begin{minipage}{\textwidth}
		\centering
		\raisebox{18mm}{\rotatebox[origin=c]{90}{\footnotesize\textbf{Fixed (10\%) error}}}\quad%
		\begin{subfigure}[t]{\sfw\textwidth}
			\centering
			\includegraphics[width=\textwidth]{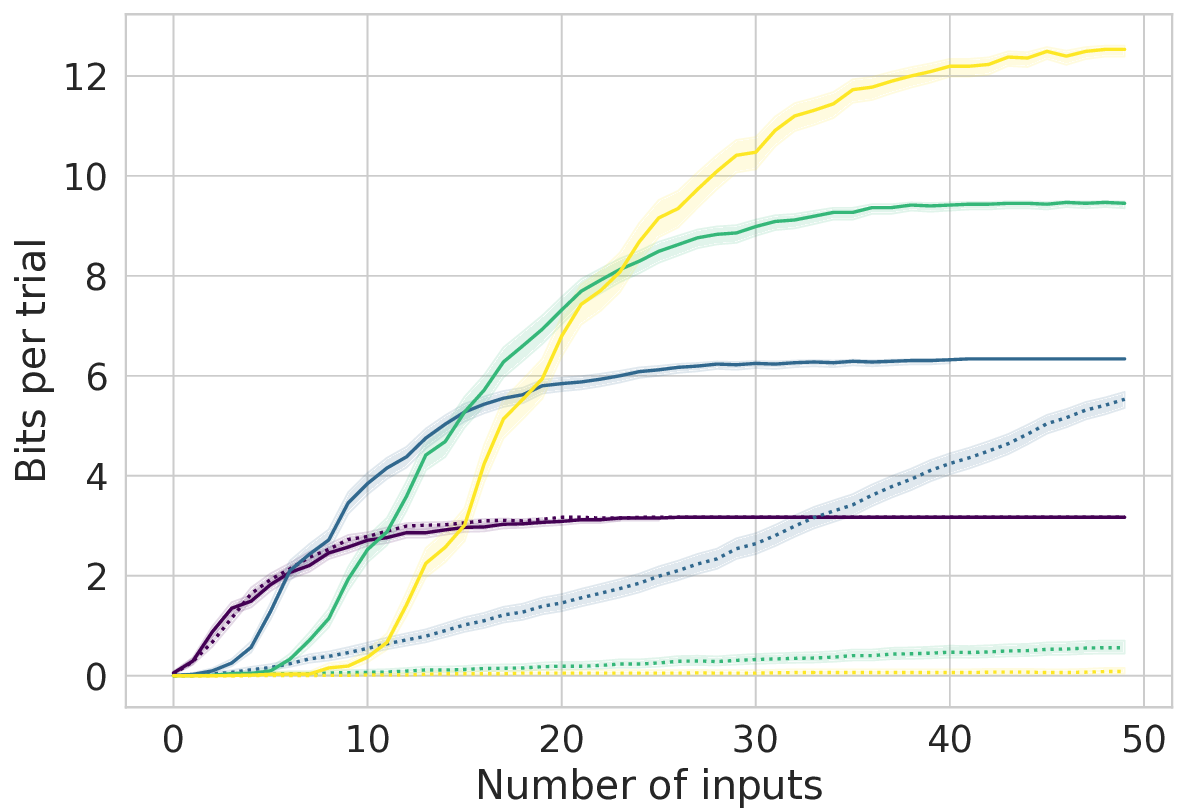}
			\caption{}
			\label{subfig:staticitr}
		\end{subfigure}%
		\hfill
		\begin{subfigure}[t]{\sfw\textwidth}
			\centering
			\includegraphics[width=\textwidth]{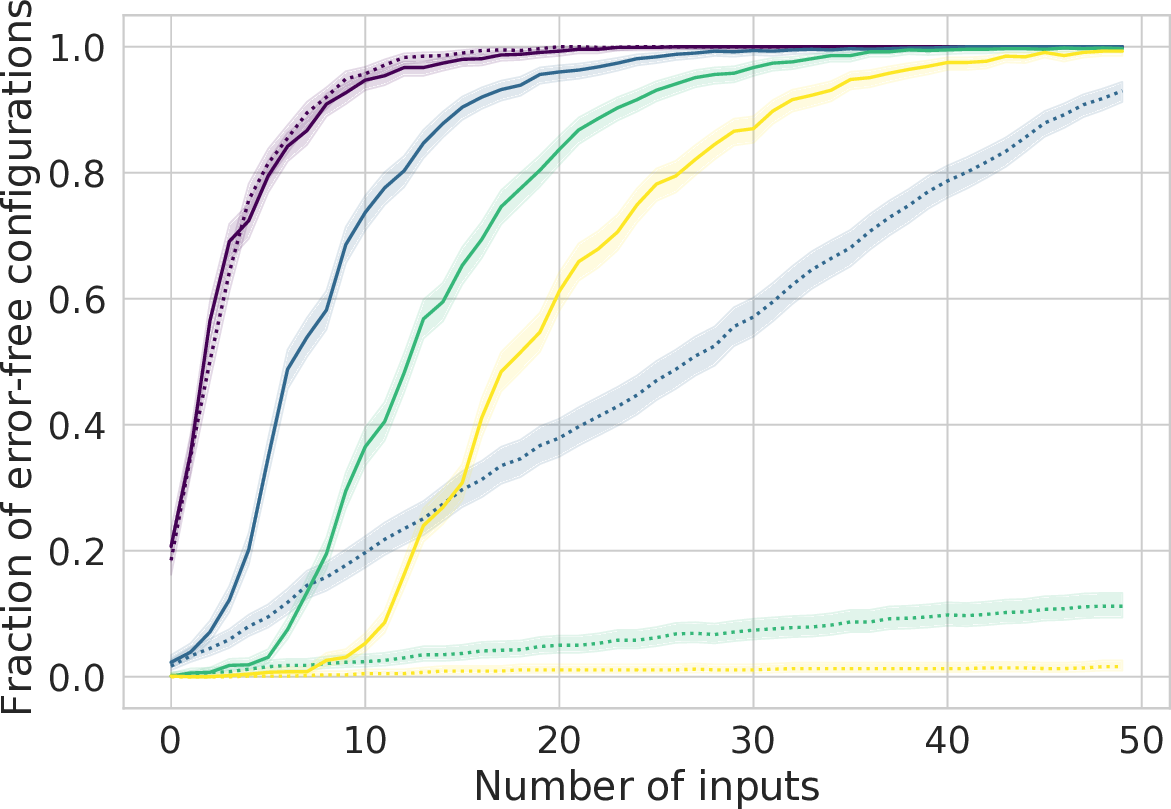}
			\caption{}
			\label{subfig:staticacc}
		\end{subfigure}%
		\hfill
		\begin{subfigure}[t]{\sfw\textwidth}
			\centering
			\includegraphics[width=\textwidth]{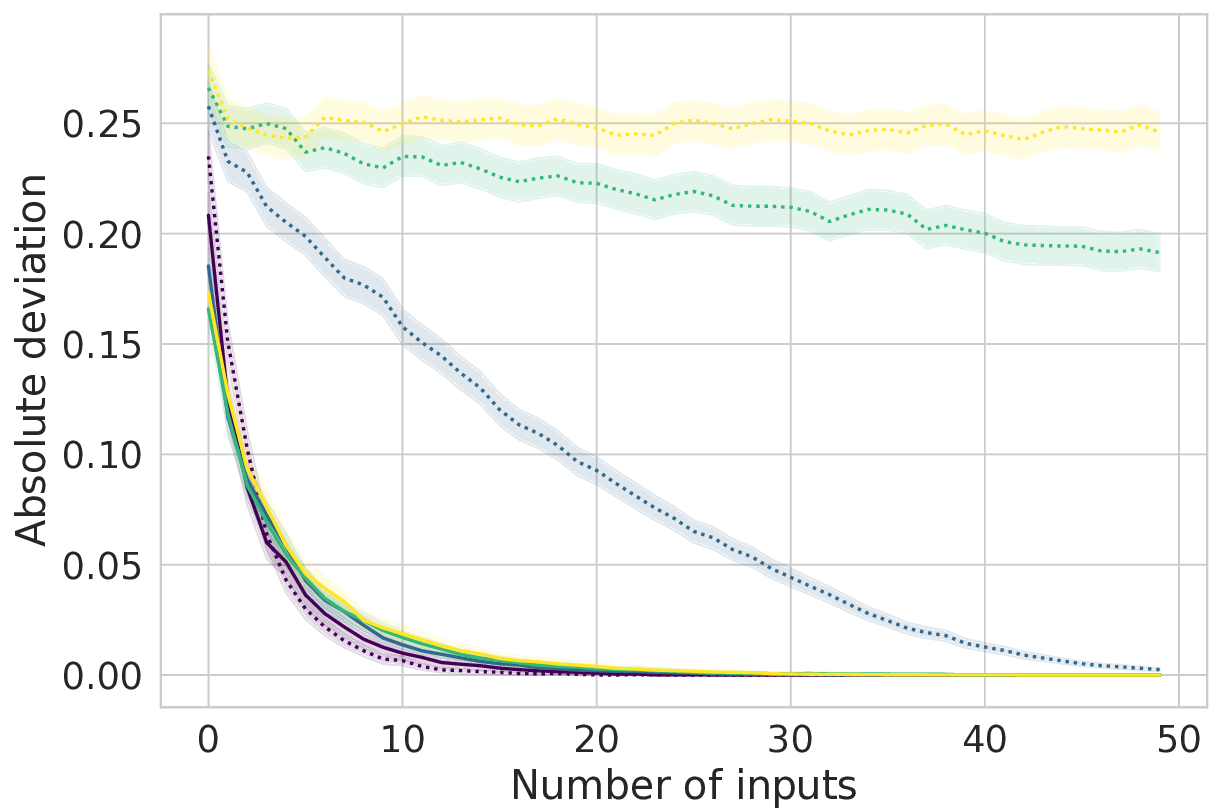}
			\caption{}
			\label{subfig:staticL1}
		\end{subfigure}
		\\
		\vspace{2mm}
		\raisebox{17mm}{\rotatebox[origin=c]{90}{\footnotesize\textbf{Non-stationary errors}}}\quad%
		\begin{subfigure}[t]{\sfw\textwidth}
			\centering
			\includegraphics[width=\textwidth]{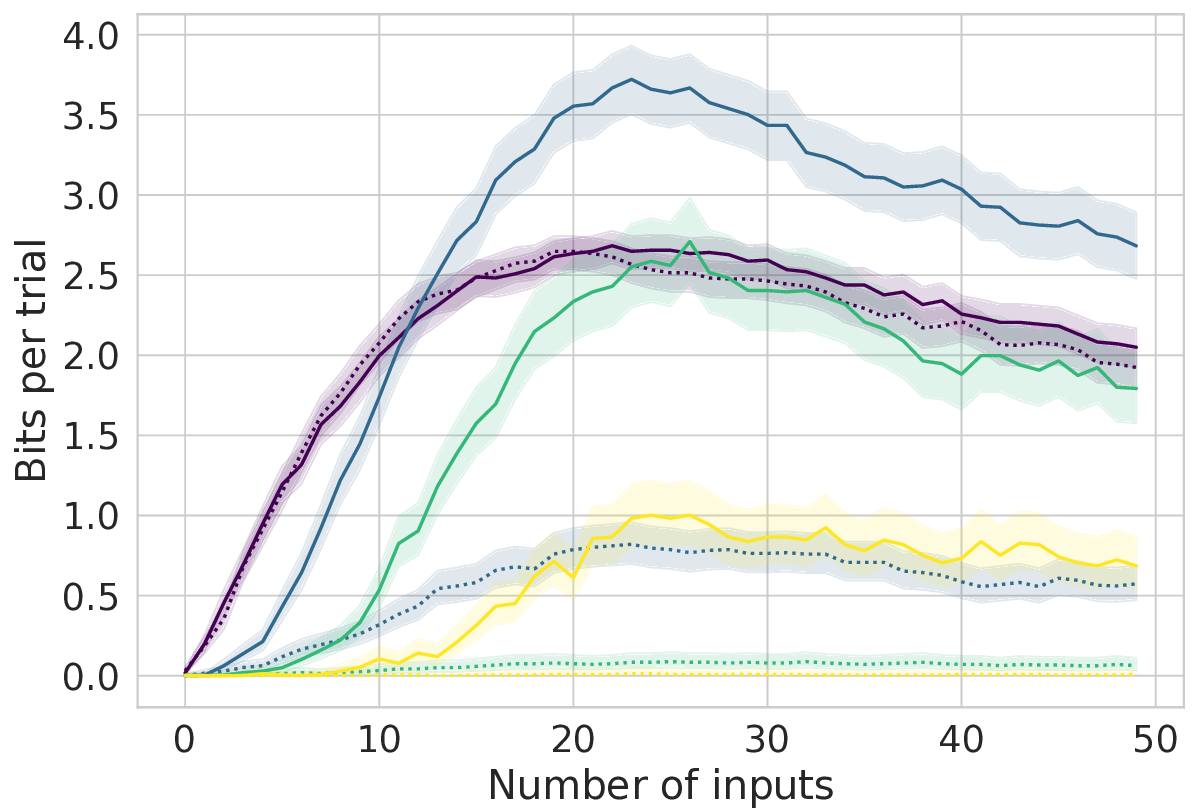}
			\caption{}
			\label{subfig:dynamicitr}
		\end{subfigure}%
		\hfill
		\begin{subfigure}[t]{\sfw\textwidth}
			\centering
			\includegraphics[width=\textwidth]{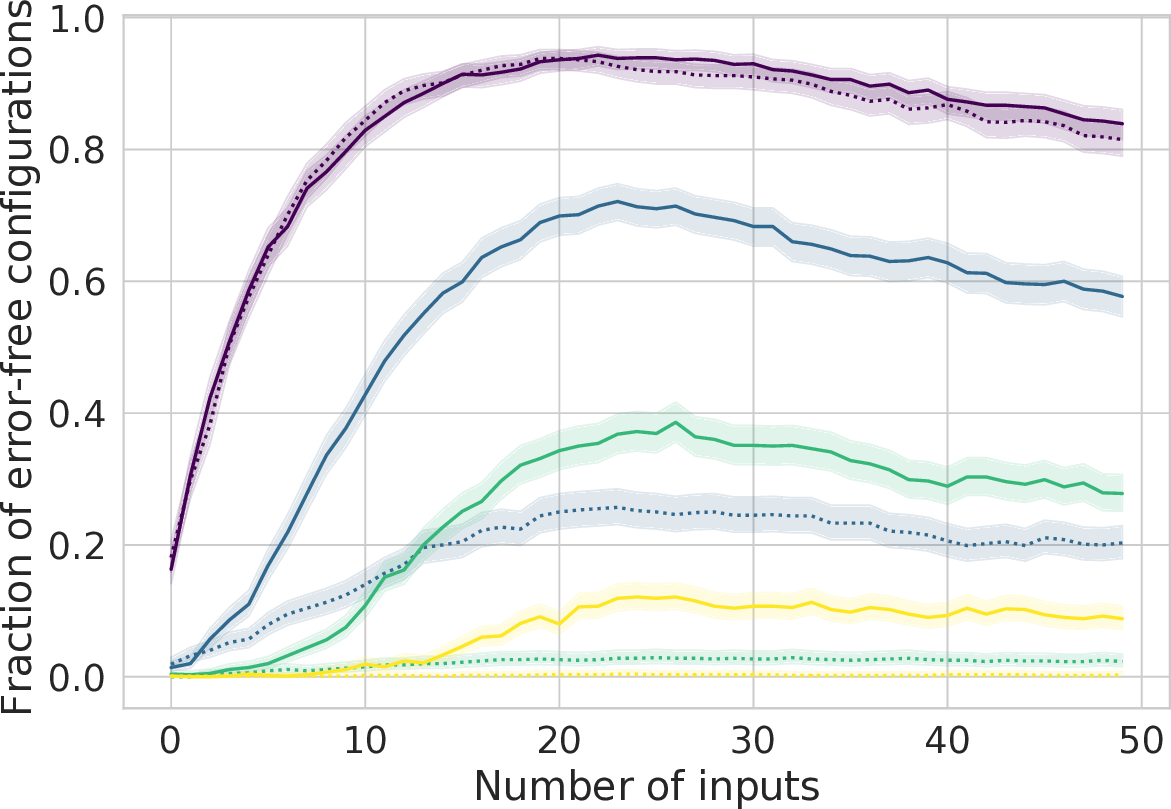}
			\caption{}
			\label{subfig:dynamicacc}
		\end{subfigure}%
		\hfill
		\begin{subfigure}[t]{\sfw\textwidth}
			\centering
			\includegraphics[width=\textwidth]{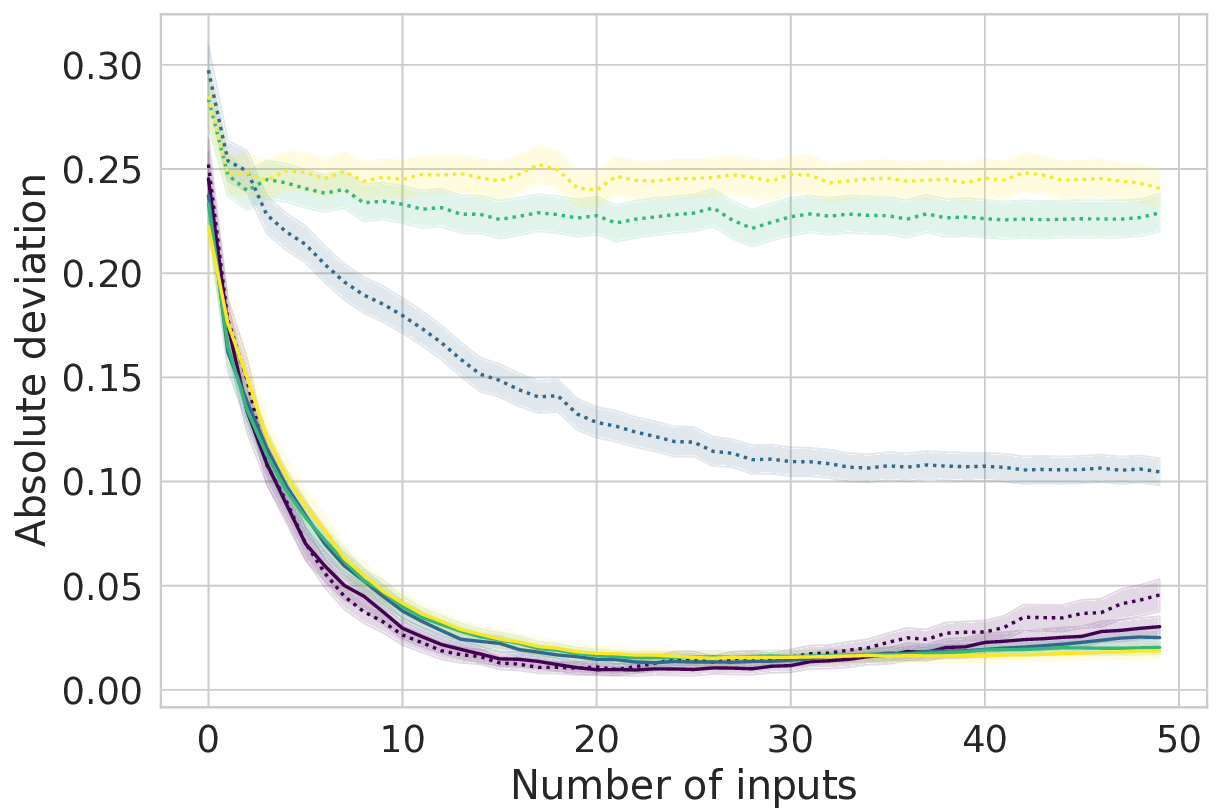}
			\caption{}
			\label{subfig:dynamicL1}
		\end{subfigure}%
	\end{minipage}
	\caption[Performance as a function of number of inputs and dictionary size.]{\textbf{Performance as a function of number of inputs and dictionary size.} We simulated the proposed posterior matching approach (solid lines) and 
	\emph{stepwise search} akin to discrete menu selection (dotted lines) over various dictionary sizes corresponding to different (estimated) end effector degrees of freedom. We evaluate both algorithms over two input error profiles: a fixed 10\% crossover probability similar to prior reported decoding performance (\textbf{a-c}), and more adverse non-stationary errors generated according a model of our physical experiments (\textbf{d-f}). 	
		\textbf{a,d,} Information transfer rate (ITR) \cite{wolpaw2002brain}, measuring the amount of information specified by a trial's inputs with respect to dictionary size. Error bars are calculated as the ITR of the corresponding accuracy limits in \textbf{b} and \textbf{e}.
		\textbf{b,e,} Fraction of estimated configurations that are a perfect match with the target configuration, with 95\% Wilson confidence intervals.
		\textbf{c,f} Absolute deviation (dictionary distance) between estimated and target configurations, where error bars depict 95\% bootstrap confidence intervals over 10,000 samples (separate resampling for every number of inputs).
		By all metrics (\textbf{a-f}), posterior matching greatly outperforms stepwise search, with differences in performance becoming more drastic for larger dictionary sizes. Additionally, posterior matching obtains high configuration accuracies at high ($>4$) estimated degrees of freedom with a modest number of inputs.}
	\label{fig:generalizing}
\end{figure}

In \Cref{subfig:staticitr,subfig:staticacc,subfig:staticL1}, a fixed 10\% crossover probability was assumed by each algorithm during posterior updating, and the same crossover probability was used to generate input errors. In \Cref{subfig:dynamicitr,subfig:dynamicacc,subfig:dynamicL1}, as in the comparison against virtual swarm trials the PCHIP error profile of \Cref{subfig:crossover} was used to generate Bernoulli input errors at each number of inputs, while each algorithm assumed a crossover probability of 21.8\% for posterior updating. Each simulation --- posterior matching or stepwise search, each run with fixed or non-stationary crossover probabilities --- was repeated for 1,000 trials, with the target configuration selected uniformly at random from the dictionary at the beginning of each trial.

To evaluate performance on various dictionary sizes, the dictionary size parameter $N_d$ in posterior matching and stepwise search was varied over simulations (see \Cref{sec:PMimp} for parameter definition). Each setting of $N_d$ corresponds to a different number of controllable dictionary degrees of freedom. To establish this relationship, we consider a dictionary with $b$ characters for each of $r$ alphabets, corresponding to $r$ degrees of freedom. The total number of strings in the dictionary is then $N_d = b^r$. To select an alphabet size $b$, we used the rounded harmonic mean of our alphabet sizes (i.e., 5,2,3,2) which evaluates to 3 characters. We generated dictionaries with $r = 2,4,6,8$ degrees of freedom, corresponding to sizes of $N_d = 9,~81,~729,$ and 6,561 respectively. Note that each algorithm operates on the total order of strings in the dictionary without regard to individual alphabets, and the only parameter that affects simulation results is the dictionary size, rather than the exact alphabet size or degrees of freedom. However, formulating the dictionary size parameter in terms of alphabet size and degrees of freedom allows us to draw connections as in \Cref{fig:generalizing} between these various parameters. We also performed the same experiments using an alphabet size of $b=5$ (see \Cref{fig:generalizing_condof}) and resulting dictionary sizes of $N_d = 25,~625,~15,625,$ and 390,625. This experiment corresponds to a more conservative relationship between degrees of freedom and dictionary size; keeping degrees of freedom fixed and increasing alphabet size results in a larger dictionary, and therefore more strings to search over (and more user inputs required) to control the same number of degrees of freedom. For this reason, we call the $b=3$ case the ``standard'' degrees of freedom estimate, and $b=5$ the ``conservative'' degrees of freedom estimate.

In \Cref{subfig:staticitr,subfig:dynamicitr}, ITR was calculated from the error-free accuracy in \Cref{subfig:staticacc,subfig:dynamicitr} respectively. At $k$ inputs issued, let $P_k$ denote the error-free accuracy, which is calculated as the number of trials where the $k$th instantaneous estimate (i.e., $\bm{z}_{j^*}$) equals the ground truth target configuration, divided by the total number of simulation trials (1,000). ITR, denoted after $k$ inputs as $R_k$, is then calculated in units of bits as \cite{wolpaw2002brain}
\[R_k = \log_2{N_d} + P_k\log_2{P_k}+(1-P_k)\log_2{\frac{1-P_k}{N_d-1}}.\]
ITR represents the aggregate amount of information about the target configuration conveyed after $k$ inputs from the user to the swarm. ITR can be also interpreted mathematically as the bit rate over a discrete memoryless channel where the target is selected with probability $P_k$, and any remaining configuration is erroneously selected with an equal probability of $\frac{1-P_k}{N_d-1}$.

To calculate absolute deviation in \Cref{subfig:staticL1,subfig:dynamicL1}, let $Z_{j^*}(k)$ and $Z_t$ denote the unit interval representations (see \Cref{sec:PMimp}) of the MAP estimate after $k$ inputs and the target configuration, respectively. Then absolute deviation, or ``dictionary distance,'' is calculated as $\lvert Z_{j^*}(k) - Z_t \rvert$, and averaged over all trials for each simulation.

\FloatBarrier

\section{Conclusion}
The results in this work demonstrate how our proposed SCINET interaction algorithm significantly expands the capabilities of low-complexity BCIs to efficiently, robustly, and scalably control high-complexity effectors, while requiring no more than currently available signal acquisition hardware already in widespread development and use. The success of human users in learning and sorting a heterogeneous shape dictionary supports the use of pairwise string sorting as a simple-to-use and tractable interface design that scales well with the complexity of the end effector system. When tested in a physical system, SCINET can perform well despite the presence of non-stationary input errors, validating the deployment of posterior matching control over a heterogeneous dictionary in a practical setting. By extending our experimental results to a range of dictionary sizes and input mechanism fidelities through realistic simulations, we find that posterior matching both outperforms a baseline algorithm comparable to discrete menu selection and exhibits the ability to control a large number of estimated degrees of freedom with only a modest number of inputs.

While posterior matching with a heterogeneous dictionary was implemented here for the control of robot swarms, the general technique is applicable to any setting where each effector parameter can be assigned its own ordered alphabet. Importantly, our approach has the flexibility for a system designer to select a dictionary size based on their effector's behavioral specifications such as allowable number of user inputs, minimum configuration accuracy, and number of effector parameters (i.e., degrees of freedom). Once the designer decides on a fixed number of dictionary elements, they can then distribute this fixed number of elements among their degrees of freedom in a customized manner by tuning the size of each character's alphabet, allowing for variable resolutions between parameters. More generally, by iteratively refining effector behavior through a sequence of low-complexity inputs rather than requiring a single high-fidelity measurement to instantaneously extract a total system state from the BCI user, SCINET complements years of research devoted to improving the input mechanisms of BCIs by instead fundamentally redesigning how inputs are utilized.

\section{Acknowledgments}
This work was supported in part through National Science Foundation CAREER award CCF-1350954, by grant number FA9550-13-1-0029 from the US Air Force Office for Scientific Research, and the Center for Advanced Brain Imaging at the Georgia Institute of Technology. We thank Chethan Pandarinath, Adam Willats, and other colleagues for their comments.
	
\bibliographystyle{IEEEtran}
\bibliography{refs}

\clearpage

\appendix

\section{Appendix}

\subsection{Ancillary Files}

\noindent\textbf{userstudy.pdf:} exact user study HIT instructions and task, as they appeared on Amazon Mechanical Turk. The HIT was originally viewed in a web browser as an HTML page, and the page sizing here is an artifact of conversion to a pdf file type.

\noindent\textbf{threshold\_table.csv:} lookup table to choose BZ algorithm stopping threshold. The leftmost column contains crossover probability values, and the top row contains maximum tolerable expected number of inputs. To choose a threshold, the BCI user finds the row corresponding to the trained motor imagery classifier’s crossover probability (i.e., misclassification probability) or the next highest value in the table if it does not appear exactly, and finds the column corresponding to their maximum tolerable expected number of inputs. The value at this row and column entry is used as a BZ stopping threshold. In the virtual swarm trials, the first author used a tolerable expected number of inputs of 25, and this threshold selection process was automated with computer code.

\noindent\textbf{physical\_swarm\_1.mov:} video of the first SCINET trial for the control of a physical robot swarm. The top section displays a side (left) and overhead (right) system view. The target configuration is illuminated on the arena, and is accentuated as a blue outline in post-production. The bottom section presents additional post hoc analysis (not presented to the user during the trial). On the bottom left, the correct input (according to the rules of posterior matching) is displayed as either “L” (left) or “R” (right), along with the actual issued input. On the bottom right, after each input a posterior probability distribution is displayed over the configuration dictionary. The mass point corresponding to the correct target configuration is highlighted in blue, and the mass point corresponding to the current estimate is highlighted in red. The stopping threshold is depicted as a dashed horizontal line.

\noindent\textbf{physical\_swarm\_2.mov:} video of the second SCINET trial for the control of a physical robot swarm. See physical\_swarm\_1.mov for a video description. The side view and audio is missing for the final 18 seconds of the video, which is after swarm convergence. The overhead view still depicts the swarm idling on the correct configuration for this duration after convergence.
 
\subsection{Additional Methods}
\label{sec:suppmethods}

\subsubsection{Cheating Detection}
\label{sec:cheat}

We now describe the procedure used to flag and examine online user study participants suspected of selecting answers at random. To automatically flag suspicious sets of responses, we established a set of benchmarks evaluating response time, overall accuracy, accuracy per character, and accuracy over the course of the HIT; a participant failing any of these benchmarks resulted in a manual approval or rejection of their HIT. A participant was flagged for suspicious duration if they completed the entire task (including reading instructions and submitting answers) in under 25 minutes, which may indicate that a sincere effort was not made to answer each query carefully. This duration threshold was estimated by the authors as the approximate time our HIT might take to complete at a reasonable completion pace.

When evaluating cheat detection queries (6 total), a participant was flagged if they correctly answered exactly 2, 3, or 4 queries. Conversely, a participant passed this benchmark if they answered 5 or 6 cheat detection questions correctly --- performing as expected --- or if they answered only 0 or 1 test queries correctly, as might be the case if they made sincere efforts but had a reversed understanding of the dictionary ordering.

A participant was flagged for suspicious overall accuracy if their p-value for a two-tailed hypothesis test exceeded a threshold of 10\%, with a null hypothesis of chance selection (50\% probability of correct selection per query). We estimated this p-value with a normal approximation to a binomial distribution with a 50\% bias over 150 queries (including cheat detection queries). The p-value is computed as the probability of an overall accuracy at or further from chance than the measured accuracy. Specifically, for $c$ correct responses over 150 shape pair queries and letting $\Phi$ denote the standard normal cumulative distribution function (c.d.f.), this p-value is calculated as
\begin{equation}
p_1(c) = 2\min\Biggl(\Phi\Biggl(\frac{\frac{c}{150} - 0.5}{\sqrt{\frac{0.25}{150}}}\Biggr),1-\Phi\Biggl(\frac{\frac{c}{150} - 0.5}{\sqrt{\frac{0.25}{150}}}\Biggr)\Biggr).\label{eq:pchance1}
\end{equation}

Two additional benchmark's were evaluated involving p-value evaluation per critical character, and error performance across the HIT duration. In the former, p-values were calculated as in \cref{eq:pchance1} when binning trials (including the six test questions) by critical character. These p-values were multiplied to form a ``net'' p-value, which flagged a participant when it exceeded 10\%. To evaluate error performance throughout the duration of a HIT, an ordinary least-squares regression was fit to the cumulative number of correct responses throughout the course of the HIT. If the $r^2$ value of the linear model fell below 0.64, then the participant was flagged. The motivation behind this test was to detect participants who performed well initially, but then decided to select random answers for the remainder of the HIT; such a participant would show a highly nonlinear error performance over the course of the HIT, unlike participants who answered consistently according to their understanding of the task. Additionally, such nonlinear behavior would not be accounted for by a varying difficulty level over multiple queries, since the sequence of queries was presented in a shuffled order. No participants were flagged for this linear model test, and so we omit its details from the discussion here. Only one participant (participant 145) was flagged for their ``net'' p-value surpassing 10\%, but this participant was also flagged for total accuracy (\cref{eq:pchance1}) and so we only discuss the latter.
 
Participants 2, 131, and 145 were flagged for cheat detection query responses (each answering 4 test queries correctly out of 6), and participants 131 and 145 were additionally flagged for overall accuracy. Participant 2 had a duration of 54 minutes, and so was approved due to an assumption of genuine effort due to their extensive completion time. Participants 131 and 145 had completion times of 85 and 58 minutes respectively, and so were also approved due to extensive completion time. Participant 33 was flagged for total accuracy only; due to a duration of 65 minutes and answering all 6 cheat detection queries correctly, they were approved. Participants 4, 41, 43, 51, 61, 63, 68, 71, 80, 86, 95, 106, 125, 139, 142, and 144 were flagged for duration only. None of these participants had an accuracy below 96\% (over all 150 queries), and all had a duration of at least 14 minutes. Since the durations were still significant and all performed at high accuracy, these participants flagged for durations were approved since their behavior did not indicate random selection. Overall, all 150 participants were approved in this study.

\subsubsection{Common Spatial Patterns Filtering}
\label{sec:CSP}

Common Spatial Patterns (CSP) filtering is a supervised spatial filtering method that maximizes the difference in filtered signal variances between two classes \cite{ramoser2000optimal, muller1999designing}. This separation is useful for motor imagery detection, which uses signal power (i.e., signal variance) as the classification feature. We briefly summarize our implementation of the CSP algorithm \cite{ramoser2000optimal,muller1999designing} below, with notation and derivations drawn largely from Ramoser et al.\ \cite{ramoser2000optimal}.

Let $\{X_{c,i}\}_{i=1}^{N_c}$ denote the training set of $N_c$ temporally filtered EEG signals for class $c \in \{l,r\}$ (for ``left'' and ``right''), where each $X_{c,i}$ is a $T \times d$ matrix of $T$ EEG samples over $d$ channels. Let $\mu_{c,i} = \frac{1}{T}\sum_{t=1}^T X_{c,i}[t,:]$ denote the spatial mean of signal block $i$ for class $c$, where $X_{c,i}[t,:]$ denotes the $t$th row of matrix $X_{c,i}$. We define the zero-mean signal matrix $\widetilde{X}_{c,i}$ by subtracting the spatial mean from each sample, i.e., $\widetilde{X}_{c,i}[t,:] = X_{c,i}[t,:] - \mu_{c,i}$. Then, we compute the averaged covariance matrices for both classes as
\begin{equation*}
C_l = \frac{1}{N_l}\sum_{i=1}^{N_l} \frac{\widetilde{X}_{l,i}^T \widetilde{X}_{l,i}}{\operatorname{Tr}(\widetilde{X}_{l,i}^T \widetilde{X}_{l,i})} \qquad C_r = \frac{1}{N_r}\sum_{i=1}^{N_r} \frac{\widetilde{X}_{r,i}^T \widetilde{X}_{r,i}}{\operatorname{Tr}(\widetilde{X}_{r,i}^T \widetilde{X}_{r,i})}
\end{equation*}
and form the composite covariance matrix as
\begin{equation}
C = C_l + C_r .\label{eq:compositecov}
\end{equation}

We then factor $C$ into its eigendecomposition $C = U \Lambda U^T$, where the eigenvalues in diagonal matrix $\Lambda$ are sorted in descending order and the columns of $U$ are orthogonal eigenvectors. Note that $\Lambda$ only has at most $d-1$ positive eigenvalues. To see this, note that since the raw EEG signals are processed with CAR referencing, each $X_{c,i}$ has rank at most $d-1$ since by definition its columns are linearly dependent, i.e., $\sum_{k=1}^d X_{c,i}[t,k] = 0$ for all $1 \le t \le T$. $\widetilde{X}_{l,i}$ also has rank at most $d-1$ since
\begin{align*}
\sum_{k=1}^d \widetilde{X}_{c,i}[t,k] &= \sum_{k=1}^d \bigl[X_{c,i}[t,k] - \mu_{c,i}[k]\bigr] \\
&= \sum_{k=1}^d \biggl[X_{c,i}[t,k] - \frac{1}{T}\sum_{j=1}^T X_{c,i}[j,k]\biggr] \\
&= \sum_{k=1}^d X_{c,i}[t,k] - \frac{1}{T}\sum_{j=1}^T \biggl(\sum_{k=1}^d X_{c,i}[j,k]\biggr) \\
&= 0 - \frac{1}{T}\sum_{j=1}^T 0 \\
&= 0.
\end{align*}

Next, consider the $d \times T(N_l + N_r)$ matrix given by the horizontal concatenation of all $\{\widetilde{X}_{c,i}^T\}_{i=1}^{N_c}$, $c \in \{l,r\}$, i.e., $\widetilde{X} = [\widetilde{X}_{l,1}^T,\widetilde{X}_{l,2}^T,\dots \widetilde{X}_{l,N_l},\widetilde{X}_{r,1}^T,\widetilde{X}_{r,2}^T,\dots \widetilde{X}_{r,N_r}]$. Clearly $\widetilde{X}$ has rank at most $d-1$, since $\sum_{j=1}^d \widetilde{X}[j,:] = 0$ by construction. Therefore the column space of $\widetilde{X}$, i.e., the span of $\{\widetilde{X}_{c,i}^T\}_{i=1}^{N_c}$, $c \in \{l,r\}$, has dimension at most $d - 1$. By construction, the columns of $C$ (\cref{eq:compositecov}) lie in this same column space, and so $C$ has rank at most $d-1$.

Therefore, before the whitening stage of CSP, we truncate $\Lambda$ to the top $d-1$ eigenvalues, resulting in $d-1 \times d-1$ matrix $\widehat{\Lambda}$, and truncate $\widehat{U}$ as the first $d-1$ columns of $U$. We then form the whitening transformation $P = \sqrt{\widehat{\Lambda}^{-1}} \widehat{U}^T$, so that $P C P^T = \sqrt{\widehat{\Lambda}^{-1}} \widehat{U}^T \widehat{U} \widehat{\Lambda} \widehat{U}^T \widehat{U} \sqrt{\widehat{\Lambda}^{-1}} = I$. Let $S_l = P C_l P^T$ and $S_r = P C_r P^T$. Then $S_l$ and $S_r$ share eigenvectors with eigenvalues that sum to unity, i.e.,
\[S_l = B \lambda_l B^T \qquad S_r = B \lambda_r B^T \qquad \lambda_l + \lambda_r = I.\]

To see this, note that $S_l + S_r = P(C_l + C_r)P^T = P C P^T = I$, and suppose that $b$ is an eigenvector for $S_l$ with eigenvalue $\lambda_l$. Then $S_l b = \lambda_l b$, and so
\[b = Ib = (S_l + S_r)b = S_l b + S_r b = \lambda_l b + S_r b \implies S_r b = (1 - \lambda_l)b\]
and therefore $b$ is an eigenvector for $S_r$ with eigenvalue $1 - \lambda_l$. This implies that projecting onto the eigenvectors of $S_l$ with the largest eigenvalues will result in variance separation between whitened data from classes $l$ and $r$: whitened data from $l$ will remain mostly unattenuated, while whitened data from class $r$ will have attenuated signal energy. The reverse relationship is true when projecting onto the top eigenvectors of $S_r$ instead.

The final CSP filtering matrix is then constructed as $W = B^T P$, where the columns of $B$ are sorted in decreasing order from largest to smallest corresponding eigenvalues of $S_l$. This transformation whitens incoming data and projects it onto each eigenvector of $S_l$, such that the resulting vector has components with separated energy levels for each class. This $d-1 \times d$ filter is applied to a $T \times d$ signal $X$ as $Z = XW^T$, where $Z$ is of size $T \times d-1$. In SCINET, we apply a truncated CSP filter that only uses the top two spatial filters for each class. Letting $w_i^T$ denote the $i$th row of $W$, we use only rows $1,2,d-2,d-1$ of the filter (corresponding to the top two eigenvectors of each class), i.e., we apply filter
\[\widehat{W} = \begin{bmatrix} \horzbar & w_1^T &  \horzbar \\ \horzbar & w_2^T &  \horzbar \\ \horzbar & w_{d-2}^T &  \horzbar \\ \horzbar & w_{d-1}^T &  \horzbar \end{bmatrix}.\] 

\subsubsection{Posterior Matching Implementation}
\label{sec:PMimp}
In this section, our notation and conventions for dictionary construction are inspired from Omar et al.\ \cite{omar2010feedback}, and the mathematical algorithm is drawn from Castro and Nowak \cite{castro2008active}. Let $\bm{z}_j = \{\sigma_j^{h},\sigma_j^{v},\sigma_j^{n},\sigma_j^{s}\}$ denote the $j$th swarm configuration in the dictionary, where $\sigma_j^{h},\sigma_j^{v},\sigma_j^{n}$, and $\sigma_j^{s}$ respectively denote character indices in the horizontal position (h), vertical position (v), number of sides (n), and size (s) alphabets. Letting $N_a$ denote the number of characters in alphabet $a \in \{h,v,n,s\}$, we have $\sigma_j^{a} \in 1,2,\dots N_a$ with alphabet precedence corresponding to character index ordering, i.e., character $\sigma_j^{a}$ precedes character $\sigma_k^{a}$ in alphabet $a$ if and only if $\sigma_j^{a} < \sigma_k^{a}$. Letting $N_d = N_hN_vN_sN_s$ denote the total number of strings in the dictionary, string $\bm{z}_j$ precedes string $\bm{z}_k$ in the total dictionary ordering if and only if $j < k$, where $j,k \in 1,2,\dots N_d$. Equivalently, $\bm{z}_j$ precedes $\bm{z}_k$ if and only if $\sigma_j^{a^*} < \sigma_k^{a^*}$, where $a^*$ is the first character position where $\bm{z}_j$ and $\bm{z}_k$ differ. 

With this dictionary notation established, let $Z_j = \frac{\sigma_j^{h}-1}{N_h} + \frac{\sigma_j^{v}-1}{N_hN_v} + \frac{\sigma_j^{n}-1}{N_hN_vN_s} + \frac{\sigma_j^{s}-1}{N_hN_vN_sN_s} = {(j-1)/N_d}$ denote a real number representation of the $j$th dictionary string; it is straightforward to show that $Z_j \in [0,1)$ and that string $\bm{z}_j$ precedes $\bm{z}_k$ if and only if $Z_j < Z_k$. Each $Z_j$ corresponds to the start of a $1/N_d$ length interval, creating a mapping between the configuration dictionary and equally sized intervals that uniformly partition $[0,1)$. In particular, $Z_j \in \{0, 1/N_d, 2/N_d, \dots 1-1/N_d\}$, each corresponding to the start of interval $I_j = [(j-1)/N_d,j/N_d)$ with length $1/N_d$.

With this notation and real number mapping defined, posterior matching can be used as an interaction algorithm to convey the user's desired configuration. Posterior matching is a general feedback coding scheme for communication across memoryless channels that allows for optimal communication of arbitrary messages on the unit interval \cite{shayevitz2011optimal}. While we use the term ``posterior matching'' here to remain consistent with previous feedback information-theoretic BCI literature \cite{omar2010feedback}, the mathematical algorithm we use in this work is a discrete variation of posterior matching known as the Burnashev-Zigangirov (BZ) algorithm \cite{castro2008active, burnashev1974interval}. We use this variation since our swarm dictionary is discrete and finite, and therefore our message set corresponds to a finite partition of the unit interval rather than spanning the entire interval. Still, in this work we refer to the BZ algorithm interchangeably with ``posterior matching'' since the differences between the two algorithms are minor.

The BZ algorithm searches for one of $N_d$, length $1/N_d$ intervals on the unit interval by taking adaptive one-bit measurements of points on the set $\{0, 1/N_d, \dots 1-1/N_d, 1\}$ and updating a probability distribution over interval set $\{I_j\}$. We adapt this algorithm to searching over a finite, discrete dictionary by utilizing the mapping described above between each string and a subinterval $I_j$, and ``measuring'' configurations $Z_j \in \{0, 1/N_d, 2/N_d, \dots 1-1/N_d\}$ by presenting the corresponding configuration as a swarm behavior to the user. Mathematically speaking, the user indicates if their configuration's real number representation is less or greater than the guess's real number value. Unlike the BZ algorithm formulation, in our setting there exists no measurement $Z_j = 1$; we address this point below. While the original BZ algorithm tracks a probability distribution over intervals, for clarity in our adaption below we describe the posterior distribution over configurations directly, rather than the intervals they correspond to.

\paragraph{Initialization} Let the $j$th swarm string in the dictionary for $1 \leq j \leq N_d$ have a posterior probability after $k$ user inputs given by $\alpha_j(k) \in [0,1]$, corresponding to a probability distribution over the set of intervals $\{I_j\}$. We initialize this probability distribution with a uniform prior, with $\alpha_j(0)=\frac{1}{N_d}$.

\paragraph{Guess selection} After $k$ user inputs, define the posterior median $N(k) \in \{1\dots N_d\}$ as the dictionary index such that
\begin{equation}
\sum_{j=1}^{N(k)-1} \alpha_j(k) < 1/2, \quad \sum_{j=1}^{N(k)} \alpha_j(k) \ge 1/2.
\end{equation}
Denoting the swarm guess after $k$ inputs as $\bm{\hat{z}}(k)$, we set $\bm{\hat{z}}(k)$ to be an adjusted version of this median string. Specifically, let
\begin{equation*}
\nu_1(k)= \sum_{j=N(k)}^{N_d} \alpha_j(k) -  \sum_{j=1}^{N(k)-1} \alpha_j(k)
\end{equation*}
and
\begin{equation*}
\nu_2(k)= \sum_{j=1}^{N(k)} \alpha_j(k) -  \sum_{j=N(k)+1}^{N_d} \alpha_j(k).
\end{equation*}
The adjusted median configuration index, denoted $n(k)$, is set to $N(k)$ with probability  $\pi_1(k)=\nu_2(k)/(\nu_1(k)+\nu_2(k))$, or $N(k)+1$ with probability ${\pi_2(k)=1-\pi_1(k)}$. The swarm configuration is then updated as $\bm{\hat{z}}(k) = \bm{z}_{n(k)}$. For the edge case of $N(k) = N_d$, we set $n(k)=N_d$ with probability 1. This adjustment is due to the fact that, unlike the original BZ algorithm, there is no measurement at $Z_{N_d+1}=1$ available.

\paragraph{Noisy user input} The algorithm receives $Y_{k+1} = \operatorname{BSC}(X_{k+1},p)$ from the user, i.e., the output of a binary symmetric channel ($Y_{k+1}  \in \{0,1\}$) with crossover probability $0 \leq p < 1/2$, where $X_{k+1}$ is issued as a left-hand motor imagery input ($X_{k+1}=0$) if the target precedes the guess $\bm{z}_{n(k)}$, or a right-hand motor imagery input ($X_{k+1}=1$) if the target succeeds \emph{or equals} the guess. Mathematically, if $\bm{z}_t$ is the target configuration with real number representation $Z_t$, then $X_{k+1}=0$ if $Z_t < Z_{n(k)}$, and $X_{k=1}=1$ if $Z_t \ge Z_{n(k)}$.

\paragraph{Update posterior} Let $q = 1 - p$, and define
\begin{equation}
\nu= \sum_{j=1}^{n(k)-1} \alpha_j(k) -  \sum_{j=n(k)}^{N_d} \alpha_j(k).
\end{equation}

\noindent If $j < n(k)$, then
\begin{equation}
\alpha_j(k+1)=
\begin{cases}
\frac{2q}{1+\nu(q-p)}\alpha_j(k) & Y_{k+1}=0\\
\frac{2p}{1-\nu(q-p)}\alpha_j(k) & Y_{k+1}=1
\end{cases}.
\end{equation}
Otherwise, if $j \geq n(k)$ then
\begin{equation}
\alpha_j(k+1)=
\begin{cases}
\frac{2p}{1+\nu(q-p)}\alpha_j(k) & Y_{k+1}=0\\
\frac{2q}{1-\nu(q-p)}\alpha_j(k) & Y_{k+1}=1
\end{cases}.
\end{equation}

\paragraph{Algorithm notes}
We can opt to run the BZ algorithm with a stopping criterion. To do so, a priori the BCI user must choose a threshold parameter $\tau \in [0,1]$ which corresponds to a convergence confidence threshold. Noting that $\alpha_j(k)$ is the posterior probability after $k$ inputs of swarm configuration $j$ being the user's desired configuration, the algorithm halts at the first instance of $\alpha_j(k) \geq \tau$ for any $j,k$. Let $j^*$ denote the configuration whose posterior probability crosses the threshold, and $k^*$ be the number of user inputs received at this point. We say that the algorithm ``converges'' to swarm $j^*$ after $k^*$ user inputs, and $j^*$ is selected as the final estimate of the user's configuration. Note that $1-a_{j^*}(k^*)$ is the posterior probability of the ground truth target being a configuration other than $j^*$ --- in other words, the probability of a configuration convergence error. Therefore, $1-\tau$ can be interpreted as a maximum error tolerance for convergence, or conversely $\tau$ is a threshold for minimum convergence accuracy.

When simulating posterior matching, for simplicity the simulation can operate directly on the unit interval, rather than needing to maintain and operate on the full set of characters $\{\sigma^h,\sigma^v,\sigma^n,\sigma^s\}$ for each dictionary string. Instead, we can simply track the posterior distribution $\alpha_j(k)$ over the dictionary strings directly, and simulate user responses by comparing the real number representations of the current guess $Z_{n(k)}$ and the target $Z_t$. Note that once the dictionary size $N_d$ is specified, such a simulation can be run without explicit knowledge of the characters that each string corresponds to, alphabets, alphabet sizes, or number of degrees of freedom. This is because the total order between configurations is fully captured by their representations $Z_j$ on the unit interval, obviating the need to make comparisons between specific characters. Such comparisons are only relevant when a human user issues commands, since the dictionary representation is crucial for a human to be able to sort according to the total ordering. However, when running simulations on a computer we can forgo this step and operate directly on the unit interval.

\subsubsection{Stepwise Search}
\label{sec:step}

Stepwise search is a Bayesian algorithm that tracks a probabilistic estimate of the user's desired configuration by incrementing or decrementing guesses one string at a time to navigate the dictionary. The initialization of stepwise search is identical to that of posterior matching.

\paragraph{Guess selection}
At $k=0$, the swarm guess is initialized as $n(0) = \lfloor N_d / 2 \rceil$ and $\bm{\hat{z}}(0) = \bm{z}_{n(k)}$, where the $\lfloor x \rceil$ operator rounds $x$ to the nearest integer. For $k > 0$, the guess is updated as
\[n(k) = \max(\min(n(k-1) + (2Y_k - 1),N_d),1)\qquad \bm{\hat{z}}(k) = \bm{z}_{n(k)}.\]
Recalling that $Y_k \in \{0,1\}$, this guessing rule is equivalent to incrementing or decrementing the guessed configuration by one string position in the direction indicated by the received input, while accounting for edge cases (receiving $Y_k=0$ for $n(k-1)=1$ maintains $n(k)=1$, and receiving $Y_k=1$ for $n(k-1)=N_d$ maintains $n(k)=N_d$). The noisy user input and posterior update stages are identical to those in posterior matching.

\FloatBarrier

\subsubsection{Additional Figures}

\begin{figure}[htb]
	\centering
	\includegraphics[width=0.6\textwidth]{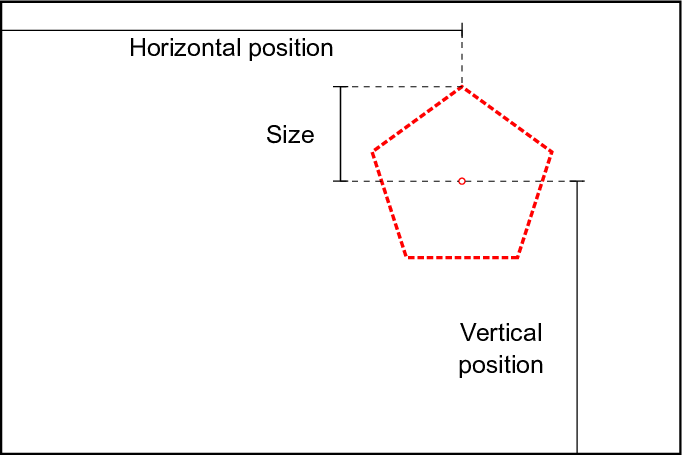}
	\caption[Polygon dictionary parameters, specified relative to the swarm arena dimensions.]{Polygon dictionary parameters, specified relative to the swarm arena dimensions.}
	\label{fig:polyparam}
\end{figure}

\begin{figure}[htb]
	\centering
	\includegraphics[width=0.6\textwidth]{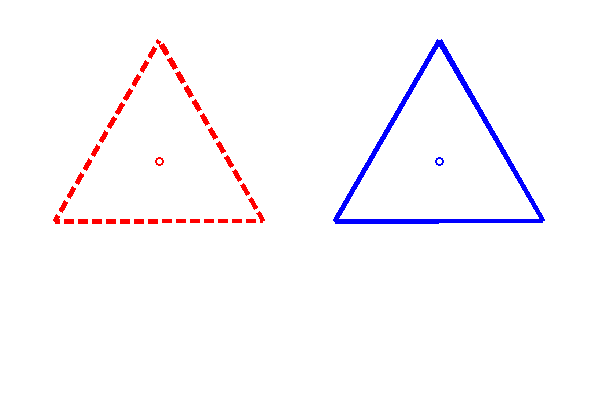}
	\caption[Shape query for HIT cheating detection.]{\textbf{Shape query for HIT cheating detection.} This query appeared 6 times in each participant's query set, randomly scattered among regular queries. Participants were not told that these queries were used for cheat detection, although they were told that random selection without a ``good faith'' effort would be detected (without stating how) and their HIT submission would then be rejected.}
	\label{fig:test_shape}
\end{figure}

\begin{figure}[htb]
	\centering
	\begin{subfigure}[t]{0.32\textwidth}
		\centering
		\includegraphics[width=\textwidth, trim={1400 1300 1300 30}, clip=true]{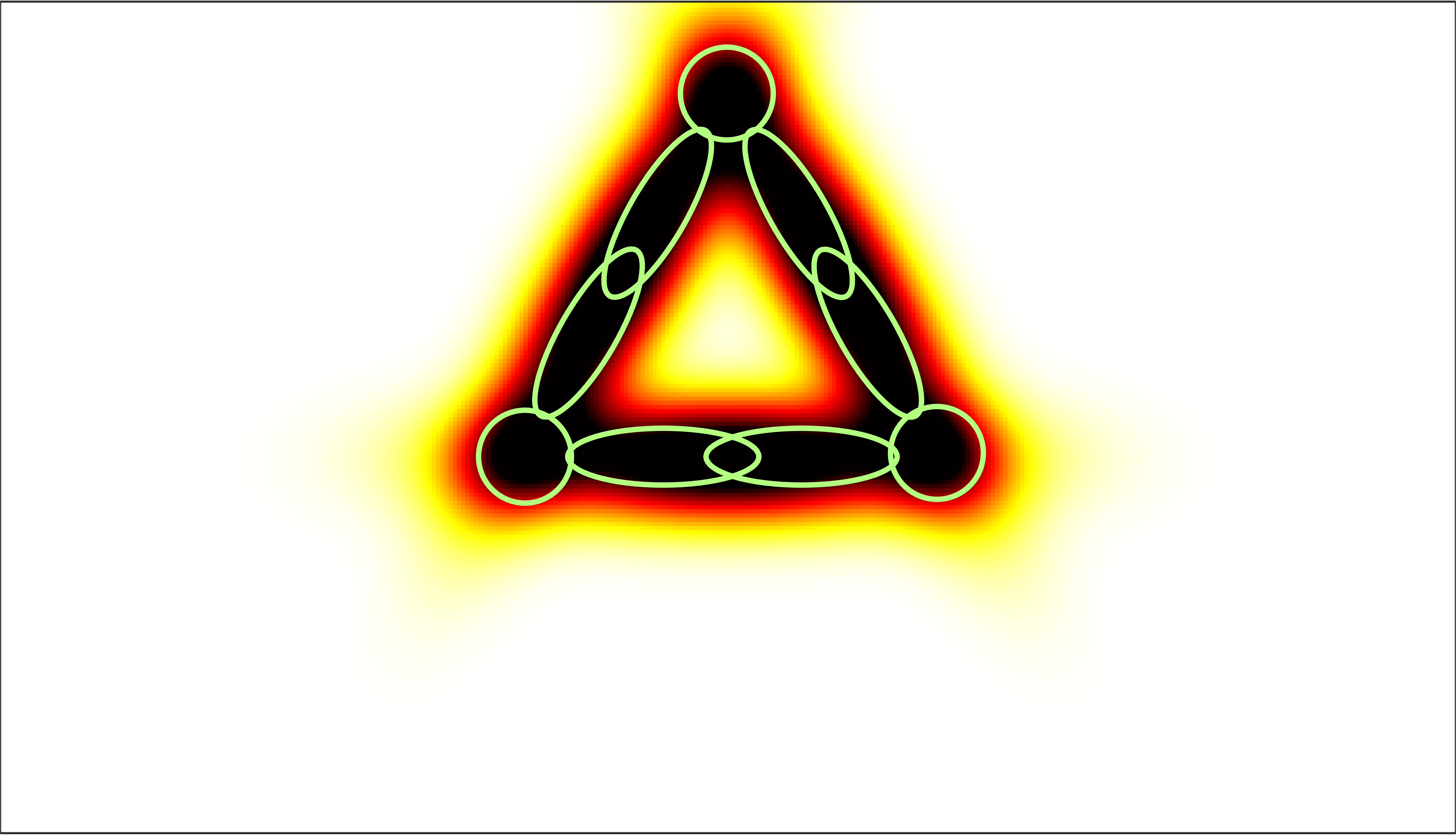}
		\caption{}
		\label{subfig:GMM-style}
	\end{subfigure}%
	\hfill
	\begin{subfigure}[t]{0.32\textwidth}
		\centering
		\includegraphics[width=\textwidth, trim={187 173 173 7}, clip=true]{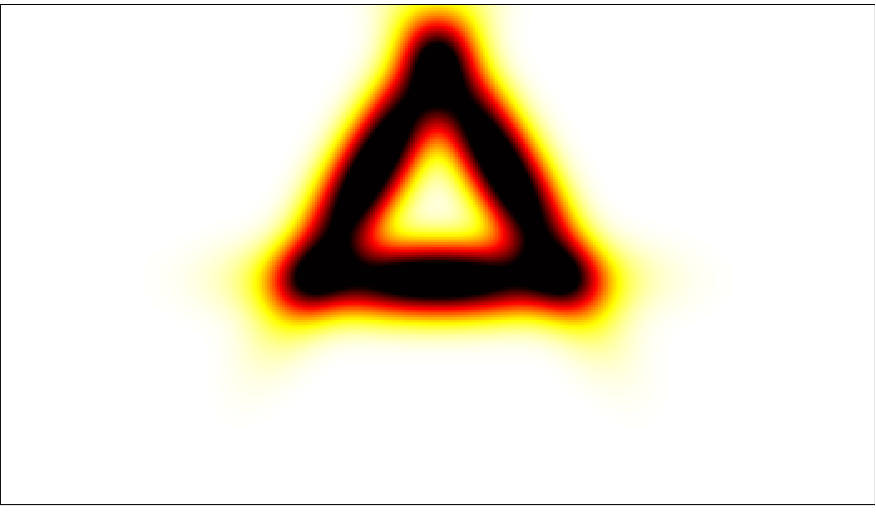}
		\caption{}
		\label{subfig:GMM-norobots}
	\end{subfigure}%
	\hfill
	\begin{subfigure}[t]{0.32\textwidth}
		\centering
		\includegraphics[width=\textwidth, trim={187 173 173 7}, clip=true]{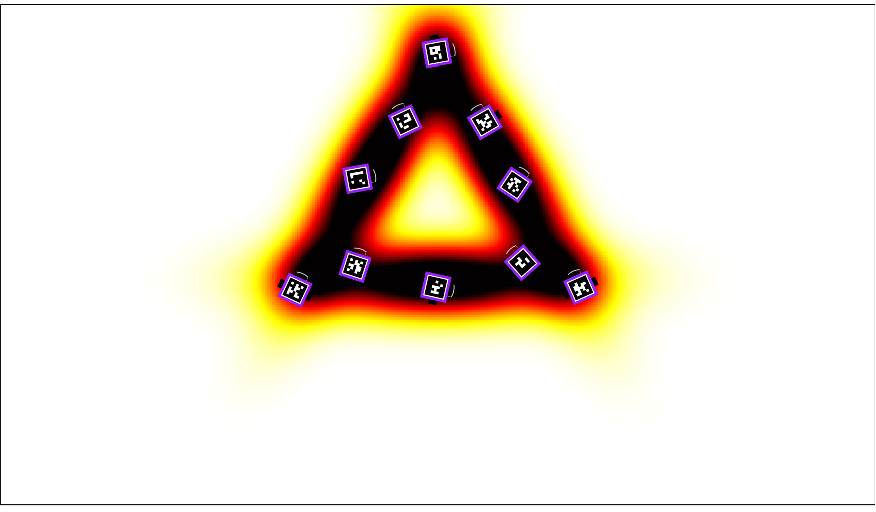}
		\caption{}
		\label{subfig:GMM-robots}
	\end{subfigure}%
	\caption[Gaussian mixture modeling for swarm density coverage.]{\textbf{Gaussian mixture modeling for swarm density coverage.} \textbf{a,} Gaussian mixture model displayed over a triangle target configuration. Each individual component of the GMM is stylized with a green outline. \textbf{b,} Gaussian mixture model density function, without stylization. As described in Diaz-Mercado et al.\ \cite{diazmercado2015distributed}, the robot swarm executes a distributed, low-level algorithm to cover the specified coverage density function, where each robot uses only local information. \textbf{c,} Gaussian mixture model density with virtual robots performing swarm coverage.}
	\label{fig:GMM}
\end{figure}

\begin{figure}[htb]
	\centering
	\def\fh{2.15in}
	\begin{subfigure}[t]{0.65\textwidth}
		\centering
		\includegraphics[height=\fh]{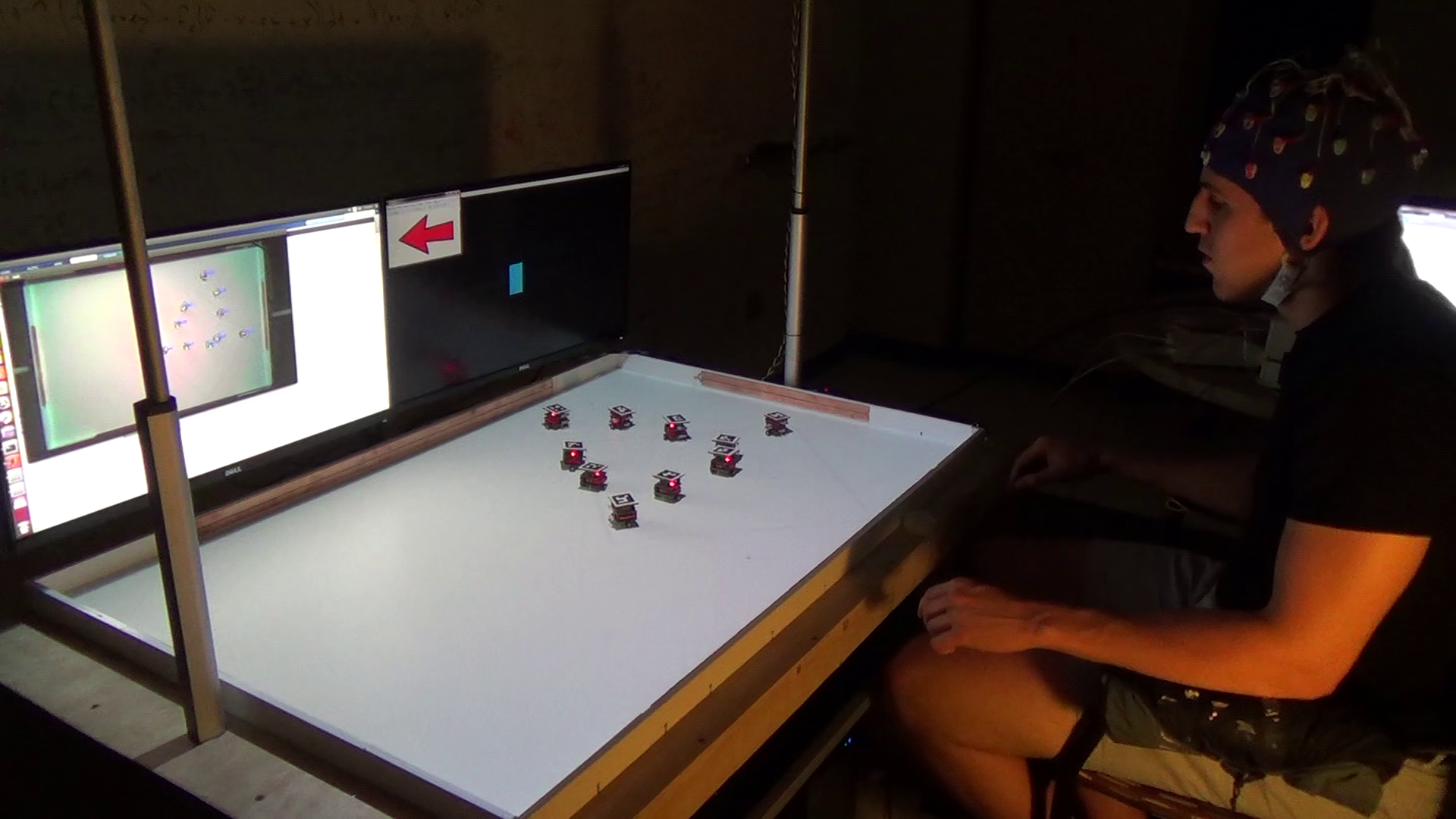}
		\caption{}
		\label{subfig:overview}
	\end{subfigure}%
	\hfill
	\begin{subfigure}[t]{0.35\textwidth}
		\centering
		\includegraphics[height=\fh]{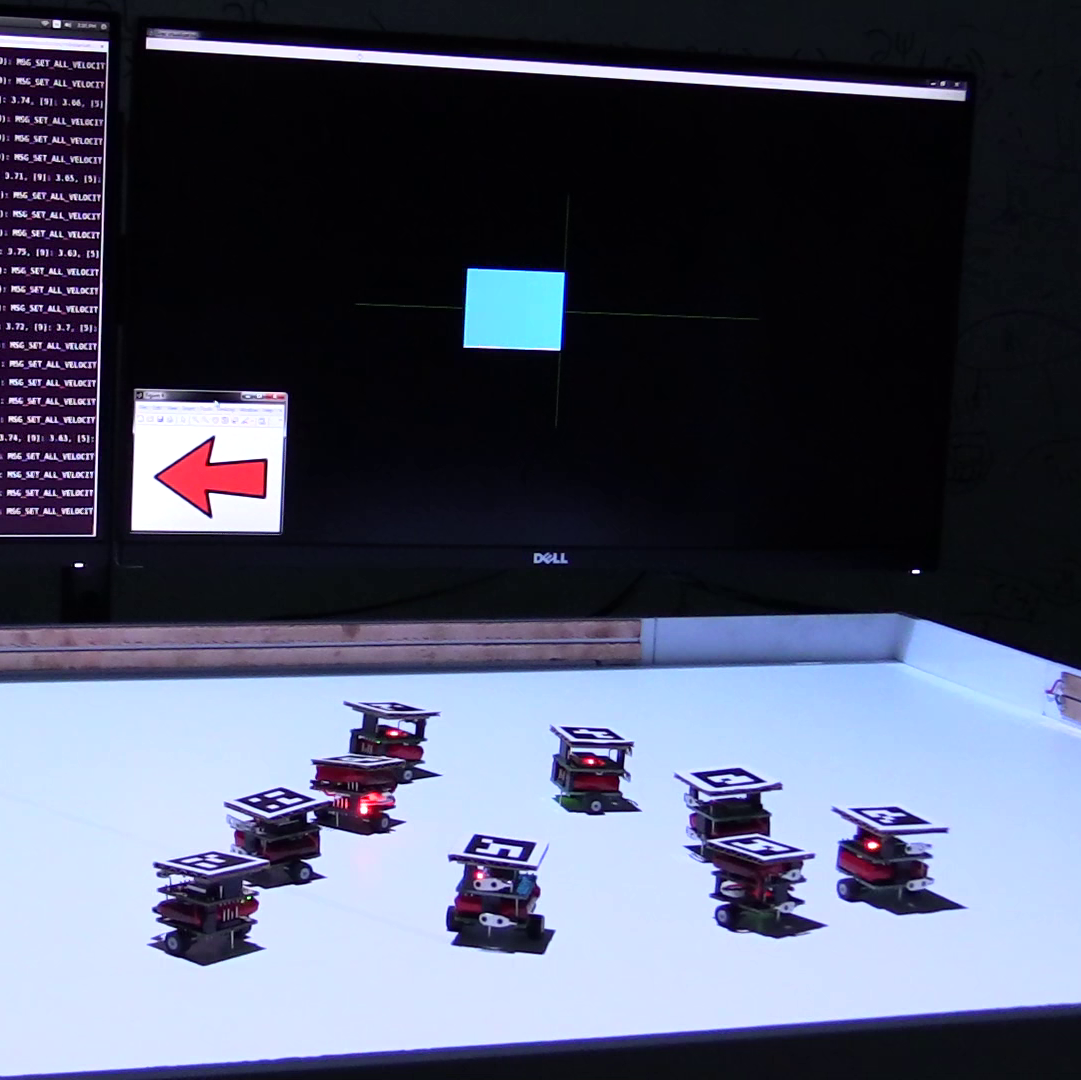}
		\caption{}
		\label{subfig:interface_cropped}
	\end{subfigure}%
	\caption[Physical swarm control.]{\textbf{Physical swarm control.} The BCI user observes the current swarm guess to decide their next motor imagery input according to the rules of posterior matching. When prompted by an audible beep, the user fixates on a blue feedback bar and imagines their intended motor imagery command; the bar responds in real-time to indicate the command being classified, which can aid the user in issuing a reliable command. After each command is issued during a synchronous window of 5 seconds, another beep sounds and an arrow indicates to the user their classified input. While the virtual swarm trials were conducted in a different room with different monitors than those shown here, the software interface used to elicit user motor imagery commands and present feedback was identical to that presented here. The first author consents to his image being published.}
	\label{fig:physical}
\end{figure}

\begin{figure}[htb]
	\centering
	\includegraphics[width=0.6\textwidth]{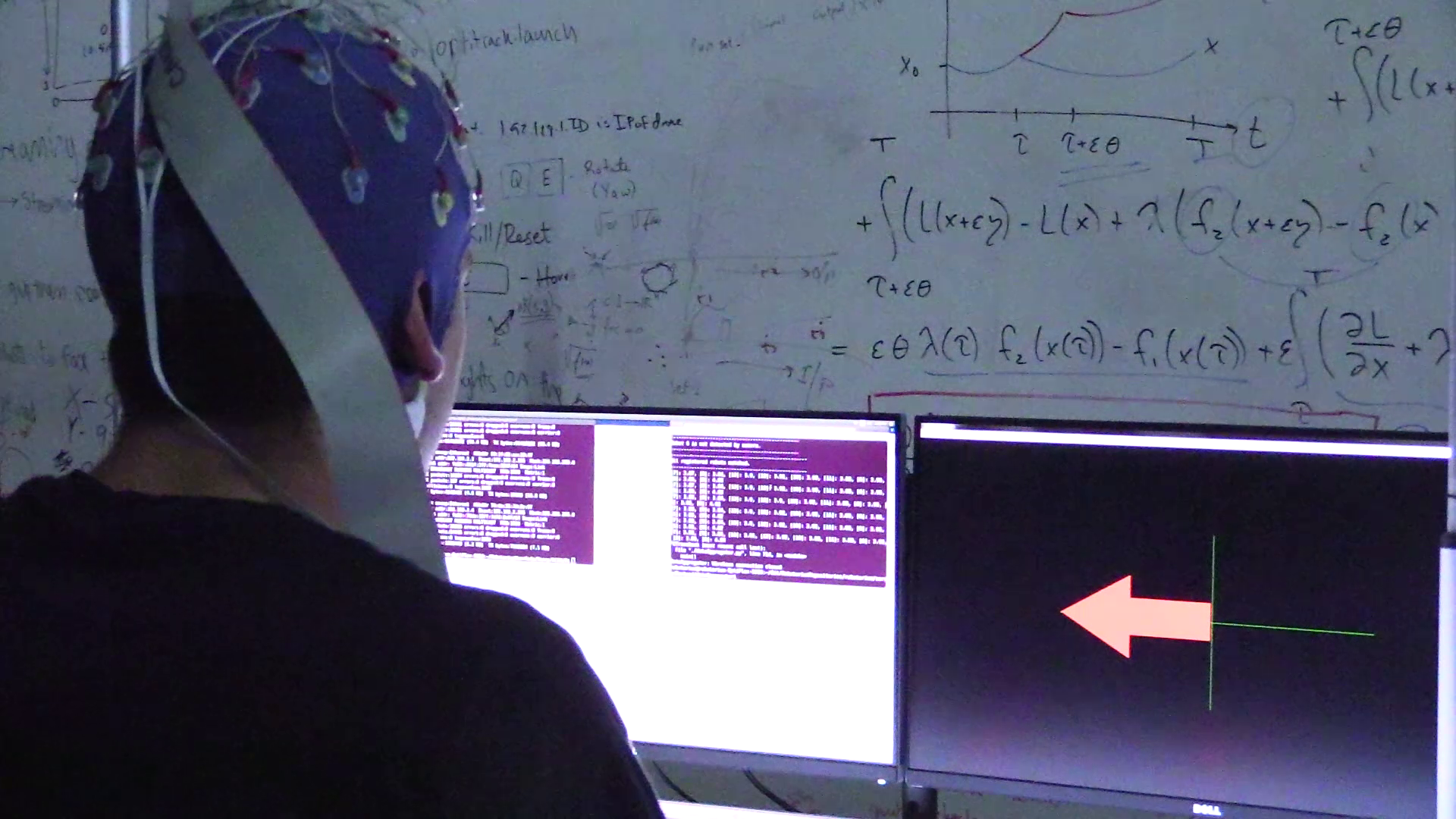}
	\caption[Motor imagery training.]{\textbf{Motor imagery training.} During motor imagery classifier training, the user imagines left or right hand motor imagery movements according to a synchronized visual cue.}
	\label{fig:training}
\end{figure}

\begin{figure}[htb]
	\centering
	\includegraphics[width=1.0\textwidth]{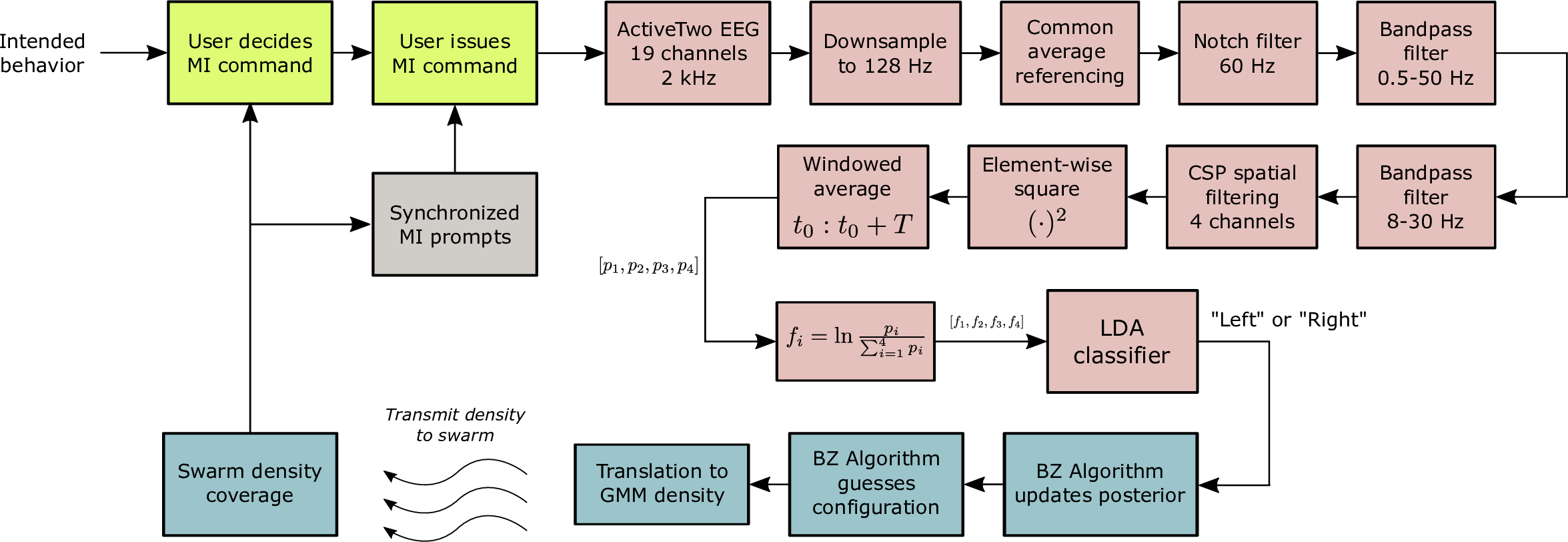}
	\caption[Full SCINET feedback system.]{\textbf{Full SCINET feedback system.} The user (green boxes, top left) observes the current swarm configuration along with synchronized motor imagery prompts (see \Cref{fig:physical}) to decide on a motor imagery command for communicating their intended behavior according to the rules of posterior matching. In a signal processing pipeline (pink boxes, top right) beginning with EEG scalp recording and ending with an LDA classifier, the user's motor imagery command is classified from raw scalp measurements. The detected motor imagery command is input into the BZ algorithm, which updates the configuration guess in the dictionary. This configuration is translated to a Gaussian mixture model density (see \Cref{fig:GMM}) and transmitted to the swarm for distributed density coverage.}
	\label{fig:pipeline}
\end{figure}

\begin{figure}[htb]
	\centering
	\begin{subfigure}[t]{0.5\textwidth}
		\centering
		\includegraphics[width=\textwidth]{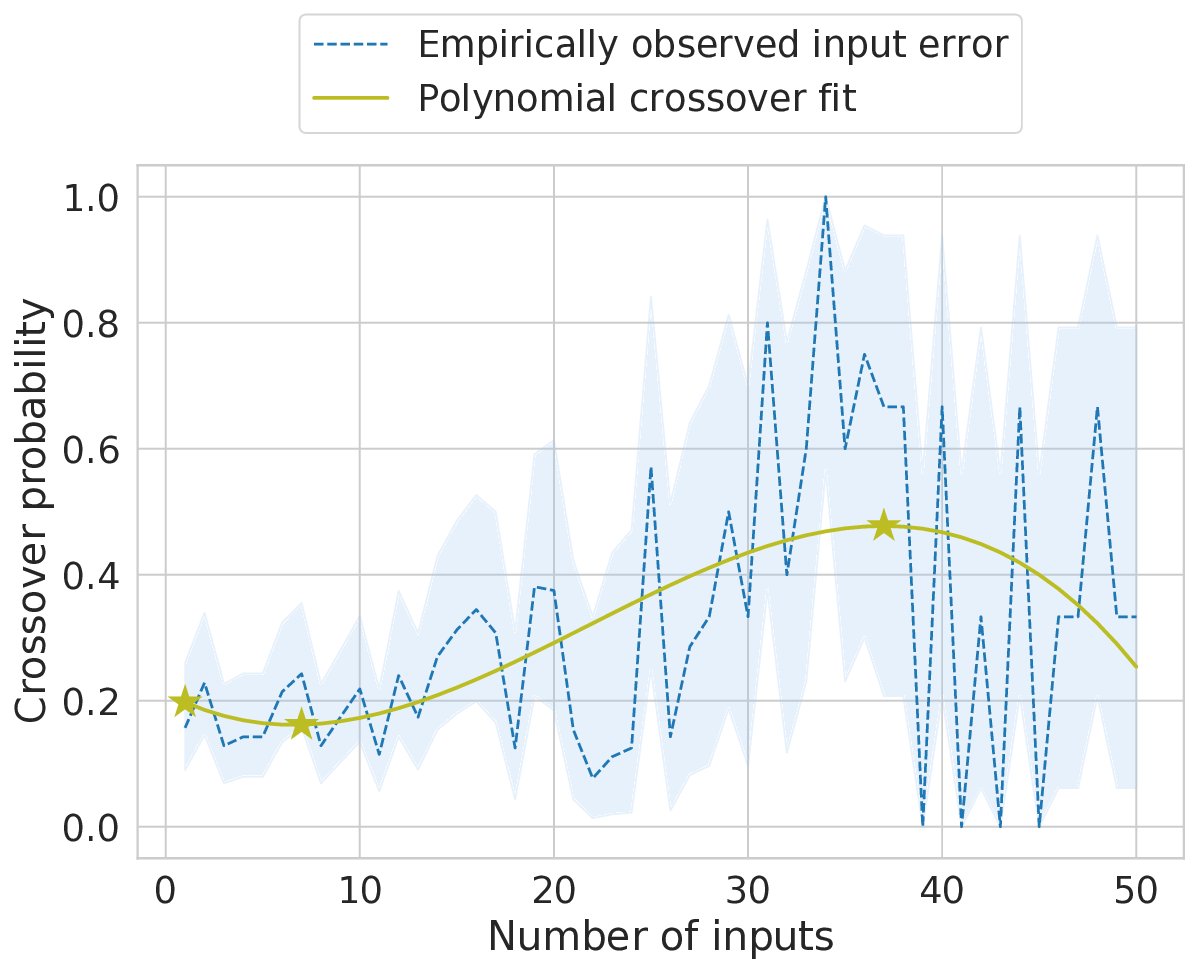}
		\caption{}
		\label{subfig:bscpoly}
	\end{subfigure}%
	\hfill
	\begin{subfigure}[t]{0.5\textwidth}
		\centering
		\includegraphics[width=\textwidth]{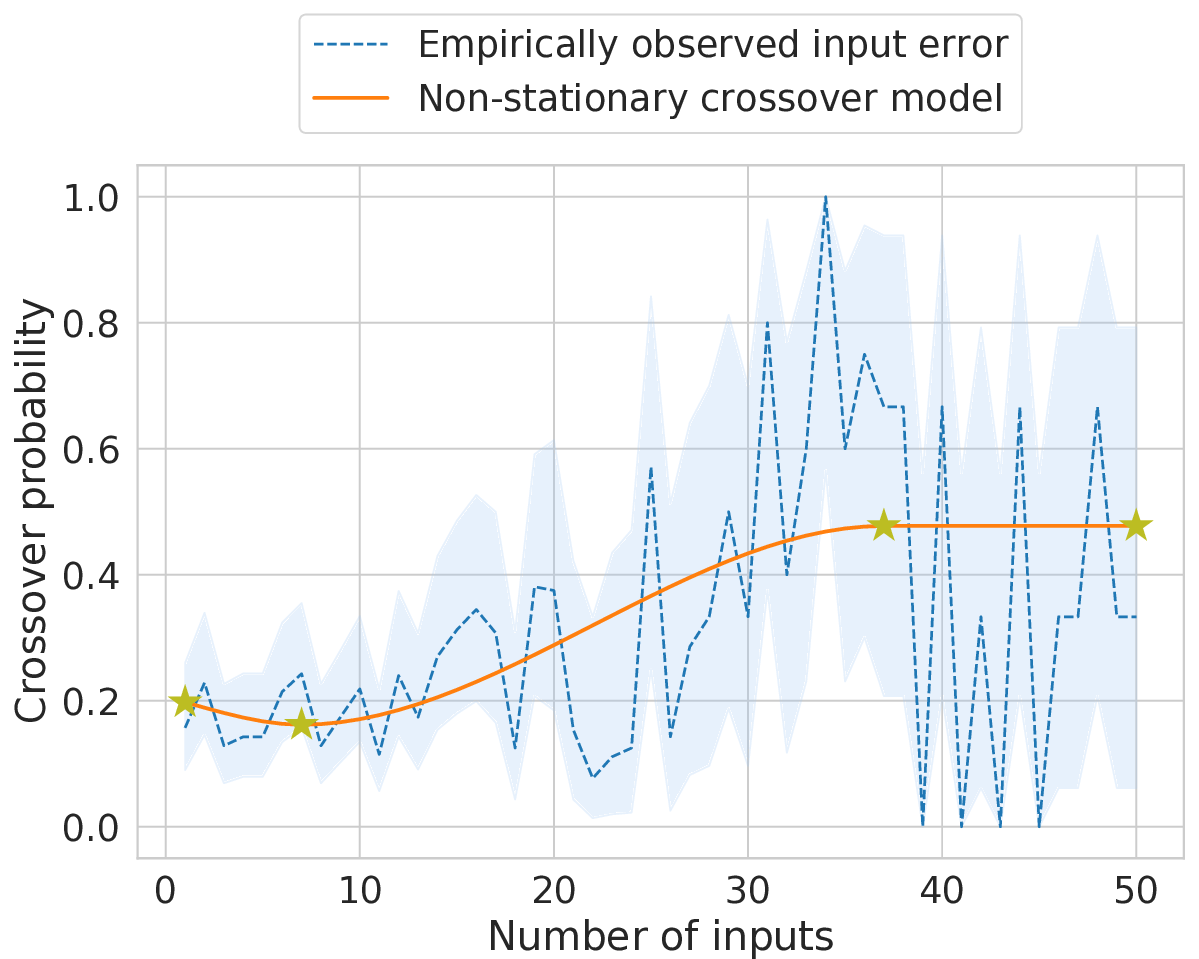}
		\caption{}
		\label{subfig:bscpchip}
	\end{subfigure}%
	\caption[Modeling a non-stationary input profile from empirical crossover data.]{\textbf{Modeling a non-stationary input profile from empirical crossover data.} \textbf{a,} Least-squares cubic fit to empirical crossover data. The model is highlighted with stars at the first input value, the minimum value, and the maximum value. \textbf{b,} A piecewise cubic Hermite interpolating polynomial (PCHIP) was fit to the first, minimum, and maximum value points from \textbf{a}. We clamped the maximum value to remain fixed until the maximum of 50 inputs, so that the resulting PCHIP error model is non-decreasing. This behavior models the fact that the BCI user may experience fatigue as the number of inputs increases, possibly resulting in non-improving input error statistics after issuing many inputs.}
	\label{fig:bscfit}
\end{figure}

\FloatBarrier

\subsection{Additional Discussion}
\label{sec:suppdisc}

\subsubsection{Online User Study Critical Character Analysis}

\begin{figure}[htb]
	\centering
	\def\fh{3in}
	\begin{minipage}{0.6\textwidth}
		\centering
		\begin{subfigure}[t]{\textwidth}
			\centering
			\includegraphics[height=\fh]{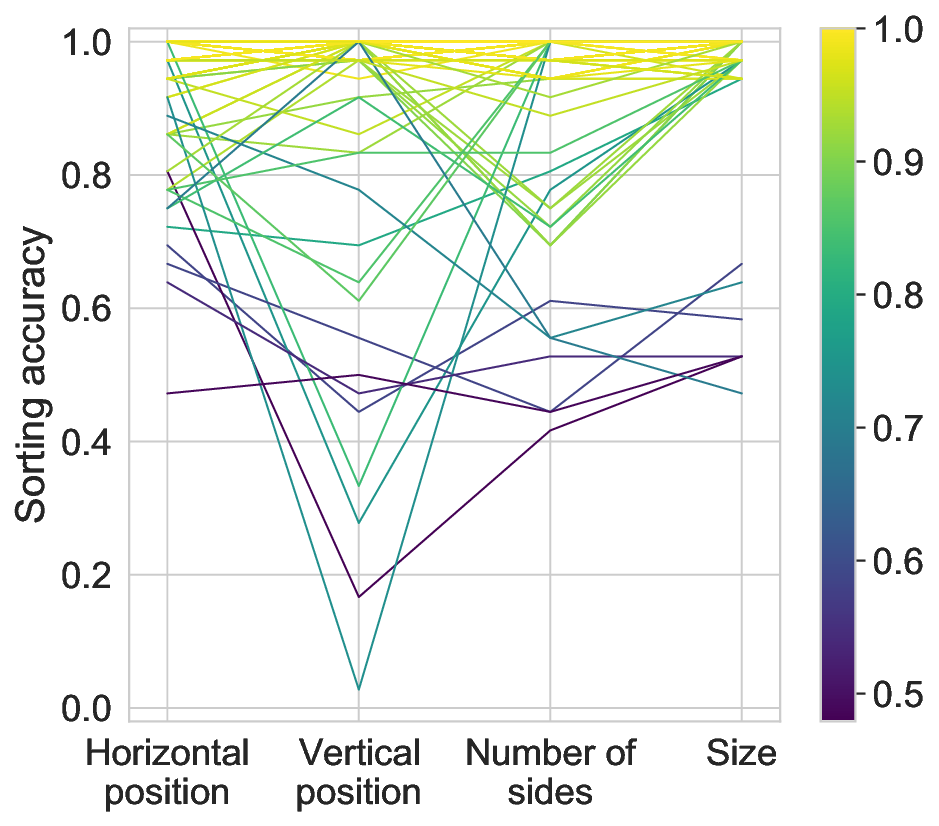}
			\caption{}
			\label{subfig:userregression-lines}
		\end{subfigure}%
		\\
		\begin{subfigure}[t]{\textwidth}
			\centering
			\includegraphics[height=\fh]{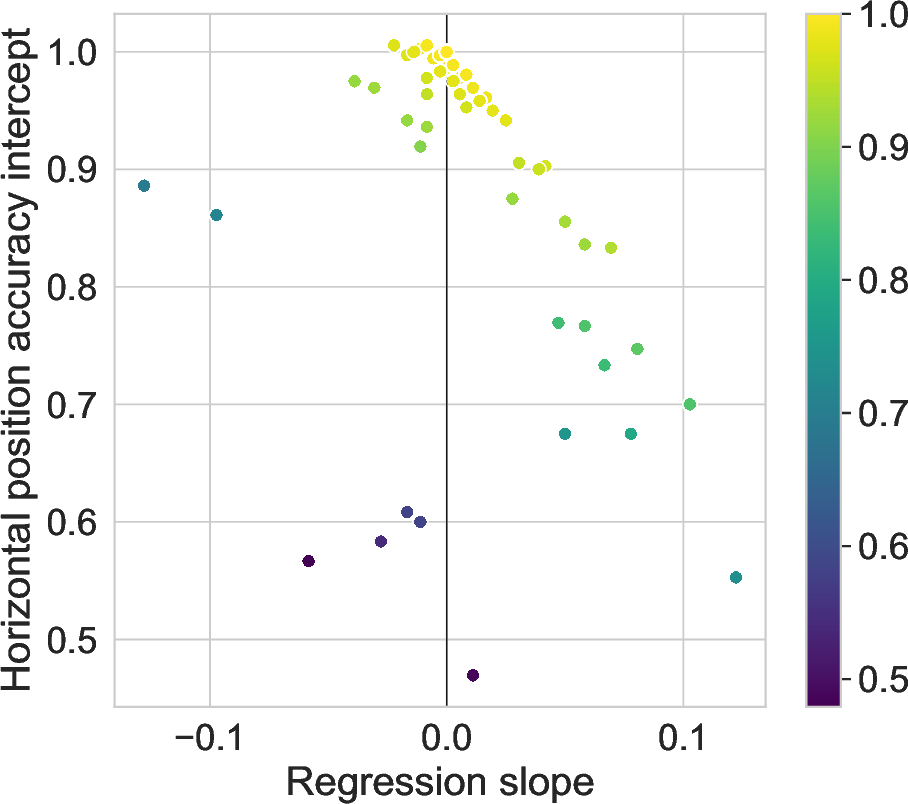}
			\caption{}
			\label{subfig:userregression-scat}
		\end{subfigure}%
	\end{minipage}%
	\begin{minipage}{0.4\textwidth}
		\centering
		\begin{subfigure}[t]{\textwidth}
			\centering
			\includegraphics[height=\fh]{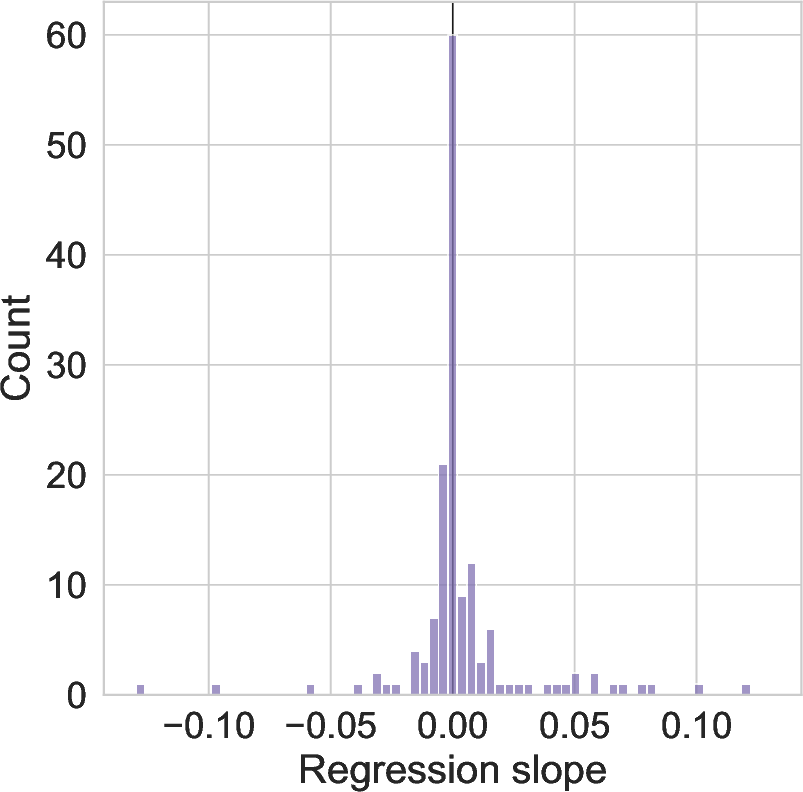}
			\caption{}
			\label{subfig:userregression-hist}
		\end{subfigure}%
		\\
		\begin{subfigure}[t]{\textwidth}
			\centering
			\includegraphics[height=\fh]{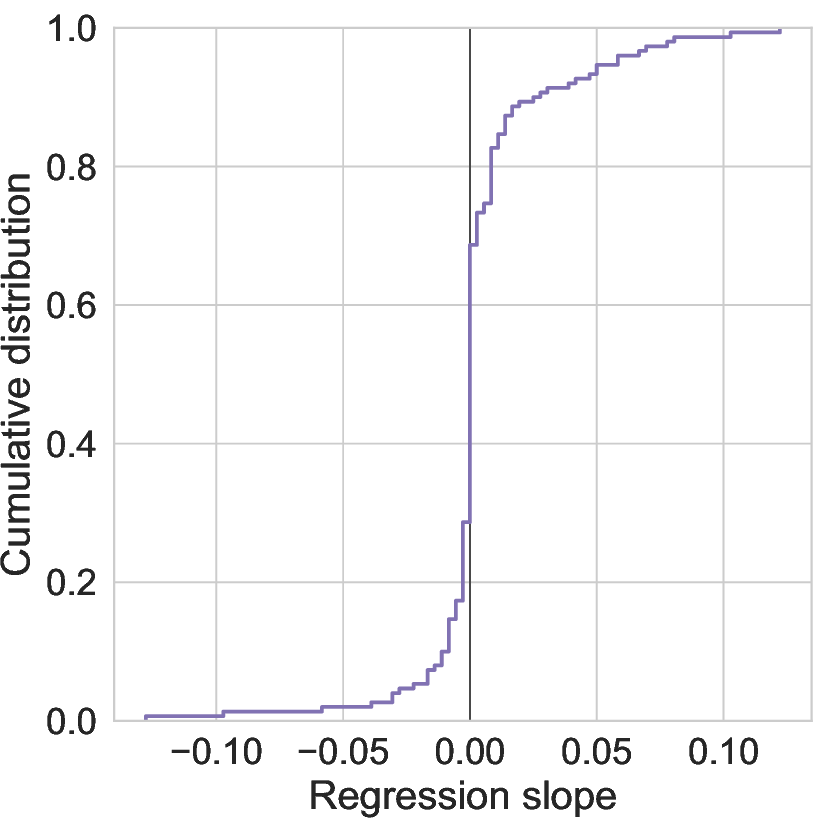}
			\caption{}
			\label{subfig:userregression-dist}
		\end{subfigure}%
	\end{minipage}%
	\caption[Trend line analysis of individual participant performance over critical character comparisons of increasing depth.]{\textbf{Trend line analysis of individual participant performance over critical character comparisons of increasing depth.}
	\textbf{a,} Each individual participant's sorting accuracy is plotted as a connected line over accuracies calculated with respect to each critical character comparison. Each line is colored by the participant's overall sorting accuracy across all queries.
	\textbf{b,} After fitting a linear least squares model to each participant's piecewise curve in \textbf{a}, we plot the slope and intercept at the horizontal position for each participant's linear regression model.
	\textbf{c,} Histogram of individual participant regression slopes.
	\textbf{d} Empirical cumulative distribution plot of individual participant regression slopes. Only approximately 30\% of participants have negative regression slopes, and only 10\% of participants have slopes of 1\% accuracy decrease or worse per additional character.}
	\label{fig:userregression}
\end{figure}

In \Cref{fig:userregression}, we analyze dictionary sorting proficiency for each individual participant in the online user study. \Cref{subfig:userregression-lines} plots each individual participant's accuracy in sorting shape pairs, binned by critical character; for each subject these values are connected by straight lines, for visualization purposes. Each participant's piecewise linear curve is colored by their \emph{overall} sorting accuracy across all 144 (not including test queries) shape pairs, as indicated by the color bar. Generally speaking, the lowest performing subjects (dark colors) performed poorly across all critical character comparisons. Interestingly, several subjects performed well for horizontal position, dropped in performance for vertical position, and increased in performance for number of sides and size comparisons. Several subjects sorted vertical position below 50\% accuracy, implying that they sorted this character consistently, but in the reversed alphabet order of precedence.

To more formally analyze the trends over each individual participant, for each participant we performed a least-squares linear fit to their accuracies across critical characters, with categorical values converted to regressors as 0 (horizontal position), 1 (vertical position), 2 (number of sides), and 3 (size). In \Cref{subfig:userregression-scat}, we present a scatter plot of the regression slope of each participant's linear model, plotted against each model's intercept at the horizontal position. As in \Cref{subfig:userregression-lines}, participants are colored coded by their overall sorting accuracy. Only two participants both sorted early string characters with high accuracy (large horizontal position intercept) and decreased in performance for deeper characters (negative regression slope). Otherwise, participants mostly performed accurately across all characters (high horizontal position intercept, flat slope), or performed moderately for early characters and increased in accuracy, with a positive regression slope. Only a few participants performed poorly for early characters and continued to perform poorly for deeper characters.

In \Cref{subfig:userregression-hist,subfig:userregression-dist}, we plot a histogram and empirical cumulative distribution function of regression slopes across all participants. As can be observed from \Cref{subfig:userregression-dist}, only approximately 30\% of participants have negative regression slopes, and only 10\% of participants have regression slopes of 1\% accuracy decrease or worse per additional character. These results collectively suggest that overall, the performance of individual participants did not decrease noticeably as character depth increased. This suggests that as the number of characters in each heterogeneous dictionary string is increased, users are still able to both identify and make accurate comparisons with respect to each critical character.

\FloatBarrier

\subsubsection{Virtual Swarm Evaluation}
\label{sec:virtualeval}

\begin{figure}[htb]
	\centering
	\begin{subfigure}[t]{0.5\textwidth}
		\centering
		\includegraphics[width=\textwidth]{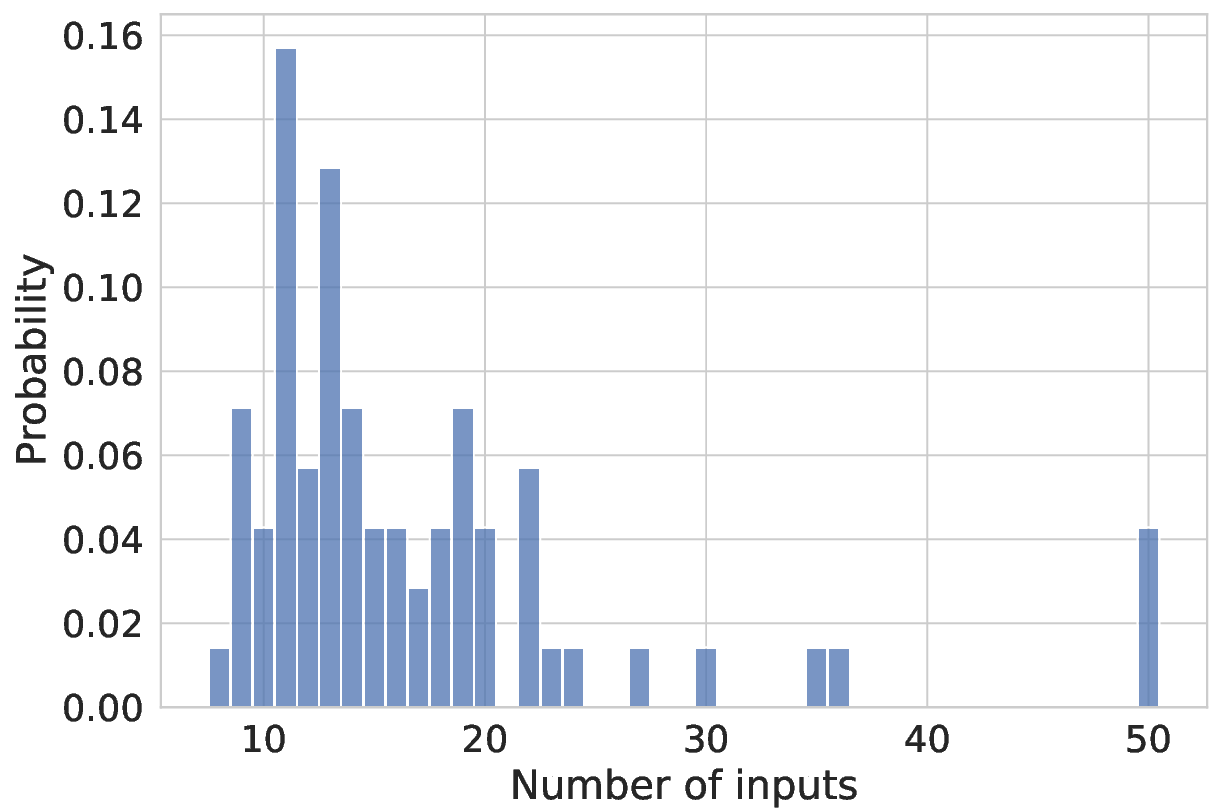}
		\caption{}
		\label{subfig:hist_exp}
	\end{subfigure}%
	\hfill
	\begin{subfigure}[t]{0.5\textwidth}
		\centering
		\includegraphics[width=\textwidth]{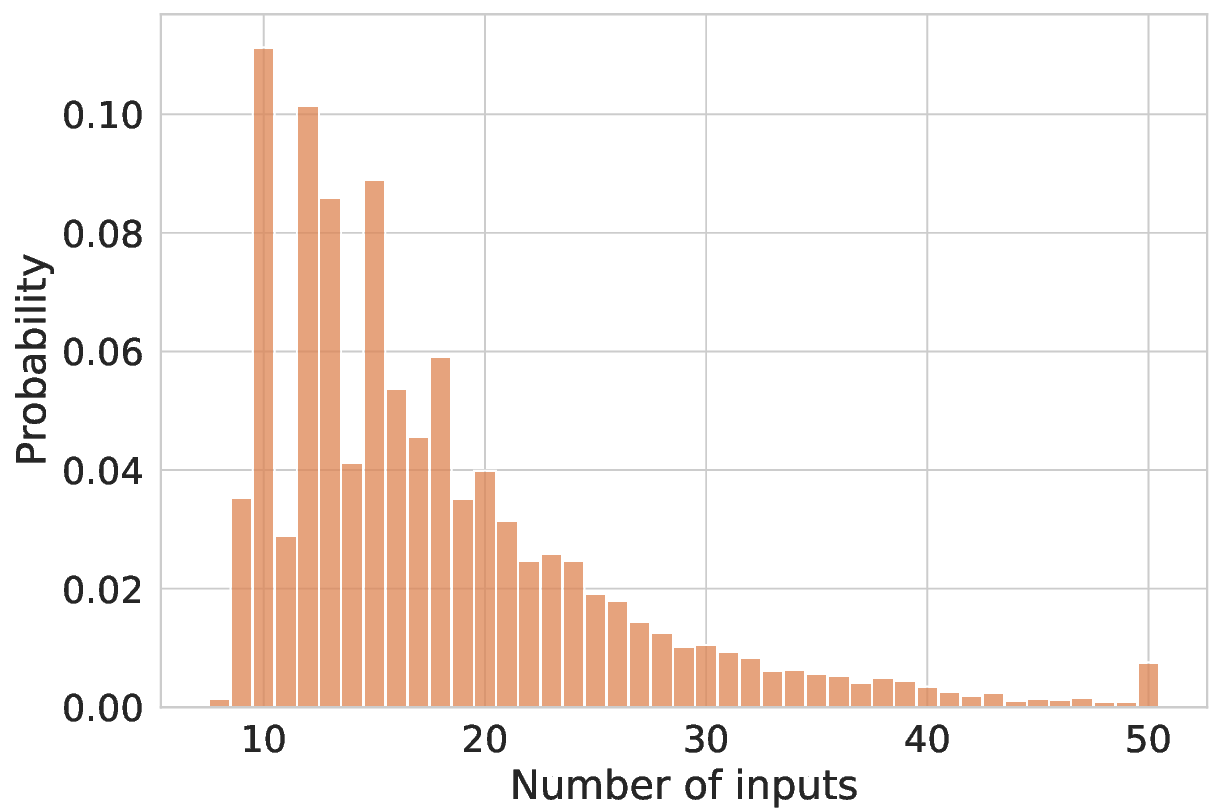}
		\caption{}
		\label{subfig:hist_sim}
	\end{subfigure}%
	\caption[Histogram of number of inputs until convergence.]{Histogram of number of inputs until convergence for virtual swarm control (\textbf{a}) and simulated (\textbf{b}) trials.}
	\label{fig:hist}
\end{figure}

\Cref{fig:hist} compares the histograms of the number of inputs required for convergence in each experimental and simulated trial. These histograms generally agree in shape, with right skewed distributions and peaks at the maximum number of inputs, reflecting that several trials ``timed out'' before convergence. This overall agreement corroborates the simulation's realistic modeling of the overall system.

\begin{figure}[htb]
	\centering
	\includegraphics[width=0.6\textwidth]{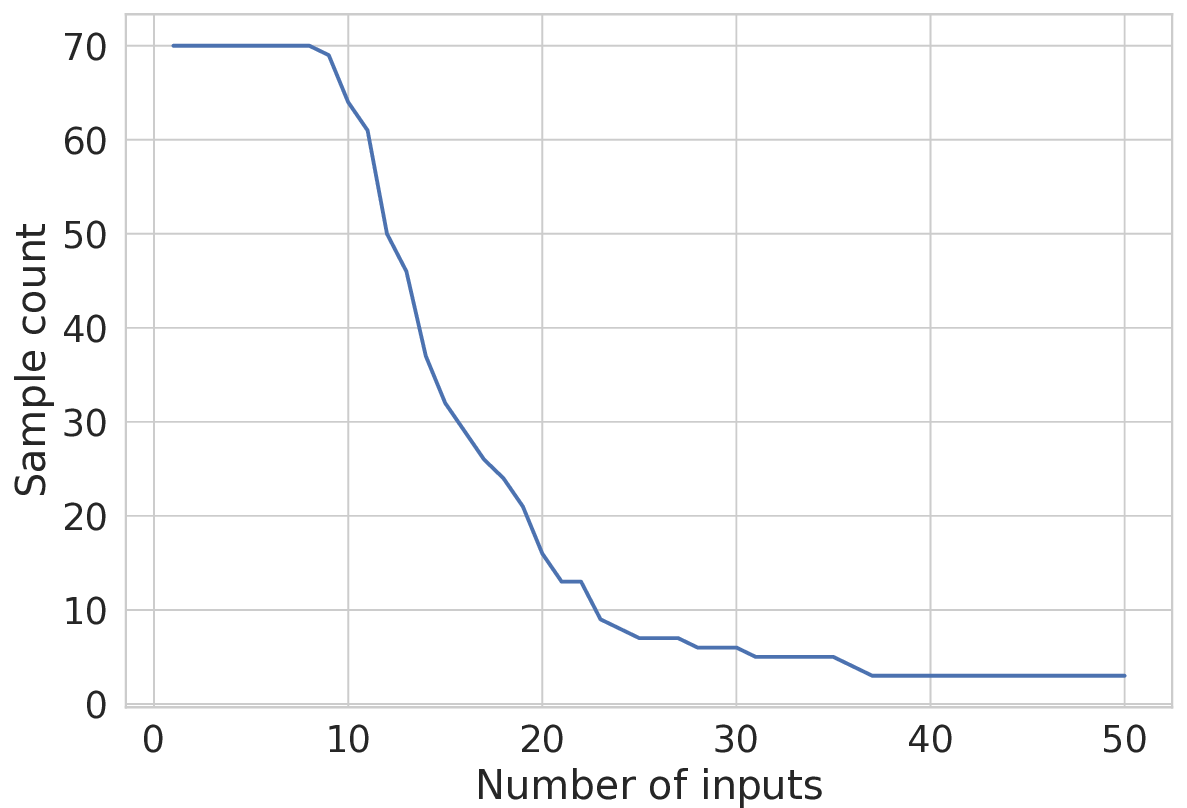}
	\caption[Number of samples at each number of issued inputs, aggregated over all virtual swarm trials.]{\textbf{Number of samples at each number of issued inputs, aggregated over all virtual swarm trials.} The number of samples observed at each number of inputs decreases as trials converge to selected configurations. This results in larger error bars in \Cref{subfig:crossover} and higher late trial variability in \Cref{fig:LDAfull,fig:LDAextra}.}
	\label{fig:samplecount}
\end{figure}

\Cref{fig:samplecount} plots the number of recorded samples at every number of inputs, across all 70 virtual swarm trials. Since no trials cross the convergence threshold until after at least 8 inputs, 70 input samples were recorded at every number of inputs at or below this point. As trials begin to converge after 8 inputs, fewer recorded samples are available for larger number of inputs due to trials converging and halting input recording. Due to this decreasing sample size at larger numbers of inputs, the empirical EEG crossover probability in \Cref{subfig:crossover} is most accurately estimated at lower numbers of inputs. As the number of inputs increases, crossover probability is less accurately estimated due to smaller sample sizes, reflected in the larger error bars in the crossover probability estimate.

\begin{figure}[htb]
	\centering
	\begin{subfigure}[t]{0.5\textwidth}
		\centering
		\includegraphics[width=\textwidth]{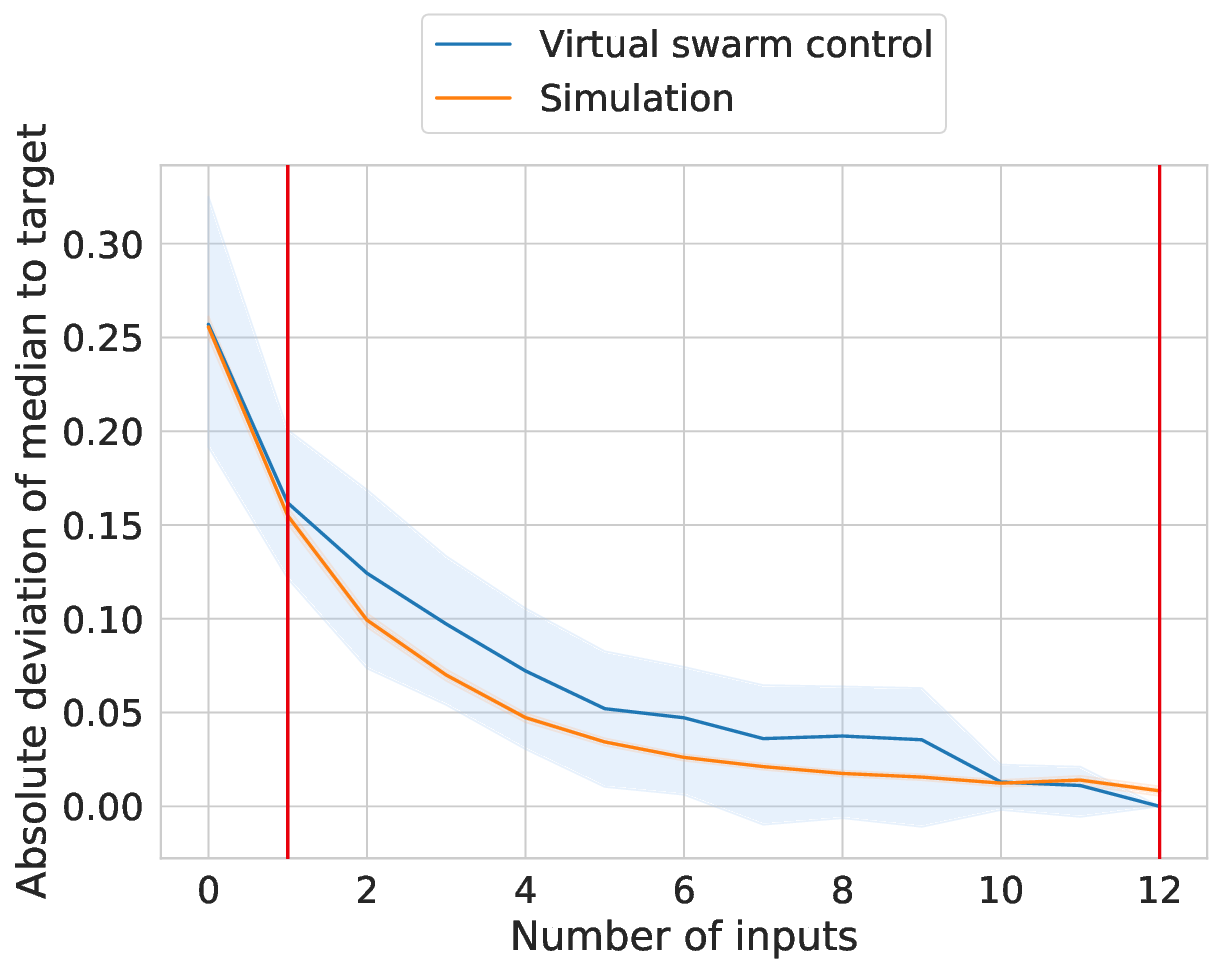}
		\caption{}
		\label{subfig:L1bin-short}
	\end{subfigure}%
	\hfill
	\begin{subfigure}[t]{0.5\textwidth}
		\centering
		\includegraphics[width=\textwidth]{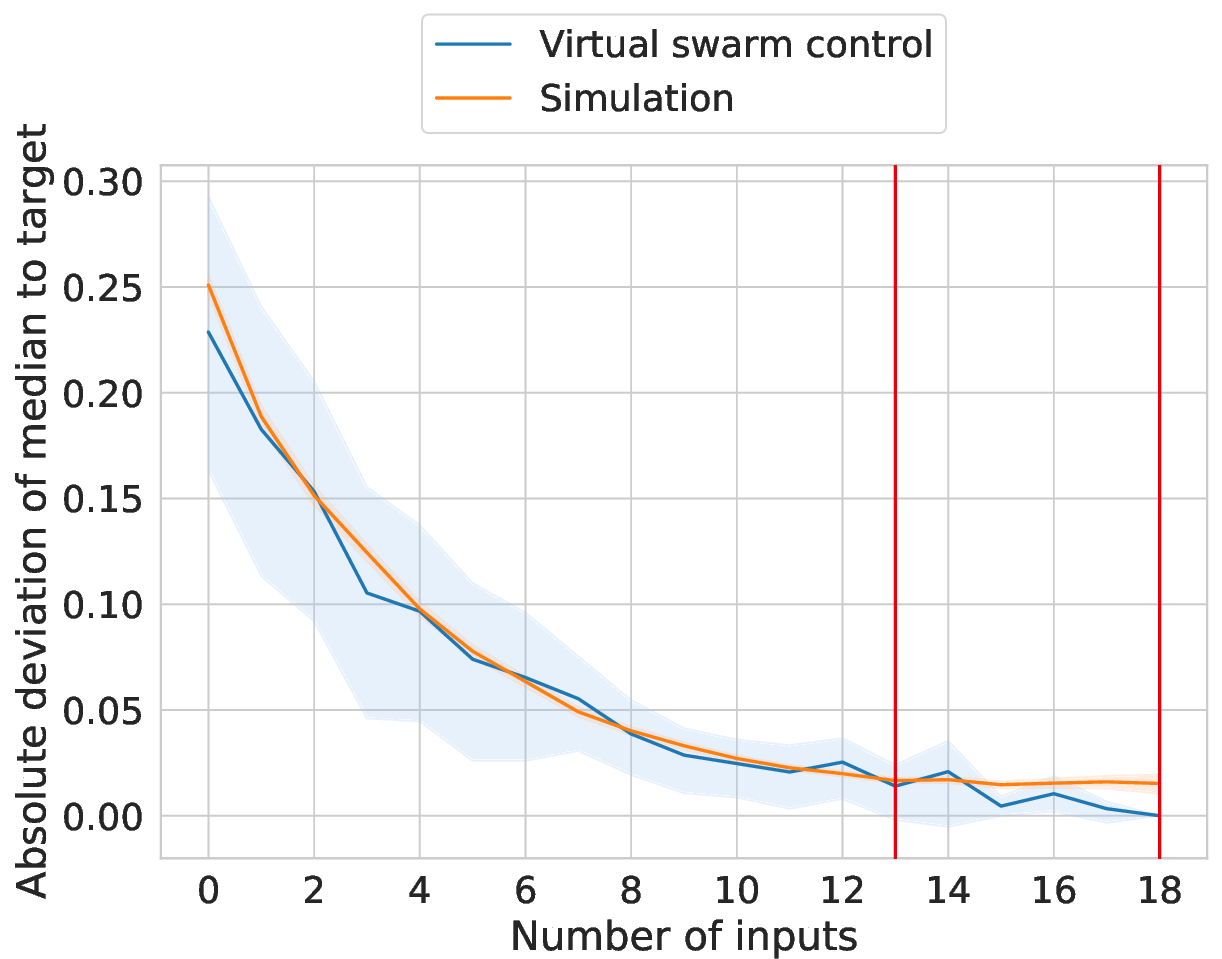}
		\caption{}
		\label{subfig:L1bin-medium}
	\end{subfigure}%
	\\
	\begin{subfigure}[t]{0.5\textwidth}
		\centering
		\includegraphics[width=\textwidth]{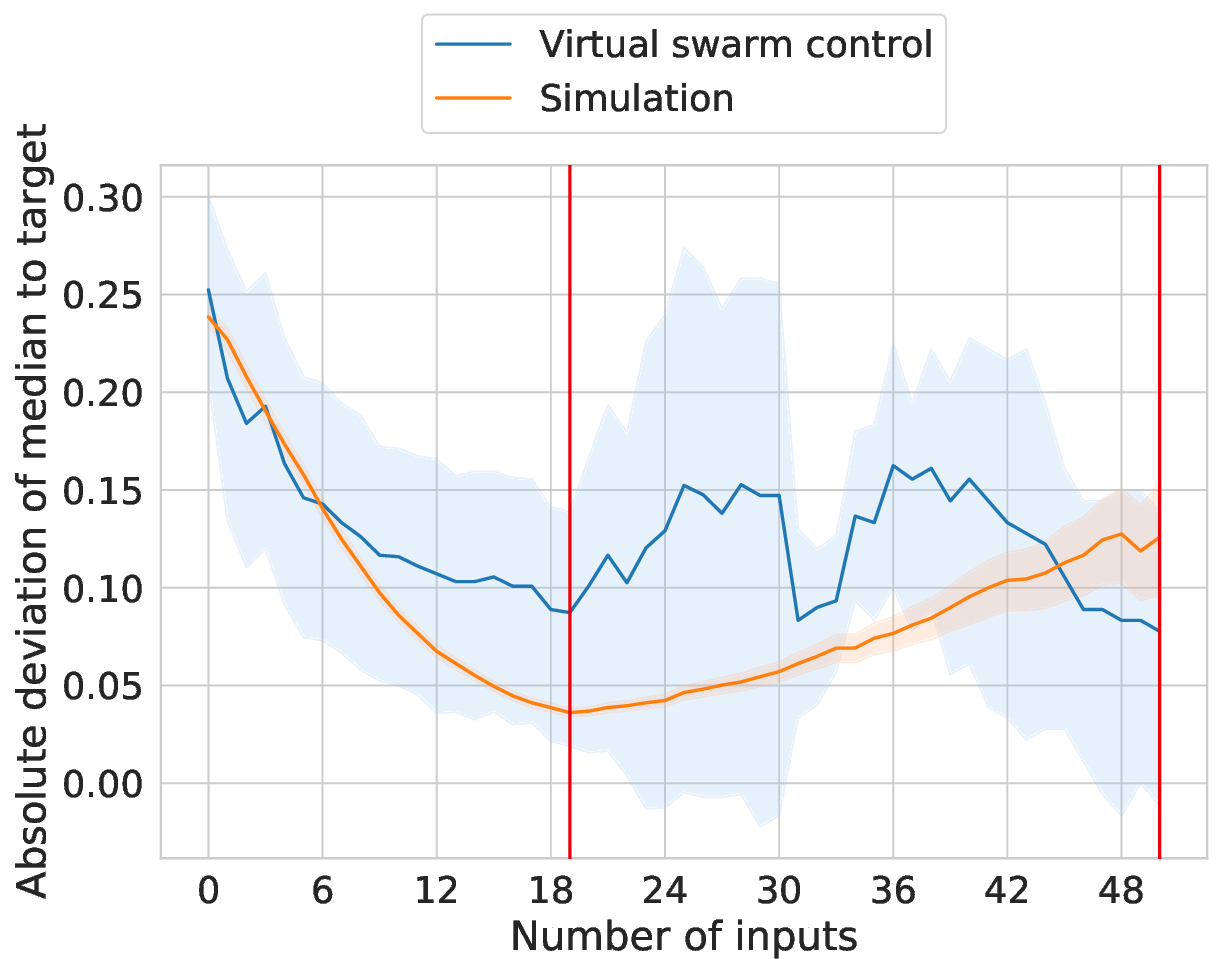}
		\caption{}
		\label{subfig:L1bin-long}
	\end{subfigure}%
	\hfill
	\begin{subfigure}[t]{0.5\textwidth}
		\centering
		\includegraphics[width=\textwidth]{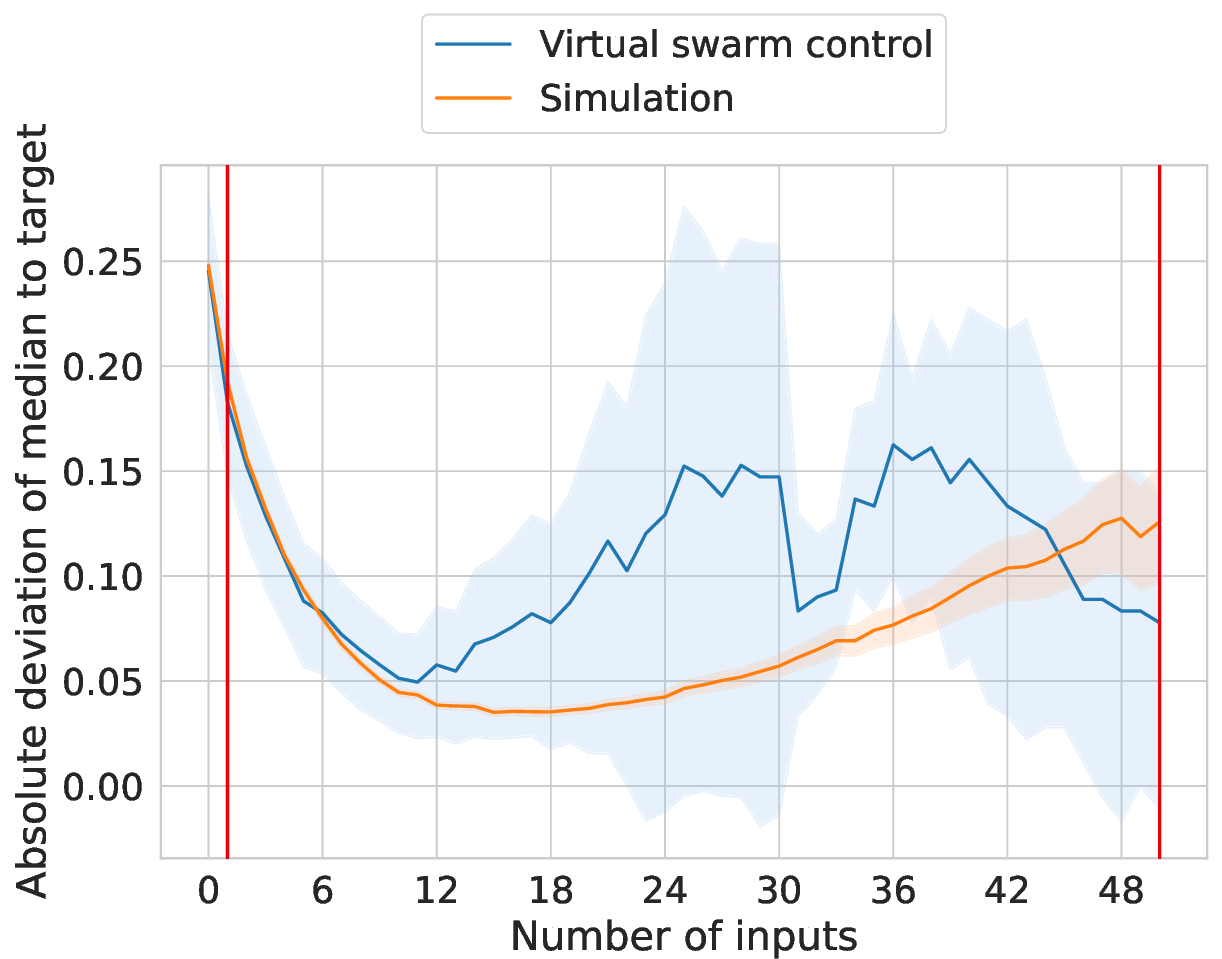}
		\caption{}
		\label{subfig:L1bin-all}
	\end{subfigure}%
	\caption[Absolute deviation between guessed swarm configuration after each number of inputs (guessed as posterior median) in comparison to target configuration, for both virtual swarm control and simulated trials.]{Absolute deviation between guessed swarm configuration after each number of inputs (guessed as posterior median) in comparison to target configuration, for both virtual swarm control and simulated trials. The crossover model from \Cref{subfig:crossover} was used to generate non-stationary errors in all 10,000 simulation trials. Each pane depicts a subset of trials binned by number of inputs until swarm convergence. The bin range for each pane is depicted through two vertical lines at the bin edges (inclusive). Each trial set, for both virtual swarm control and simulated trials, is plotted as mean absolute deviation with error bars depicting 95\% bootstrap confidence intervals over 10,000 samples (separate resampling for every number of inputs).  \textbf{a,} Trials converging between 1 and 12 inputs, inclusive. \textbf{b,} Trials converging between 13 and 18 inputs, inclusive. \textbf{c,} Trials converging between 19 and 50 inputs, inclusive. \textbf{d,} All trials. In each pane, values at 0 inputs indicate absolute error at initialization, before a trial begins.}
	\label{fig:L1bin}
\end{figure}

In \Cref{fig:L1bin}, we analyze the data in \Cref{subfig:simcomp} in terms of absolute deviation rather than error-free configuration accuracy. As in \Cref{subfig:simcomp}, virtual swarm and simulated trials were separated into the same ``Short'' (between 1 and 12 inputs until convergence, inclusive), ``Medium'' (between 13 and 18 inputs until convergence, inclusive), and ``Long'' (between 19 and 50 inputs until convergence, inclusive) trial bins. Then, we analyzed the absolute deviation trajectories within each bin by comparing a trial's target configuration to the posterior median guessed by the swarm after every input; this differs from the absolute deviation calculation in \Cref{fig:generalizing}, which instead calculates absolute deviation with respect to an instantaneous MAP configuration estimate. \Cref{subfig:L1bin-short,subfig:L1bin-medium,subfig:L1bin-long} plot absolute deviations for both virtual swarm and simulated trials within Short, Medium, and Long bins, respectively. \Cref{subfig:L1bin-all} plots absolute deviation over all trials. In each figure, vertical red lines visually indicate the span of each bin range, depicting the range of inputs in which all trials within the bin converged. The virtual swarm control and simulation trials have absolute deviations that mostly agree in the first two bins, with increased differences between empirical and simulated trials in the Long bin. Both empirical and simulated results in the Long bin experience increasing absolute deviations, due to the fact that input errors increase toward chance for larger numbers of inputs (see \Cref{subfig:crossover}). Regardless, the simulated system in its entirety matches the behavior of the experimental trials, indicating that the simulator used here can reliably account for realistic experimental factors such as user fatigue.

\FloatBarrier

\paragraph{Signal feature analysis}
In \Cref{subfig:crossover}, the empirically observed crossover probability degrades in quality (approached chance) at the number of issued inputs increases. As demonstrated in \Cref{subfig:simcomp}, this behavior can be accounted for in a posterior matching simulator that assumes a fixed crossover probability while errors are generated according to a non-stationary input error profile. Below, we further analyze the empirically observed input degradation. In particular, we analyze the behavior of the LDA classifier component of our motor imagery detection pipeline, and measure signal quality by observing the changing classifier confidence over time.

Let $f \in \mathbb{R}^4$ denote the EEG feature vector for a given input (see \Cref{fig:pipeline}), and let $\mu \in \mathbb{R}^4$ and $\tau \in \mathbb{R}$ denote the hyperplane weights and offset respectively of the trained LDA classifier \cite{hastie2009elements}. Let $Y \in \{0,1\}$ denote a classification result of left-hand ($Y = 0$) or right-hand ($Y=1$) motor imagery detection, determined by the sign of the distance to the classifier hyperplane, i.e., $Y = \operatorname{sign}(\mu^T f - \tau)$. In LDA, a standard result \cite{hastie2009elements} is that the log-ratio of the class posterior distribution is given by
\[
\log \frac{P(Y = 1 \mid f)}{P(Y = 0 \mid f)} = \mu^Tf - \tau.\]

Denote $X \in\{0,1\}$ to be the ground truth (i.e., correct) motor imagery input that the user should issue, according to the rules of posterior matching. One way to measure the quality of the classifier's decision is to evaluate the ratio of the probability that it detects the ground truth input correctly (i.e., $P(Y = X \mid f)$) over the probability that it detects the ground truth incorrectly (i.e., $P(Y \neq X \mid f)$) It is easy to show that the log of this ratio takes a convenient form:
\begin{align}
\log \frac{P(Y = X \mid f)}{P(Y \neq X \mid f)} &= X\log \frac{P(Y = 1 \mid f)}{P(Y = 0 \mid f)} + (1-X)\log \frac{P(Y = 0 \mid f)}{P(Y = 1 \mid f)}\notag\\
&= X\log \frac{P(Y = 1 \mid f)}{P(Y = 0 \mid f)} + (X-1)\log \frac{P(Y = 1 \mid f)}{P(Y = 0 \mid f)}\notag\\
&= (2X-1)\log \frac{P(Y = 1 \mid f)}{P(Y = 0 \mid f)}\notag\\
&= (2X-1)(\mu^Tf - \tau)\label{suppeq:LDadistcor}.
\end{align}

When the classifier is more confident in the correct class (which is unavailable to the classifier at classification time, but is available in post hoc analysis) than the incorrect class, then this log-ratio will be positive. Conversely, if the classifier is confident in the \emph{incorrect} decision, then this log-ratio will be negative. If the classifier is ``unsure'' in its decision, then this log-ratio will be close to 0 since the classifier assigns equal probability to both the correct and incorrect decision. These scenarios can also be interpreted geometrically by considering the equivalent log-ratio form in \cref{suppeq:LDadistcor}: the log-ratio of probability of a correct to an incorrect decision corresponds to the signed distance from the feature vector to the hyperplane, where the sign is determined by whether the classifier is correct or incorrect in its decision. Decisions that are confident and correct will have a positive signed distance, ambiguous decisions a signed distance of 0, and confident but incorrect decisions will have a negative signed distance.

\begin{figure}[htb]
	\centering
	\includegraphics[width=0.6\textwidth]{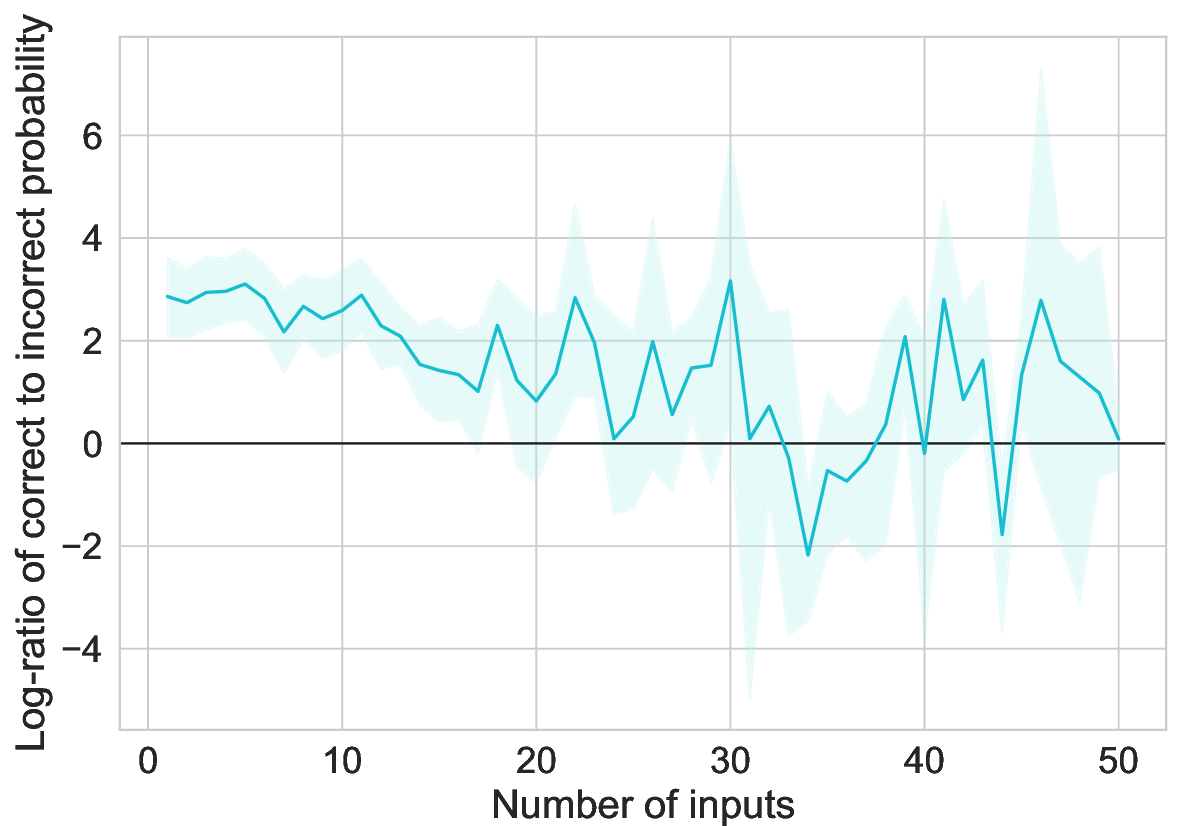}
	\caption[Log-ratio of classifier probability assigned to the correct input over the probability assigned to the incorrect input, plotted against the number of inputs issued in a trial.]{Log-ratio of classifier probability assigned to the correct input over the probability assigned to the incorrect input, plotted against the number of inputs issued in a trial. Results are aggregated over all virtual swarm trials and plotted as mean log-ratio with error bars depicting 95\% bootstrap confidence intervals over 10,000 samples (separate resampling for every number of inputs). When taken as a measure of classifier confidence, the log-ratio's steady decline indicates decreasing classifier confidence in its decisions, as the classifier's probability assignment to the correct input approaches 0.5. Equivalently, this log-ratio measures the signed distance of each feature vector to the classifier hyperplane, signed such that positive distances indicate correct classification.}
	\label{fig:LDAfull}
\end{figure}

\Cref{fig:LDAfull} depicts the distribution of this log-ratio (or signed classifier distance) over increasing numbers of inputs, on the same input data evaluated in \Cref{subfig:crossover}. After each number of inputs, we calculate the log ratio in \cref{suppeq:LDadistcor} with respect to the corresponding EEG feature vector. Initially, the classifier is confident in correct decisions, and this confidence decreases gradually towards zero as more inputs are issued. This metric is a direct measure of feature degradation, due to the direct correspondence between the log-ratio of classifier probabilities and the distance from each feature vector to the decision boundary. In other words, as the number of inputs increases, the processed EEG signal vectors are on average closer to the decision boundary, indicating that features are no longer being well separated according to the classifier geometry established during motor imagery training.

\begin{figure}[tb]
	\centering
	\begin{subfigure}[t]{0.45\textwidth}
		\centering
		\includegraphics[width=\textwidth]{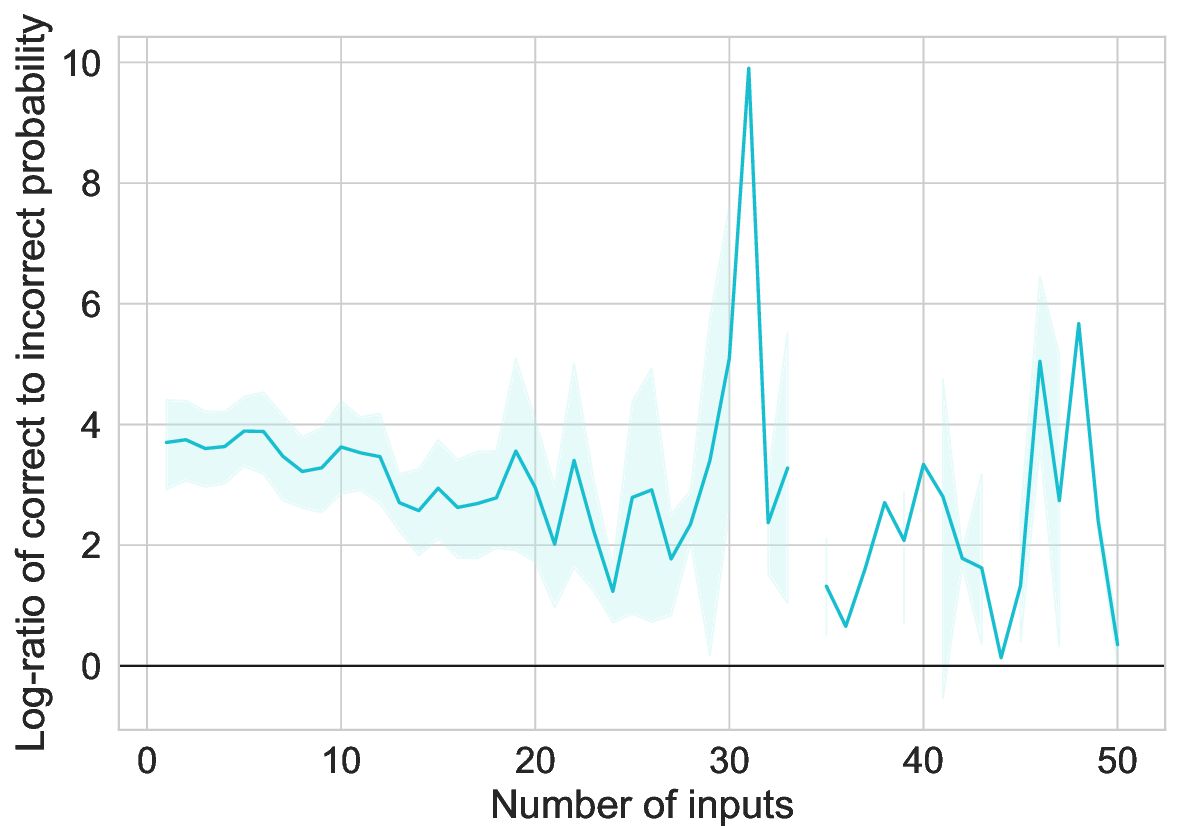}
		\caption{Inputs classified correctly}
		\label{subfig:LDAcorrect}
	\end{subfigure}%
	\hfill
	\begin{subfigure}[t]{0.45\textwidth}
		\centering
		\includegraphics[width=\textwidth]{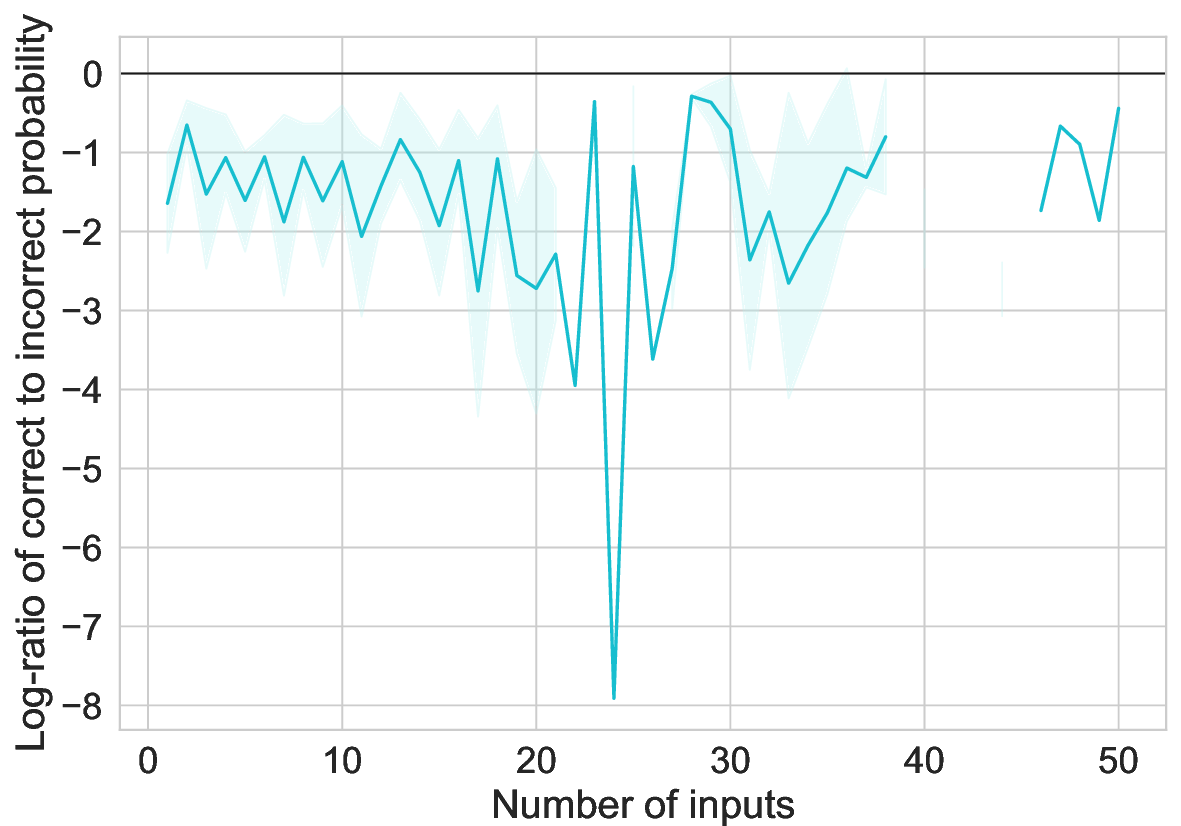}
		\caption{Inputs classified incorrectly}
		\label{subfig:LDAincorrect}
	\end{subfigure}%
	\caption[Log-ratio of classifier probability assigned to the correct input over the probability assigned to the incorrect input, grouped by inputs that were classified correctly or incorrectly.]{Log-ratio of classifier probability assigned to the correct input over the probability assigned to the incorrect input, grouped by inputs that were classified correctly (\textbf{a}) or incorrectly (\textbf{b}). Results are aggregated over all recorded inputs in each group and plotted as mean log-ratio with error bars depicting 95\% bootstrap confidence intervals over 10,000 samples (separate resampling for every number of inputs).
	\textbf{a,} When only analyzing inputs that were detected correctly by the classifier, we still observe a decline in confidence.
	\textbf{b,} We also analyze classifier confidence when grouped over inputs that were detected incorrectly.
	Both \textbf{a} and \textbf{b} have missing data, since at certain numbers of issued inputs all samples were either detected correctly or incorrectly. Error bars are omitted at numbers of inputs with only a single sample.
	}
	\label{fig:LDAextra}
\end{figure}

For completeness, we also perform the same analysis when grouping inputs by correct or incorrect classification (\Cref{fig:LDAextra}). In \Cref{subfig:LDAcorrect}, we evaluate the log-ratio of classifier probabilities only on inputs that were classified correctly. Although the classifier is making correct decisions on this data group, it is clear from the gradual log-ratio decline that the classifier's correct decisions are made with less confidence as the number of inputs increases to approximately 25 inputs. This corresponds directly to the increasing crossover probability observed in \Cref{subfig:crossover}. Conversely, in \Cref{subfig:LDAincorrect}, we evaluate log-ratio for only \emph{incorrectly} classified inputs. In this case, the log probability ratio becomes more negative for increasing numbers of inputs up to approximately 25 inputs, indicating that the classifier becomes more confident in its incorrect decisions. Interestingly, this behavior would indicate a degree of class reversal in the statistics of the extracted EEG features. Both \Cref{subfig:LDAcorrect,subfig:LDAincorrect} have missing data, since at certain numbers of inputs all signals were detected either correctly or incorrectly.

Overall, the analysis in this section indicates that during test time (i.e., when issuing inputs for swarm control, rather than for motor imagery training), the EEG feature statistics shift after increasing numbers of inputs in a way that no longer aligns with or perhaps experiences a reversal with respect to the linear classifier trained before each session. While the focus of this work is high-complexity control \emph{despite} the presence of such noisy inputs, these results suggest that the increasing crossover probability observed in \Cref{subfig:crossover} could possibly be mitigated through a calibration step during the effector control period. During such a step, effector control could be briefly halted to collect a small number of additional supervised motor imagery inputs using the same protocol during training, such that the CSP and LDA parameters can be readjusted. This calibration step could either be set ahead of time (e.g., after 25 inputs), or could be automatically triggered when the LDA feature distance (i.e., $\lvert \mu^T f - \tau \rvert$) falls below a prespecified threshold across multiple inputs, indicating a possible mismatch between the current EEG statistics and the original classification pipeline learned during session training. Finally, the analysis in this section does not take into account the possibility of user error rather than classifier error; it is possible that an ``incorrect'' trial as analyzed above is in fact the result of correct motor imagery classification, issued with an incorrect user input. To fully disambiguate these sources of error, additional data collection is necessary where after issuing each motor imagery input the user, via another means such as a keypress or speech (if they are able to do so), indicates which hand movement they intended to convey.

\FloatBarrier

\subsubsection{Extended Evaluation of Generalized Dictionary Simulations}

\begin{figure}[tb]
	\def\sfw{0.3}
	\centering
	\includegraphics[width=0.9\textwidth]{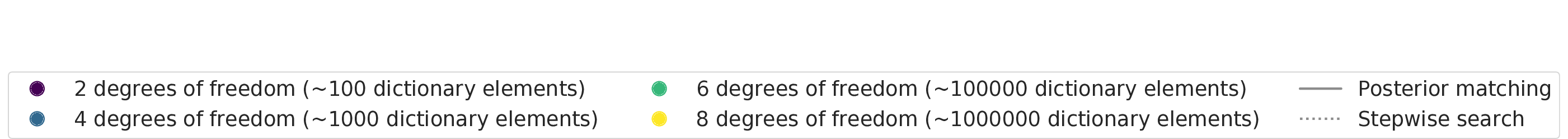}
	
	\tradeoffheadingtwo

	\begin{minipage}{\textwidth}
		\centering
		\raisebox{18mm}{\rotatebox[origin=c]{90}{\footnotesize\textbf{Fixed (10\%) error}}}\quad%
		\begin{subfigure}[t]{\sfw\textwidth}
			\centering
			\includegraphics[width=\textwidth]{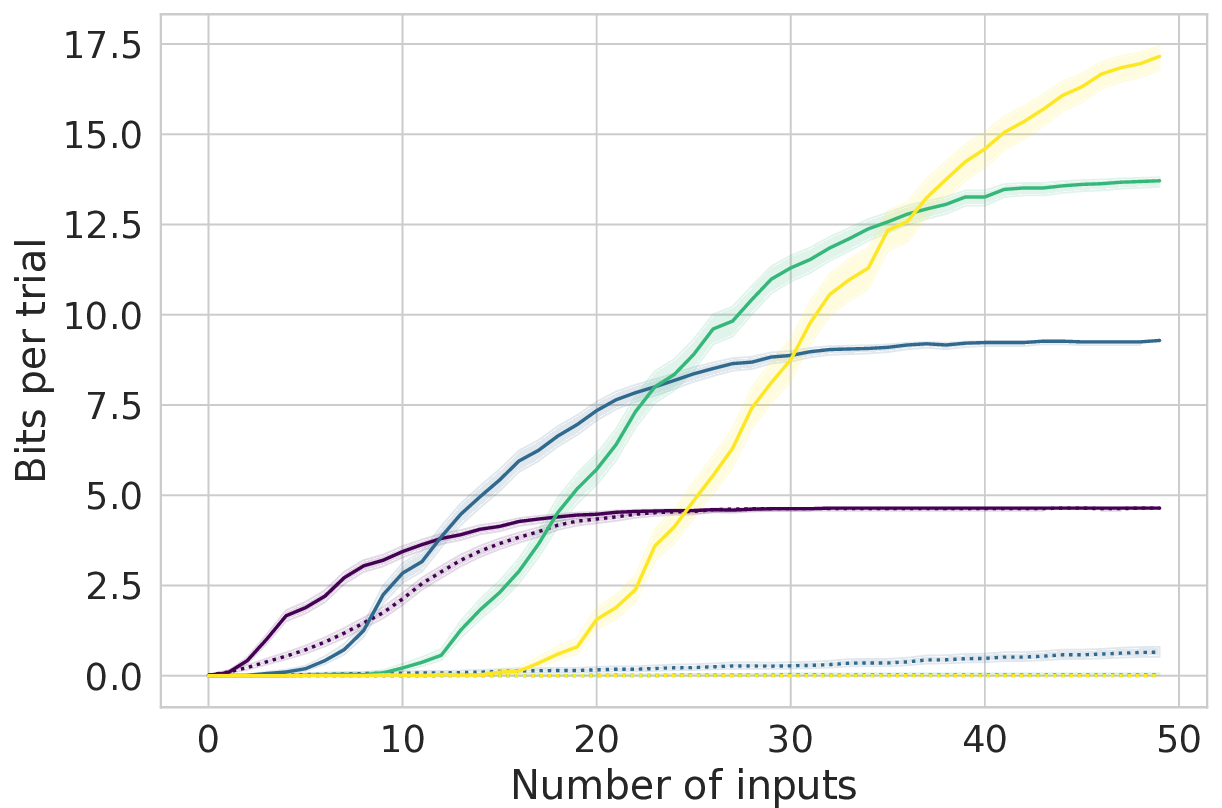}
			\caption{}
			\label{subfig:staticitr_condof}
		\end{subfigure}%
		\hfill
		\begin{subfigure}[t]{\sfw\textwidth}
			\centering
			\includegraphics[width=\textwidth]{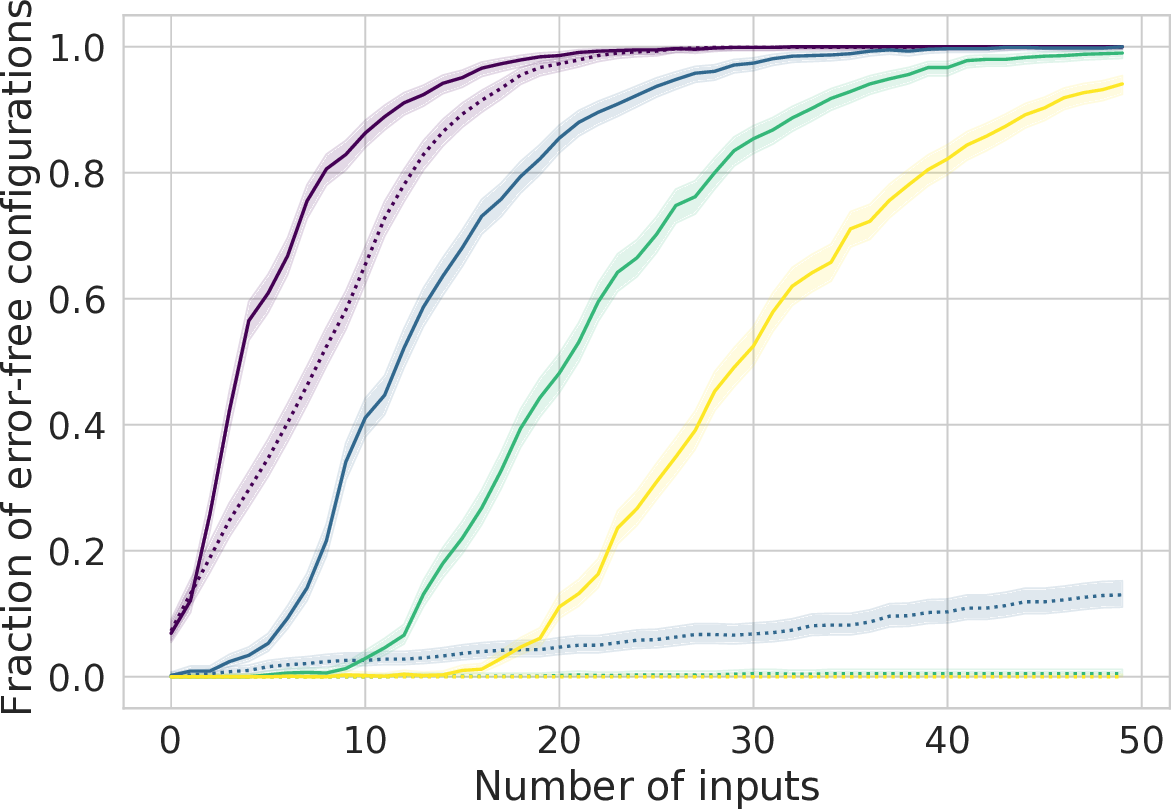}
			\caption{}
			\label{subfig:staticacc_condof}
		\end{subfigure}%
		\hfill
		\begin{subfigure}[t]{\sfw\textwidth}
			\centering
			\includegraphics[width=\textwidth]{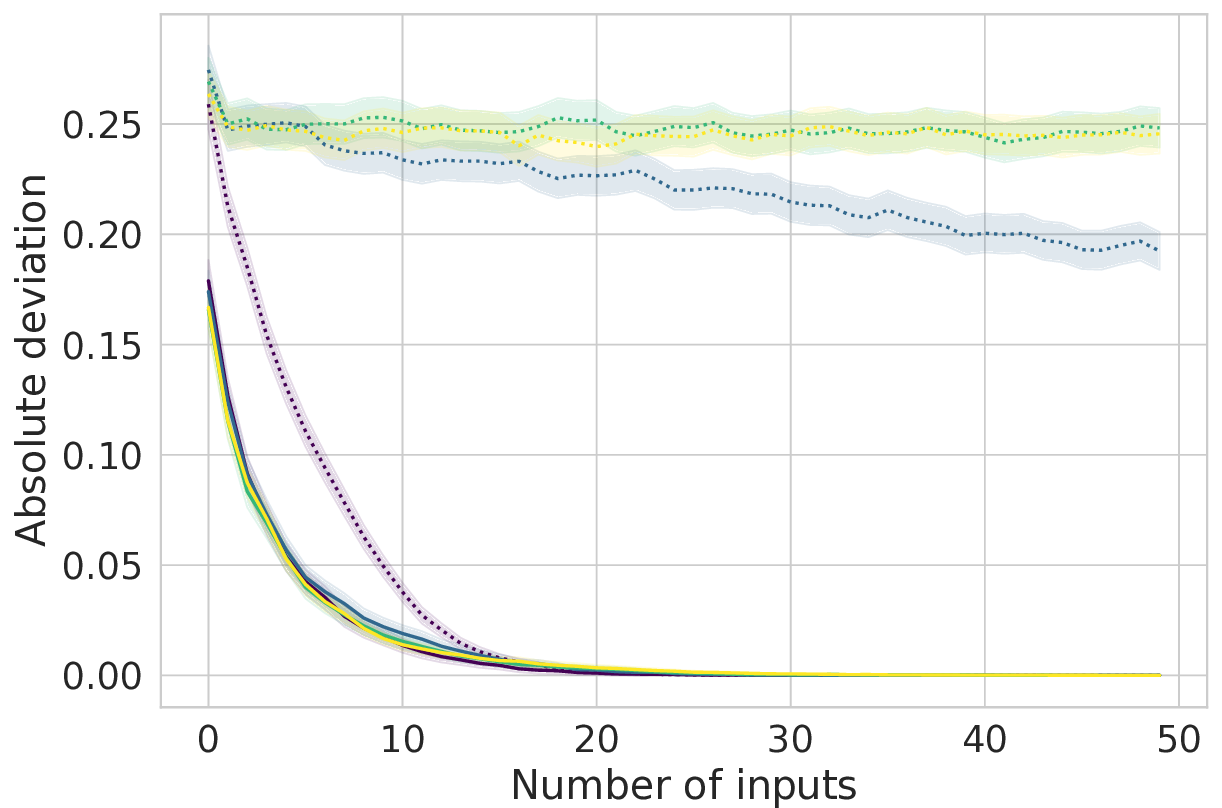}
			\caption{}
			\label{subfig:staticL1_condof}
		\end{subfigure}
		\\
		\vspace{2mm}
		\raisebox{17mm}{\rotatebox[origin=c]{90}{\footnotesize\textbf{Non-stationary errors}}}\quad%
		\begin{subfigure}[t]{\sfw\textwidth}
			\centering
			\includegraphics[width=\textwidth]{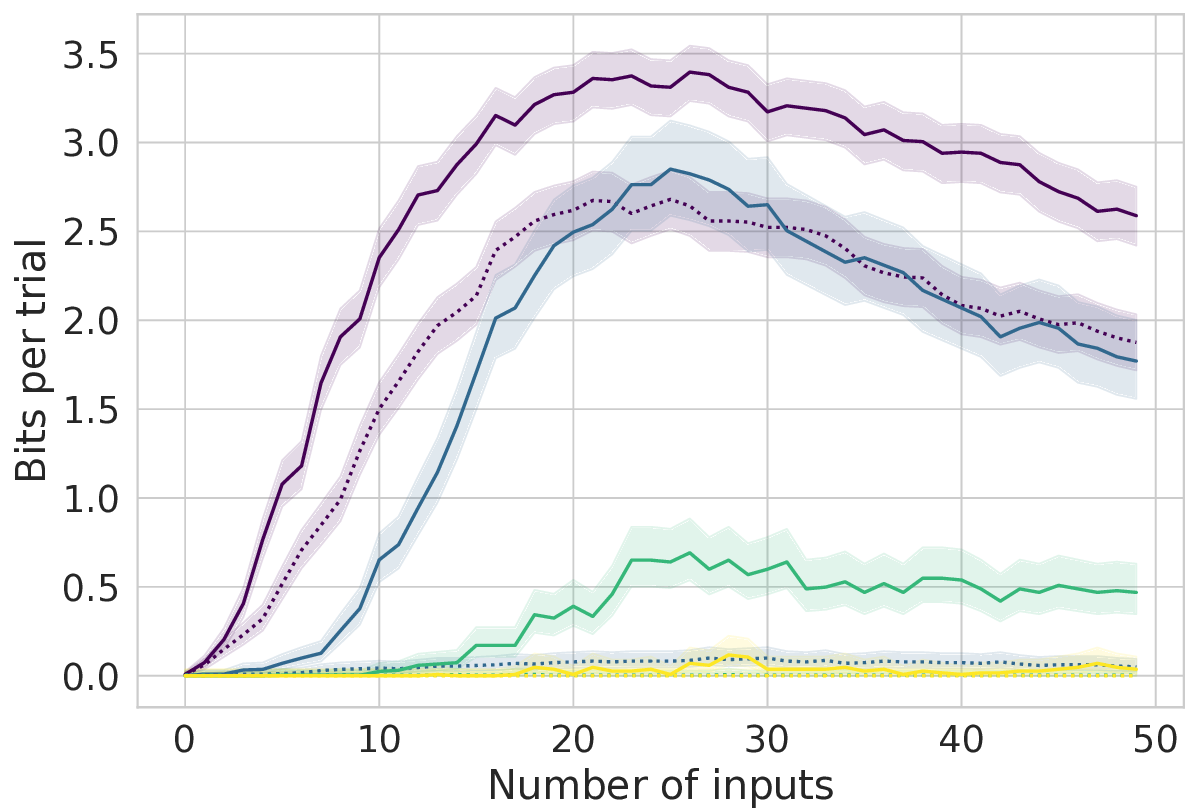}
			\caption{}
			\label{subfig:dynamicitr_condof}
		\end{subfigure}%
		\hfill
		\begin{subfigure}[t]{\sfw\textwidth}
			\centering
			\includegraphics[width=\textwidth]{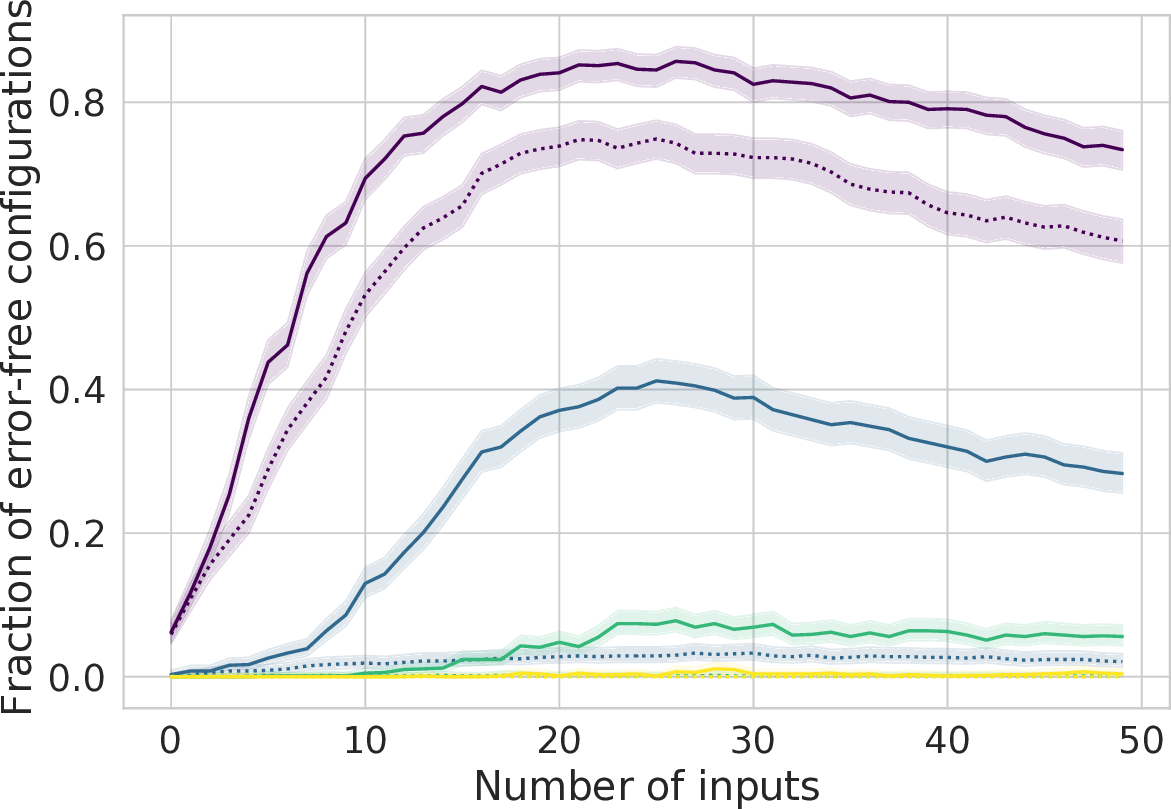}
			\caption{}
			\label{subfig:dynamicacc_condof}
		\end{subfigure}%
		\hfill
		\begin{subfigure}[t]{\sfw\textwidth}
			\centering
			\includegraphics[width=\textwidth]{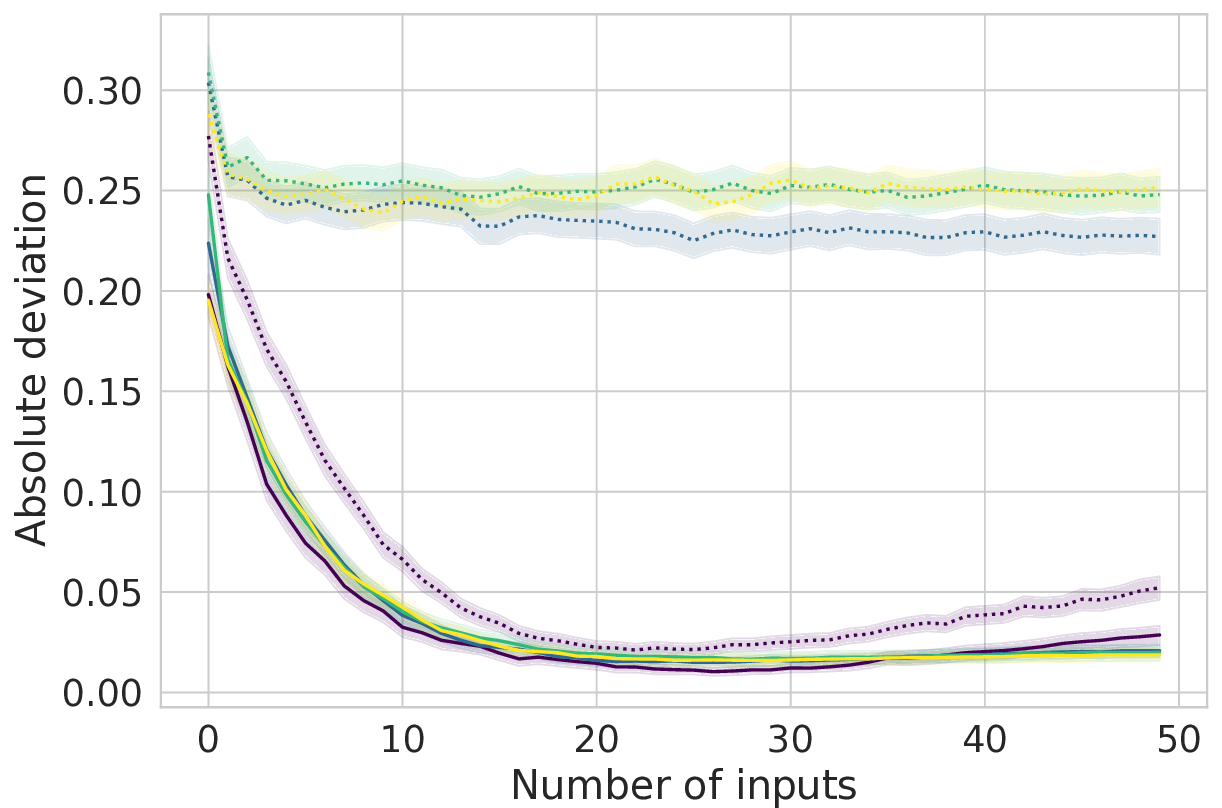}
			\caption{}
			\label{subfig:dynamicL1_condof}
		\end{subfigure}%
	\end{minipage}
	\caption[Performance as a function of number of inputs and dictionary size across both fixed and non-stationary input errors, with conservative degrees of freedom estimates.]{Performance as a function of number of inputs and dictionary size across both fixed and non-stationary input errors, with conservative degrees of freedom estimates. The same performance metrics (ITR, error-free accuracy, absolute deviation) with corresponding error bars are displayed here as in the standard degrees of freedom estimate in \Cref{fig:generalizing}.}
	\label{fig:generalizing_condof}
\end{figure}

In \Cref{fig:generalizing_condof}, we perform the same analysis as in \Cref{fig:generalizing} except with a simulated alphabet size of $b=5$ rather than $b=3$. We refer to the case of $b=5$ as the ``conservative'' degrees of freedom estimate, recalling that we refer to $b=3$ as the ``standard'' estimate. Since the posterior matching and stepwise search simulations track mathematical partitions of the unit interval and operate directly on the total dictionary order of strings rather than distinguishing between individual alphabet orders (which is performed by the human user), the only parameter affected by alphabet size relevant for simulation is the overall dictionary size $N_d$. In other words, distinct alphabet sizes $b_1$ and $b_2$ and numbers of degrees of freedom $r_1$ and $r_2$ that happen to produce the same dictionary size $N_d=b_1^{r_1}=b_2^{r_2}$ will yield the same simulation results, since both scenarios map to the same partitions of the unit interval. Because of this, \Cref{fig:generalizing_condof} with $b=5$ can be seen as mathematically equivalent to \Cref{fig:generalizing} with $b=3$, except with different dictionary sizes for each number of degrees of freedom.

What does differ between these figures is the translation of each dictionary size to estimated degrees of freedom, since a fixed number of degrees of freedom for a larger alphabet size corresponds to a larger dictionary than for smaller alphabets. For this reason, \Cref{fig:generalizing_condof} serves as a more conservative estimate of the tradeoffs involved in controlling more degrees of freedom. In particular, since each number of degrees of freedom corresponds to a larger dictionary than in \Cref{fig:generalizing}, more inputs are needed to control a conservatively estimated number of degrees of freedom than would be required for the standard estimate. While this additional cost in number of inputs is apparent in \Cref{fig:generalizing_condof} in comparison to \Cref{fig:generalizing}, the same overall trend holds that posterior matching significantly outperforms stepwise search across all metrics, and that with enough refinements posterior matching can obtain high configuration accuracies at high estimated degrees of freedom.

\begin{figure}[htb]
	\def\sfw{0.3}
	\centering
	\includegraphics[width=0.9\textwidth]{tradeoff_sns_dynamic1_condof_0_legend.png}
	
	\tradeoffheadingthree

	\begin{minipage}{\textwidth}
		\centering
		\raisebox{18mm}{\rotatebox[origin=c]{90}{\footnotesize\textbf{Fixed (10\%) error}}}\quad%
		\begin{subfigure}[t]{\sfw\textwidth}
			\centering
			\includegraphics[width=\textwidth]{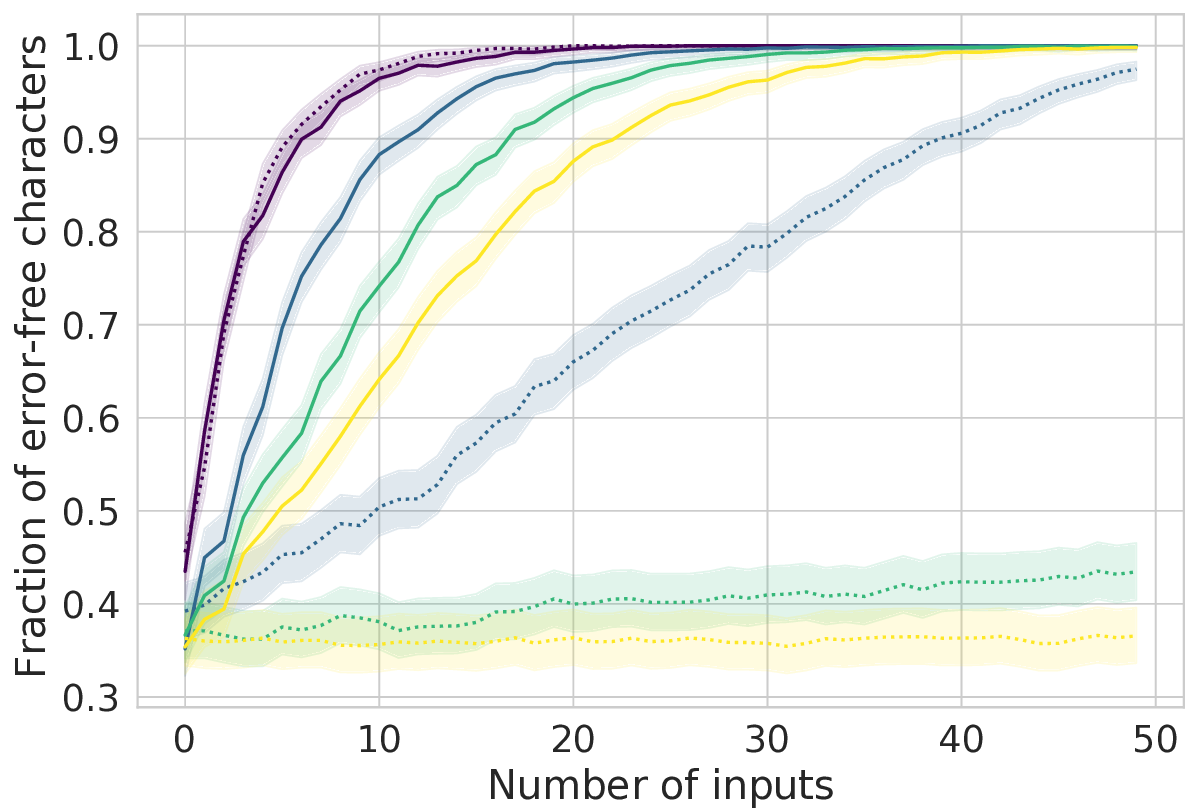}
			\caption{}
			\label{subfig:static_alphacc_stand}
		\end{subfigure}%
		\hfill
		\begin{subfigure}[t]{\sfw\textwidth}
			\centering
			\includegraphics[width=\textwidth]{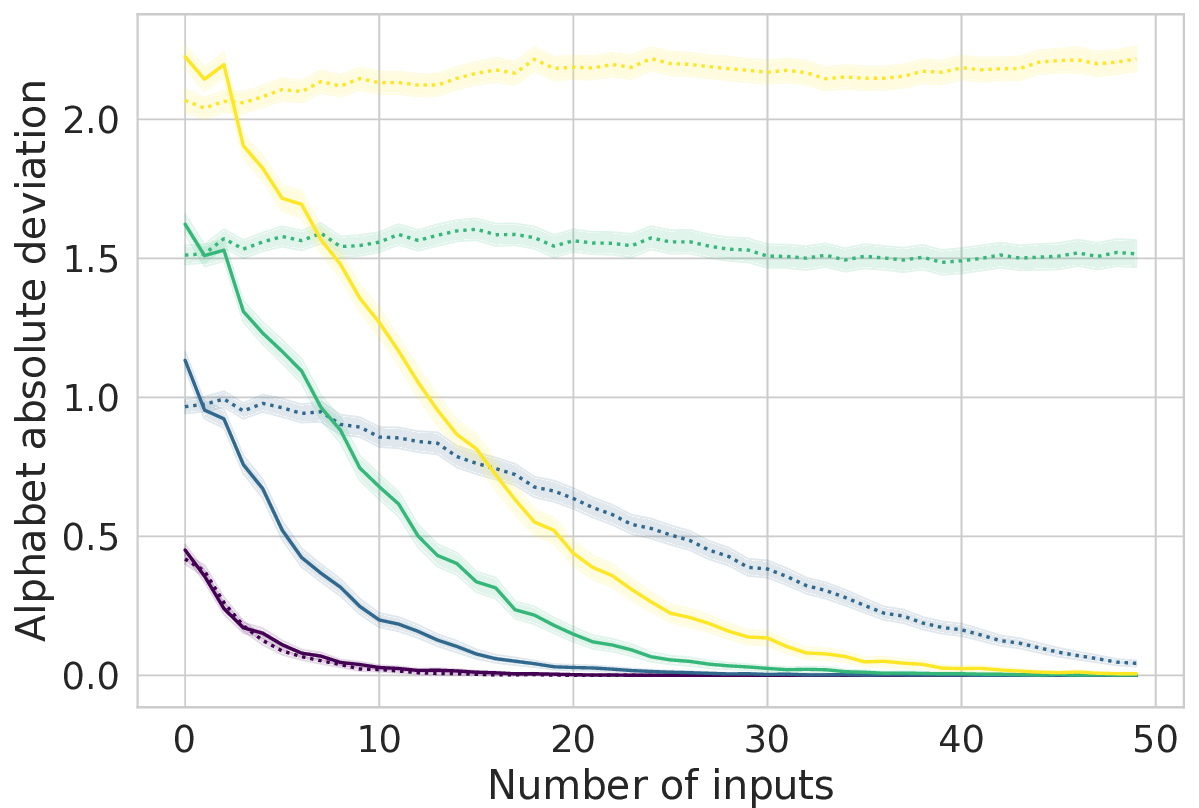}
			\caption{}
			\label{subfig:static_alphL1_stand}
		\end{subfigure}%
		\hfill
		\begin{subfigure}[t]{\sfw\textwidth}
			\centering
			\includegraphics[width=\textwidth]{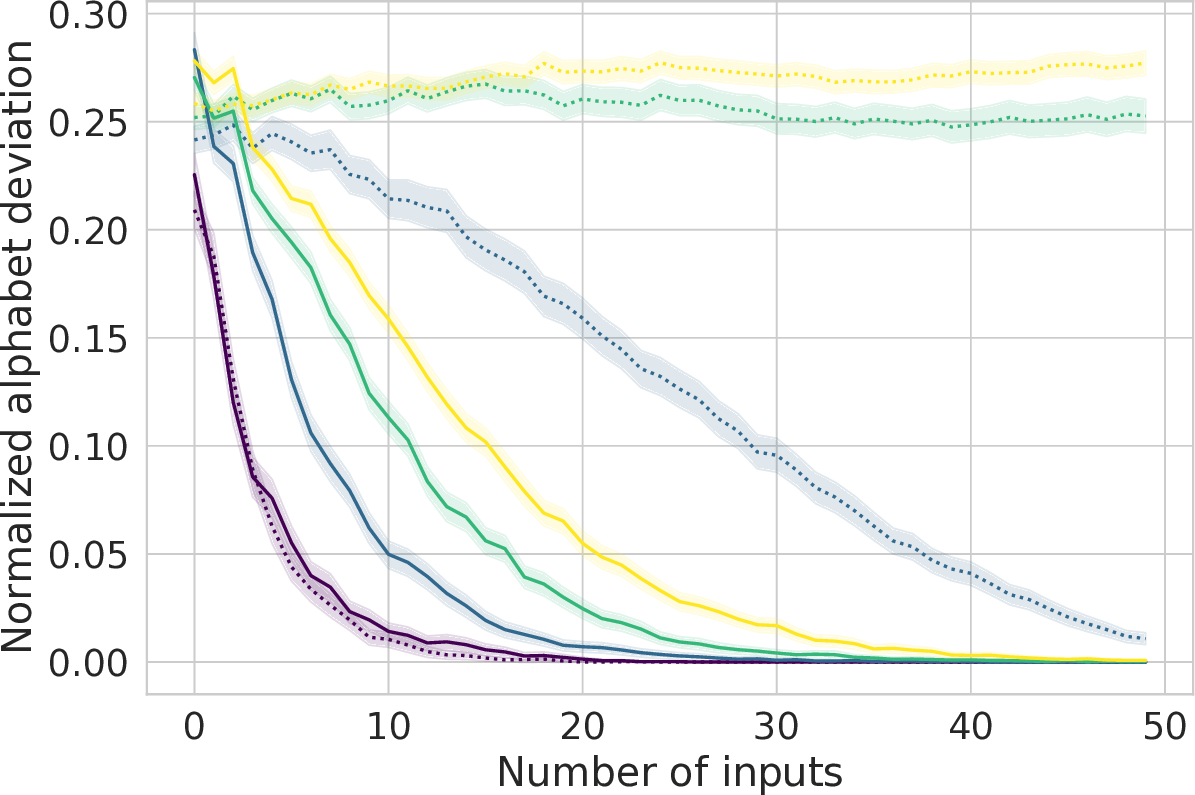}
			\caption{}
			\label{subfig:static_alphnormL1_stand}
		\end{subfigure}
		\\
		\vspace{2mm}
		\raisebox{17mm}{\rotatebox[origin=c]{90}{\footnotesize\textbf{Non-stationary errors}}}\quad%
		\begin{subfigure}[t]{\sfw\textwidth}
			\centering
			\includegraphics[width=\textwidth]{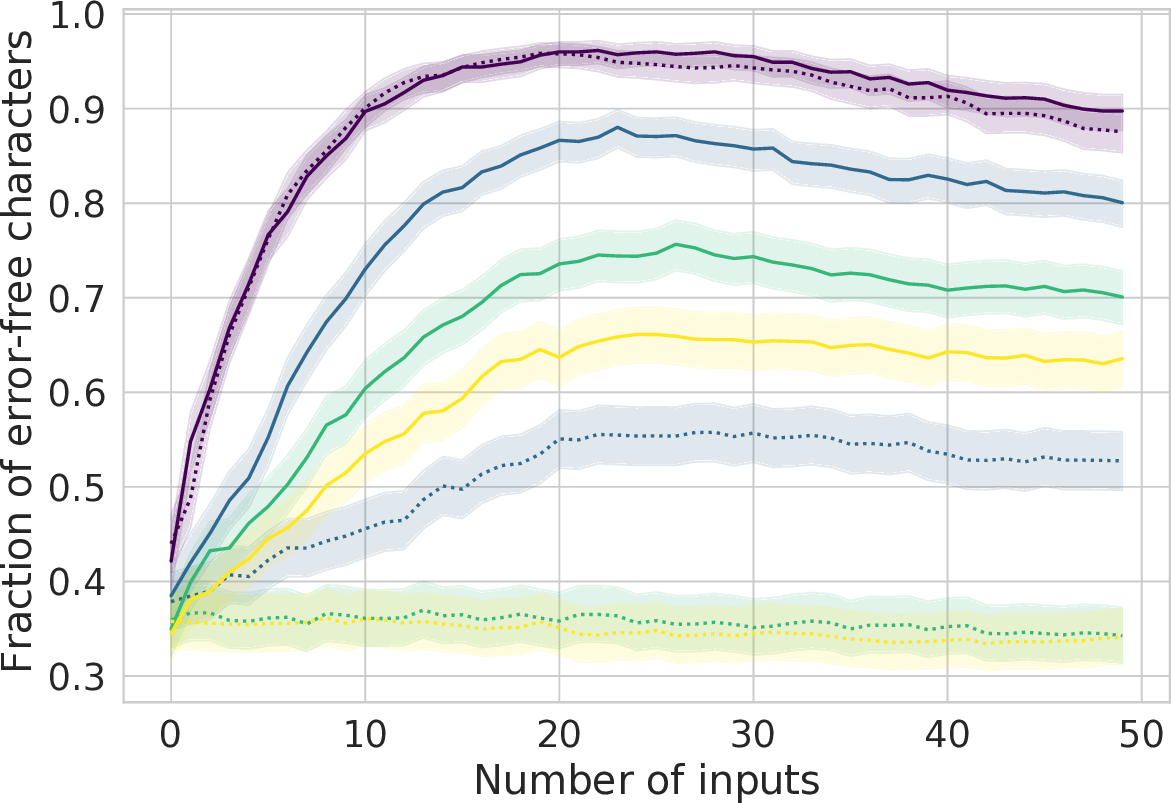}
			\caption{}
			\label{subfig:dynamic_alphacc_stand}
		\end{subfigure}%
		\hfill
		\begin{subfigure}[t]{\sfw\textwidth}
			\centering
			\includegraphics[width=\textwidth]{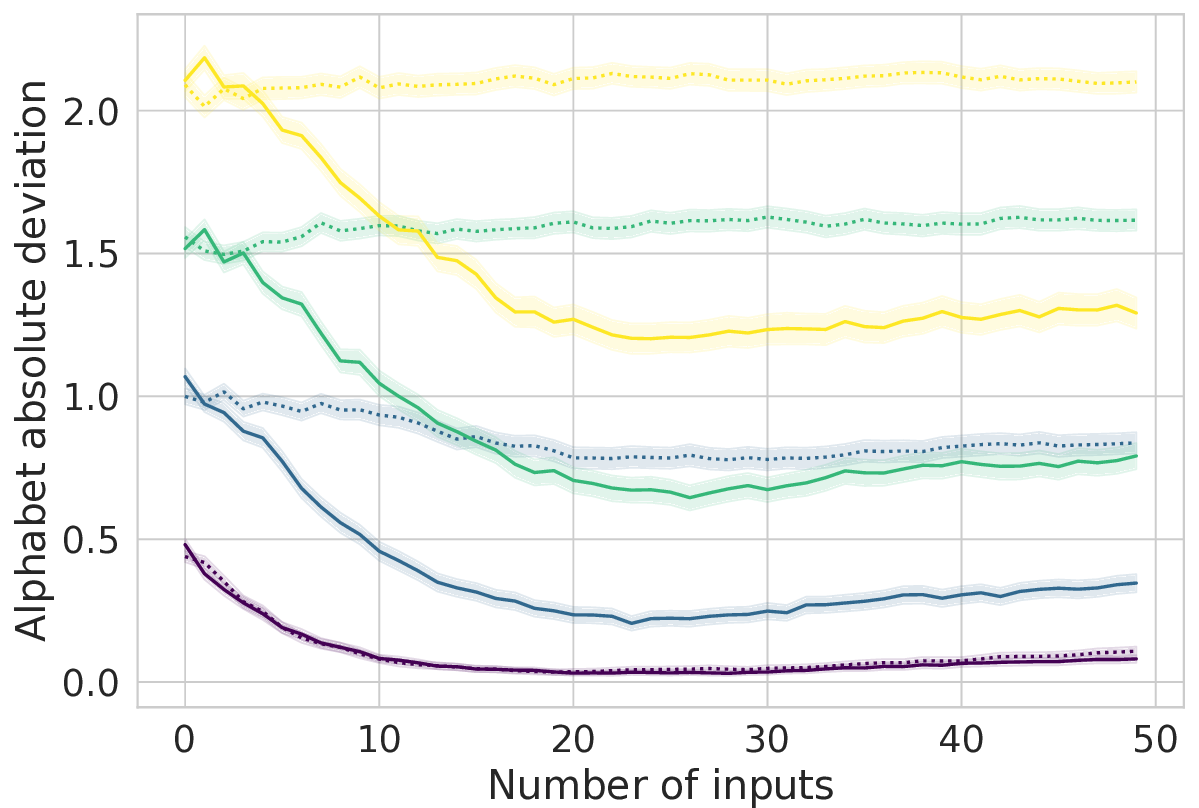}
			\caption{}
			\label{subfig:dynamic_alphL1_stand}
		\end{subfigure}%
		\hfill
		\begin{subfigure}[t]{\sfw\textwidth}
			\centering
			\includegraphics[width=\textwidth]{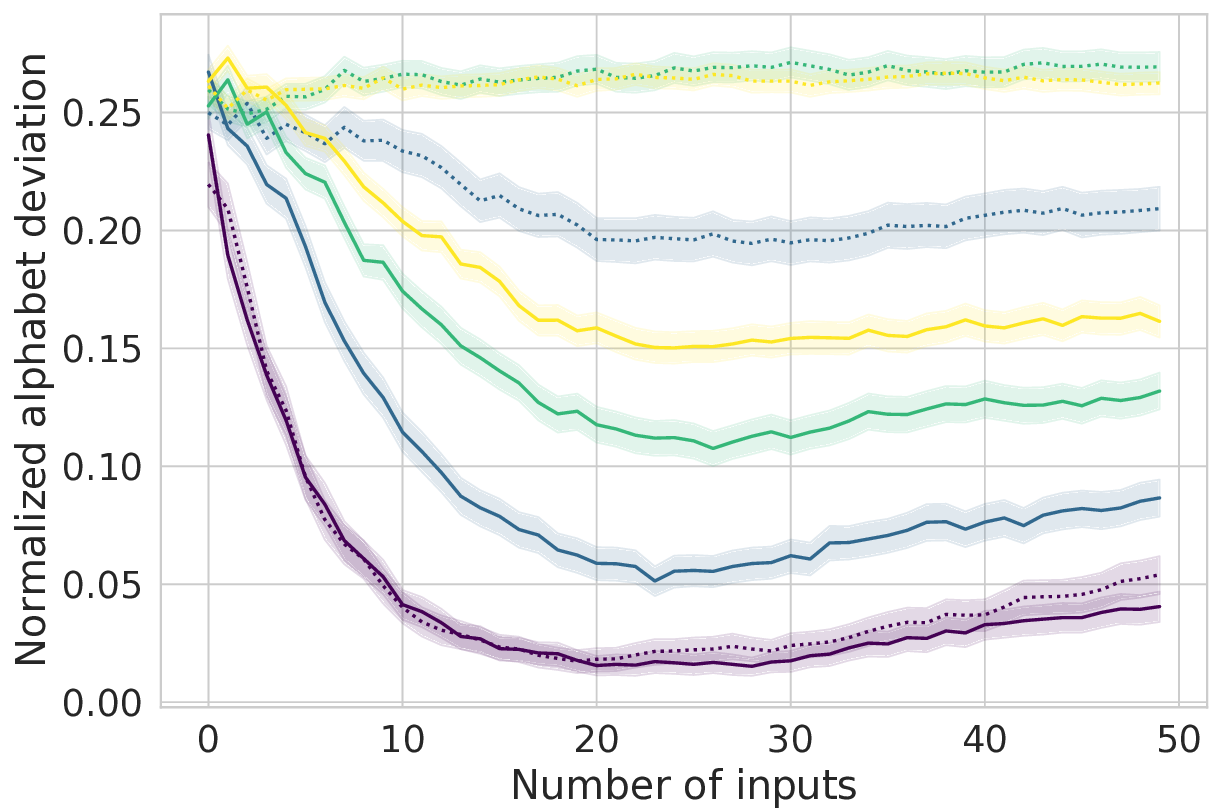}
			\caption{}
			\label{subfig:dynamic_alphnormL1_stand}
		\end{subfigure}%
	\end{minipage}
	\caption[Alphabet-wise performance metrics as a function of number of inputs and dictionary size across both fixed and non-stationary input errors, with standard degrees of freedom estimates.]{Alphabet-wise performance metrics as a function of number of inputs and dictionary size across both fixed and non-stationary input errors, with standard degrees of freedom estimates. Alphabet accuracy (\textbf{a,d}) measures the fraction of error-free guessed characters, plotted at its mean value with a 95\% Wilson confidence interval. Alphabet deviation (\textbf{b,e}) measures the sum of absolute deviations \emph{within} each alphabet of a configuration guess, plotted at its mean value with error bars depicting 95\% bootstrap confidence intervals over 10,000 samples (separate resampling for every number of inputs). Normalized alphabet deviation (\textbf{c,f}) calculates alphabet deviation, but normalizes each value by the number of degrees of freedom, plotted at its mean value with error bars depicting 95\% bootstrap confidence intervals over 10,000 samples (separate resampling for every number of inputs).}
	\label{fig:alphstats_stand}
\end{figure}

\begin{figure}[htb]
	\def\sfw{0.3}
	\centering
	\includegraphics[width=0.9\textwidth]{tradeoff_sns_dynamic1_condof_1_legend.png}
	
	\tradeoffheadingthree
	
	\begin{minipage}{\textwidth}
		\centering
		\raisebox{18mm}{\rotatebox[origin=c]{90}{\footnotesize\textbf{Fixed (10\%) error}}}\quad%
		\begin{subfigure}[t]{\sfw\textwidth}
			\centering
			\includegraphics[width=\textwidth]{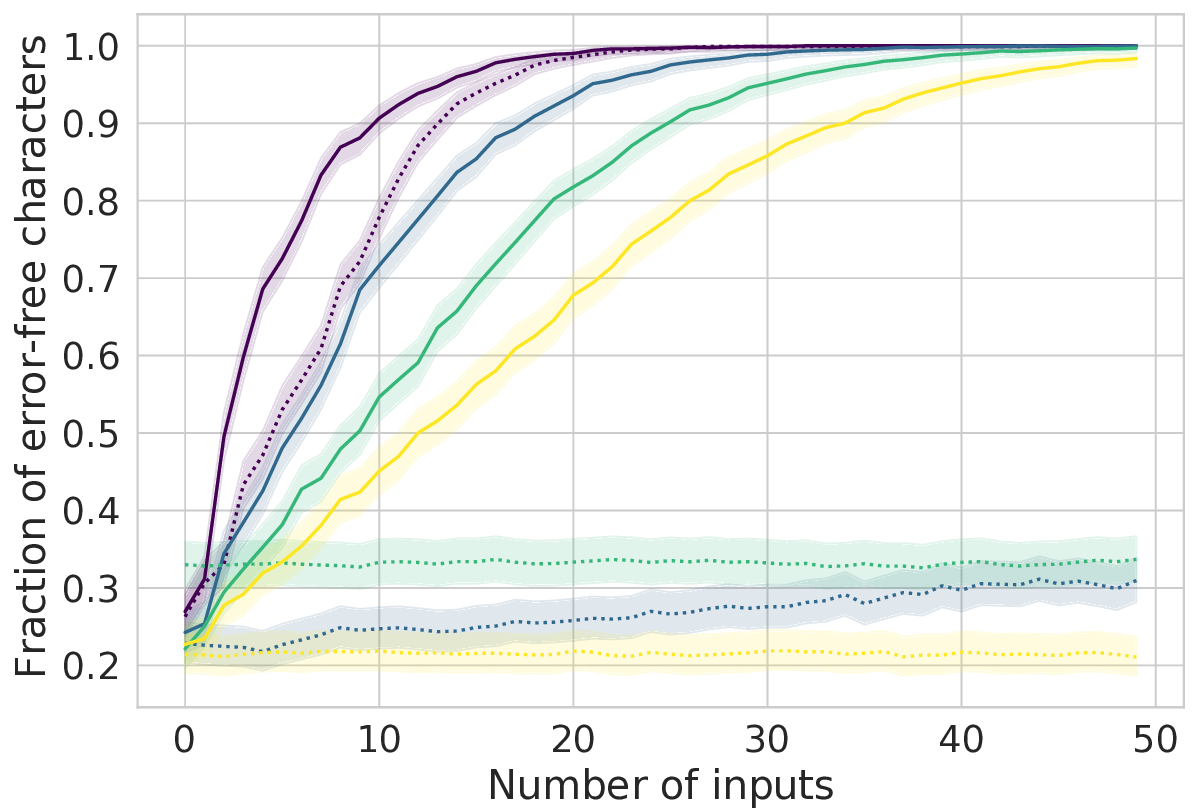}
			\caption{}
			\label{subfig:static_alphacc_cons}
		\end{subfigure}%
		\hfill
		\begin{subfigure}[t]{\sfw\textwidth}
			\centering
			\includegraphics[width=\textwidth]{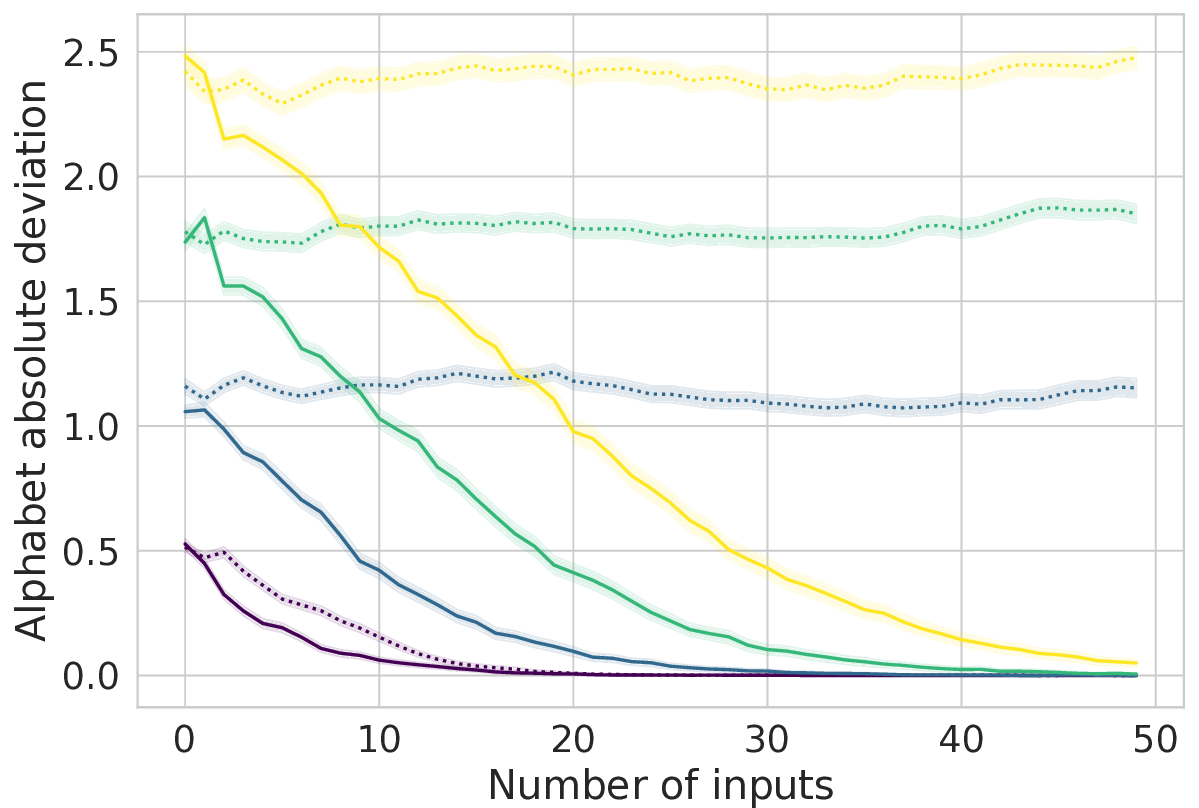}
			\caption{}
			\label{subfig:static_alphL1_cons}
		\end{subfigure}%
		\hfill
		\begin{subfigure}[t]{\sfw\textwidth}
			\centering
			\includegraphics[width=\textwidth]{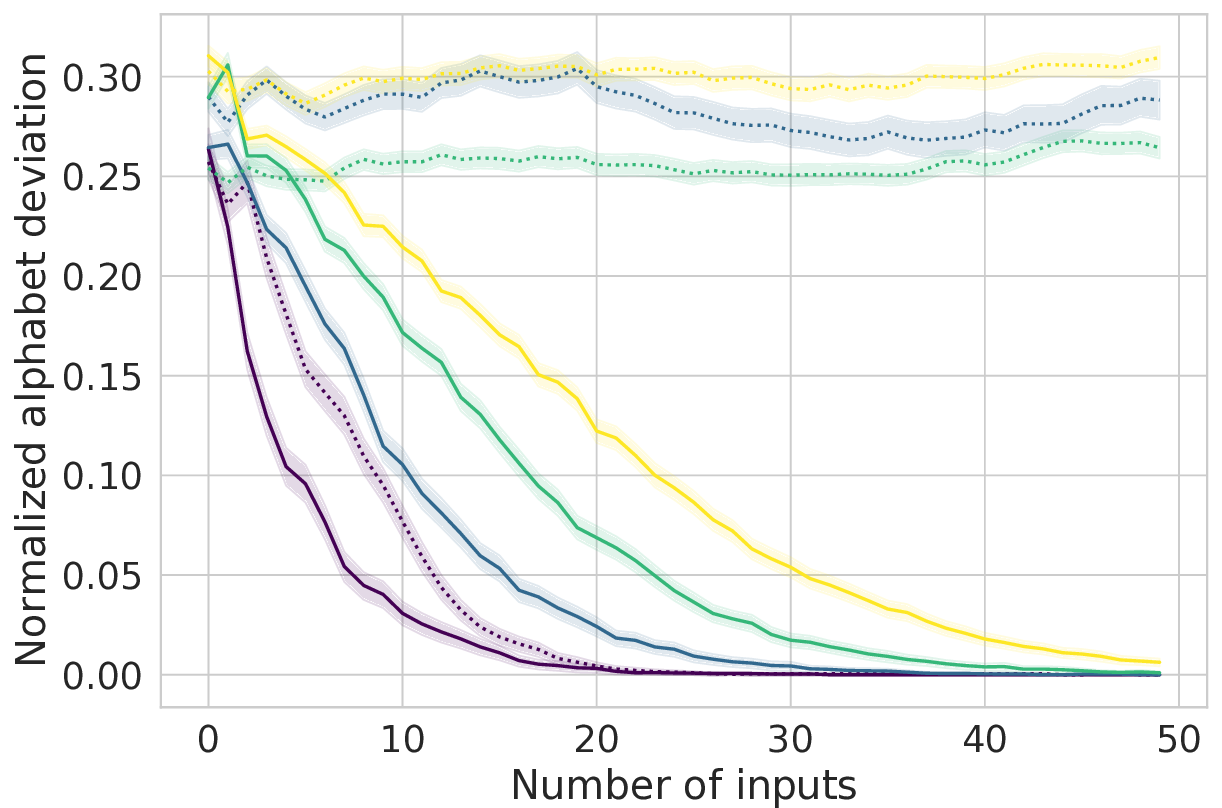}
			\caption{}
			\label{subfig:static_alphnormL1_cons}
		\end{subfigure}
		\\
		\vspace{2mm}
		\raisebox{17mm}{\rotatebox[origin=c]{90}{\footnotesize\textbf{Non-stationary errors}}}\quad%
		\begin{subfigure}[t]{\sfw\textwidth}
			\centering
			\includegraphics[width=\textwidth]{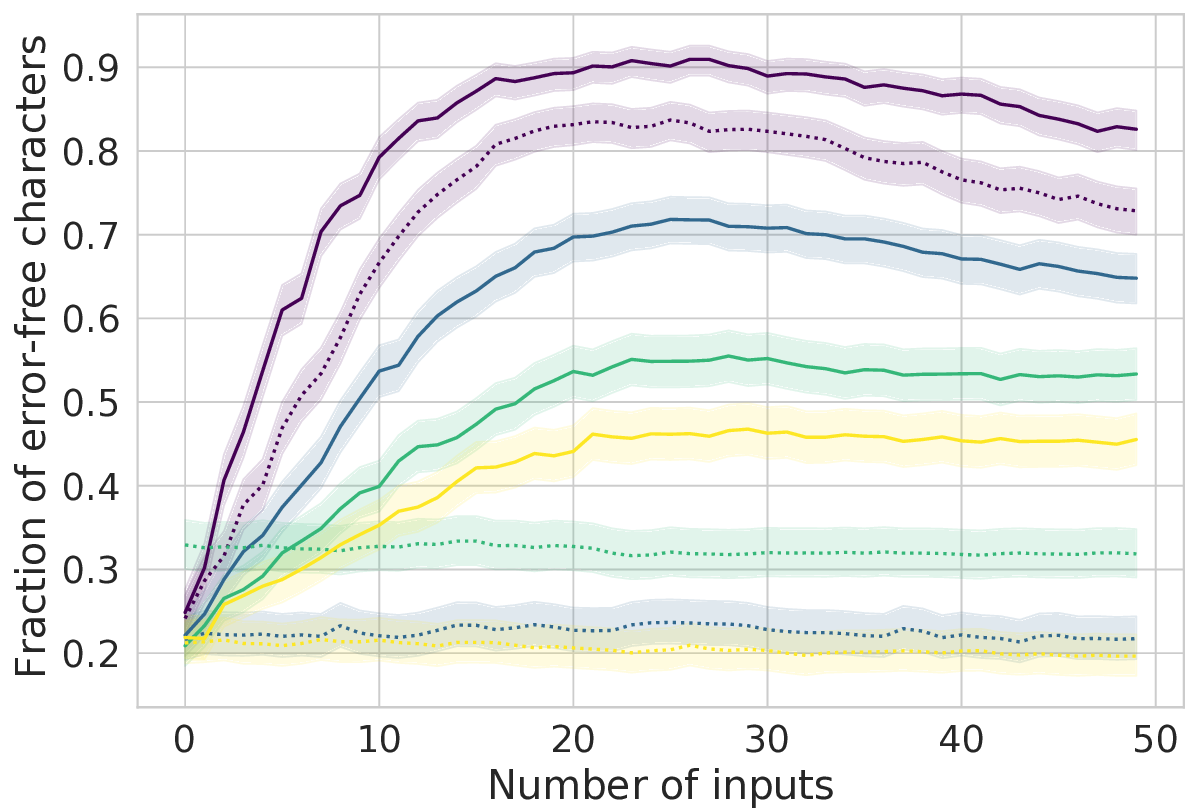}
			\caption{}
			\label{subfig:dynamic_alphacc_cons}
		\end{subfigure}%
		\hfill
		\begin{subfigure}[t]{\sfw\textwidth}
			\centering
			\includegraphics[width=\textwidth]{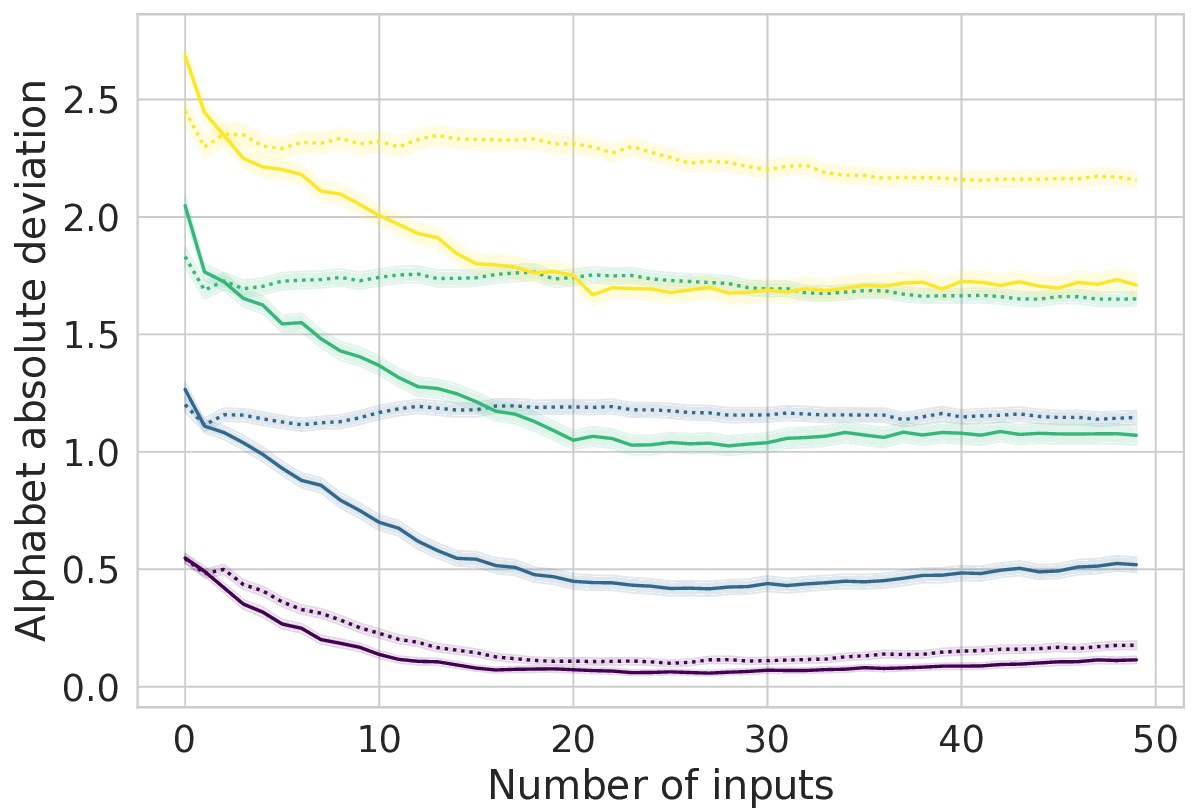}
			\caption{}
			\label{subfig:dynamic_alphL1_cons}
		\end{subfigure}%
		\hfill
		\begin{subfigure}[t]{\sfw\textwidth}
			\centering
			\includegraphics[width=\textwidth]{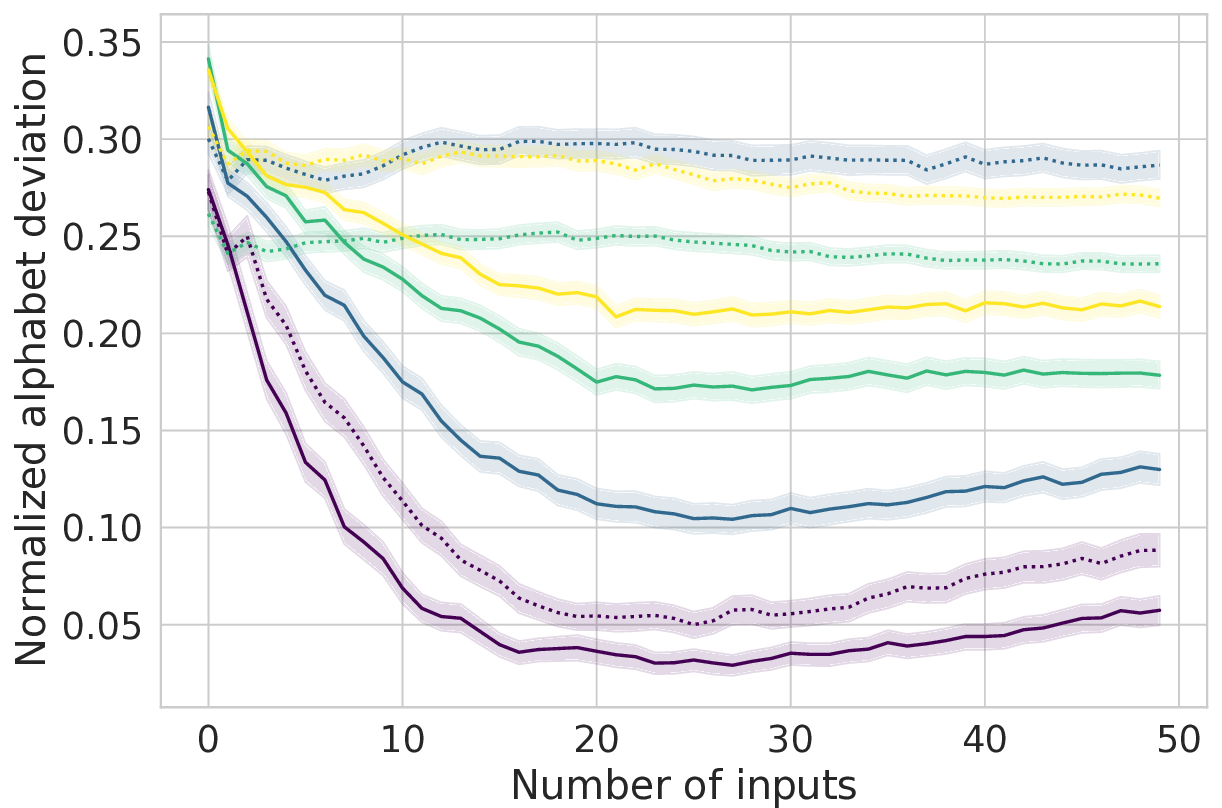}
			\caption{}
			\label{subfig:dynamic_alphnormL1_cons}
		\end{subfigure}%
	\end{minipage}
	\caption[Alphabet-wise performance metrics as a function of number of inputs and dictionary size across both fixed and non-stationary input errors, with conservative degrees of freedom estimates.]{Alphabet-wise performance metrics as a function of number of inputs and dictionary size across both fixed and non-stationary input errors, with conservative degrees of freedom estimates. All plotted metrics and error bars are otherwise identical to those in \Cref{fig:alphstats_stand}.}
	\label{fig:alphstats_cons}
\end{figure}

In \Cref{fig:alphstats_stand,fig:alphstats_cons}, we evaluate additional simulation performance metrics based on alphabet-wise convergence for both standard and conservative degrees of freedom estimates. These metrics evaluate the performance of each algorithm in driving each \emph{individual} character to its correct target rather than measuring convergence of the entire string at once, and provide another perspective on alphabet-wise convergence not captured by the latter.

Expanding on notation from the \Cref{sec:PMimp}, let $\widehat{\bm{z}} = \{\widehat{\sigma}^i\}_{i=1}^r$ denote a configuration estimate with $i$th character $\widehat{\sigma}^i \in 1,2,\dots b$, and let $\bm{z}_t = \{\sigma^i_t\}_{i=1}^r$ denote the target configuration. Both configurations have $r$ degrees of freedom each with alphabets of size $b$. Our first alphabet performance metric is ``alphabet accuracy,'' which for a single configuration estimate measures the fraction its characters that equal the corresponding characters in the target configuration:
\[\operatorname{AlphAcc}(\widehat{\bm{z}},\bm{z}_t) = \frac{1}{r}\sum_{i=1}^r \delta(\widehat{\sigma}^i,\sigma^i_t),\]
where $\delta(x,y) = 1$ if $x=y$, and 0 otherwise. \Cref{subfig:static_alphacc_stand,subfig:dynamic_alphacc_stand,subfig:static_alphacc_cons,subfig:dynamic_alphacc_cons} plot alphabet accuracy averaged over multiple trials, calculated with respect to the instantaneous MAP configuration estimate after every number of inputs.

Next, we calculate ``alphabet deviation,'' which in a similar spirit to absolute deviation (or dictionary distance) calculates the sum of absolute deviations within each \emph{alphabet}. As in dictionary distance, the alphabet deviation measures the rate of convergence of each alphabet character to its respective target character, rather than measuring a binary notion of correct or incorrect convergence. Alphabet deviation is calculated as
\[\operatorname{AlphDev}(\widehat{\bm{z}},\bm{z}_t) = \sum_{i=1}^r \biggl\lvert \frac{\widehat{\sigma}^i - \sigma^i_t}{b} \biggr\rvert.\]
We normalize the absolute deviation within each alphabet by the alphabet size $b$ such that the resulting metric has a range between $0$ and $r$ that does not depend on alphabet size. \Cref{subfig:static_alphL1_stand,subfig:dynamic_alphL1_stand,subfig:static_alphL1_cons,subfig:dynamic_alphL1_cons} plot alphabet deviation averaged over multiple trials, calculated with respect to the instantaneous MAP configuration estimate after every number of inputs.

For completeness, we also calculate a normalized version of alphabet deviation which we call ``normalized alphabet deviation.'' Normalized alphabet deviation is simply equal to alphabet deviation normalized by the number of degrees of freedom. This yields an alphabet deviation metric that scales between 0 and 1 for any number of degrees of freedom $r$, and allows for a more equalized comparison between different degrees of freedom.
\[\operatorname{NormAlphDev}(\widehat{\bm{z}},\bm{z}_t) = \frac{1}{r} \sum_{i=1}^r \biggl\lvert \frac{\widehat{\sigma}^i - \sigma^i_t}{b} \biggr\rvert.\]
\Cref{subfig:static_alphnormL1_stand,subfig:dynamic_alphnormL1_stand,subfig:static_alphnormL1_cons,subfig:dynamic_alphnormL1_cons} plot normalized alphabet deviation averaged over multiple trials, calculated with respect to the instantaneous MAP configuration estimate after every number of inputs.

\Cref{fig:alphstats_stand} depicts these alphabet-wise metrics for the standard degrees of freedom estimate across both fixed and non-stationary errors. The most direct point of comparison for alphabet accuracy in \Cref{subfig:static_alphacc_stand,subfig:dynamic_alphacc_stand} is error-free accuracy with respect to the entire string, as depicted in \Cref{subfig:staticacc,subfig:dynamicacc}. In the fixed error case, both string-wise and alphabet-wise metrics capture similar behavior. In the case of non-stationary errors, measuring alphabet accuracy appears to penalize larger numbers of degrees of freedom less harshly than accuracy with respect to the entire string. Intuitively, alphabet accuracy accounts for the fact that early characters may have converged successfully to the target while later characters are still being refined, which is a subtlety that is ignored when assessing error-free accuracy at the string level.

Alphabet deviation and normalized alphabet deviation depict similar trends to one another; we focus on normalized alphabet deviation due to its attractive normalization between different degrees of freedom. When comparing normalized alphabet deviation in \Cref{subfig:static_alphnormL1_stand,subfig:dynamic_alphnormL1_stand} to string-wise dictionary distance in \Cref{subfig:staticL1,subfig:dynamicL1}, we can observe characteristics of normalized alphabet deviation not captured by dictionary distance. In particular, since posterior matching operates through bisections of the entire dictionary, it is able to quickly refine a configuration's equivalent point on the unit interval to the correct numerical neighborhood, regardless of number of degrees of freedom (see \Cref{sec:PMimp}). This is reflected in dictionary distance by the clustering of all degrees of freedom at low deviation values in \Cref{subfig:staticL1,subfig:dynamicL1}. Unlike dictionary distance, normalized absolute deviation accounts for individual convergence within each alphabet, which may still be large for later characters even if posterior matching has converged well when measured by overall dictionary distance. As can be observed in \Cref{subfig:static_alphnormL1_stand,subfig:dynamic_alphnormL1_stand}, larger numbers of degrees of freedom require more inputs to refine character selection within a larger number of alphabets. Similar trends as these can be observed when using conservative degrees of freedom estimates in \Cref{fig:alphstats_cons} and comparing to the corresponding string-wise metrics in \Cref{fig:generalizing_condof}.

\end{document}